\documentclass [12pt] {article}
\usepackage{a4}

\usepackage{amsfonts}
\usepackage{amsmath}

\chardef\kuvioletter=\catcode`\@\catcode`\@=11
\chardef\K@aa=\catcode`\? \catcode`\?=7
\gdef\K@ab{\global\let\K@ab\kuviobye}%
\gdef\kuviobye{\gdef\D@kuviolevel{0}%
\K@ac\@\kuvioletter\K@ac\?\K@aa
\K@ac\~\K@ad\K@ac\??Z\K@ae
\K@ac\??Y\K@af\K@ac\??X\K@ag
\K@ac\??W\K@ah\K@ac\??V\K@ai
\K@ac\??U\K@aj\K@ac\??T\K@ak
\K@ac\??S\K@al\K@ac\??R\K@am
\K@ac\??Q\K@an\K@ac\??P\K@ao
\K@ac\??O\K@ap\K@ac\??N\K@aq
\K@ac\??A\K@ar
 }%
\gdef\K@as#1#2{\chardef#2=\catcode`#1 \catcode`#1=14 }%
\K@as\~\K@ad\K@as\??Z\K@ae
\K@as\??Y\K@af\K@as\??X\K@ag
\K@as\??W\K@ah\K@as\??V\K@ai
\K@as\??U\K@aj\K@as\??T\K@ak
\K@as\??S\K@al\K@as\??R\K@am
\K@as\??Q\K@an\K@as\??P\K@ao
\K@as\??O\K@ap\K@as\??N\K@aq
\K@as\??A\K@ar
\ifx\D@kuvioloaded\relax
\ifcase\D@kuviolevel\relax
\gdef\K@ab{\message{again}\kuviobye}%
\or\relax
\or\catcode`\??Z=9 \or\catcode`\??Y=9 \or\catcode`\??X=9
\or\catcode`\??W=9 \or\catcode`\??V=9 \or\catcode`\??U=9
\or\catcode`\??T=9 \or\catcode`\??S=9 \or\catcode`\??R=9
\or\catcode`\??Q=9 \or\catcode`\??P=9 \or\catcode`\??O=9
\or\catcode`\??N=9 \or\catcode`\??A=9
\fi
\else
\chardef\K@at=\catcode`\( \chardef\K@au=\catcode`\_
\chardef\K@av=\catcode`\^ \chardef\K@aw=\catcode`\<
\chardef\K@ax=\catcode`\> \chardef\K@ay=\catcode`\:
\chardef\K@az=\catcode`\/ \chardef\K@ba=\catcode`\*
\chardef\K@bb=\catcode`\| \gdef\D@kuviolevel{1}%
\catcode`\~=9 \let\D@kuvioloaded\relax\global\let\K@ab\relax
\fi
\K@ab
~\def\kuviocs#1{%
~{\escapechar=-1
~\expandafter\ifx\csname D@\string#1@k@D\endcsname\relax
~\xdef\K@bc{\string#1}\else\xdef\K@bc{D@\string#1@k@D}\fi
~}\csname\K@bc\endcsname}
~\def\nonkuviocs#1{%
~{\escapechar=-1
~\expandafter\ifx\csname D@\string#1@nk@D\endcsname\relax
~\xdef\K@bc{\string#1}\else\xdef\K@bc{D@\string#1@nk@D}\fi
~}\csname\K@bc\endcsname}
~\def\K@bd#1#2{%
~{\escapechar=-1
~\expandafter\ifx\csname\string#2\endcsname\relax
~\xdef\K@bc{\string#2}\else
~\aftergroup\K@be\aftergroup#2\xdef\K@bc{D@\string#2@k@D}\fi
~}\def\K@bf{#1}\expandafter\K@bf\csname\K@bc\endcsname}
~\def\K@bg{\K@bd\def}
~\def\K@bh{\K@bd\let}
~\def\K@bi#1#2{%
~{\escapechar=-1
~\expandafter\ifx\csname\string#2\endcsname\relax
~\else\expandafter\expandafter\expandafter\let\expandafter\expandafter
~\csname D@\string#2@nk@D\endcsname\csname\string#2\endcsname
~\aftergroup\K@bj\aftergroup#2\fi
~\xdef\K@bc{\string#2}%
~}\def\K@bf{#1}\expandafter\K@bf\csname\K@bc\endcsname}
~\def\K@bk{\K@bi\def}
~\def\K@bl{\K@bi\let}
~\def\K@bm#1#2{%
~{\escapechar=-1
~\expandafter\ifx\csname\string#1\endcsname\relax
~\xdef\K@bc{\string#1}\else
~\aftergroup\K@be\aftergroup#1\xdef\K@bc{D@\string#1@k@D}\fi
~\expandafter\ifx\csname D@\string#2@k@D\endcsname\relax
~\xdef\K@bf{\string#2}\else\xdef\K@bf{D@\string#2@k@D}\fi
~}\expandafter\expandafter\expandafter\let\expandafter\expandafter
~\csname\K@bc\endcsname\csname\K@bf\endcsname}
~\let\kuviodef\K@bg\let\rekuviodef\K@bk
~\let\kuviolet\K@bh\let\rekuviolet\K@bl
~\let\kuviodefine\K@bd\let\rekuviodefine\K@bi
~\def\kuvioforce#1{%
~{\escapechar=-1
~\expandafter\ifx\csname D@\string#1@k@D\endcsname\relax
~\else\xdef\K@bc{\let\noexpand#1\expandafter\noexpand
~\csname D@\string#1@k@D\endcsname\let\expandafter\noexpand
~\csname D@\string#1@k@D\endcsname\relax}\aftergroup\K@bc\fi
~}}
~\def\K@be#1{\immediate\write16
~{** \ifnum\D@kuviolevel=1 \else(kuvio.tex) \fi
~\string#1\K@bn already defined: use \string\kuviocs\string#1}}
~\def\K@bj#1{\immediate\write16
~{** \ifnum\D@kuviolevel=1 \else(kuvio.tex) \fi
~\string#1\K@bn redefined}}
~\def\K@bn{ }
~{\newlinechar=`\!\catcode`\ =13 \let \K@bn
~\immediate\write16{***************************************************
~!**  Diagrams for dvips users                     **
~!**  patchlevel 216, 19 Feb 1996                  **
~!**  Anders G S Svensson  <svensson@math.ubc.ca>  **
~!***************************************************}}
~\def\K@ac#1#2{\catcode`#1=#2\relax}\def\K@bo{\K@bp{2}{rotation}\global\let
~\K@bo\relax}\def\K@bq{\K@bp{3}{frames and shades}\global\let\K@bq\relax
~}\def\K@br{\K@bp{4}{figures and graphs}\global\let\K@br\relax}\def\K@bs
~{\K@bp{5}{shifts}\global\let\K@bs\relax}\def\K@bt{\K@bp{6}{diagrams}\global
~\let\K@bt\relax}\def\K@bu{\K@bp{7}{kdg files}\global\let\K@bu\relax}\def
~\K@bv{\K@bp{8}{kuv files}\global\let\K@bv\relax}\def\K@bw{\K@bp
~{9}{modification cells}\global\let\K@bw\relax}\def\K@bx{\K@bp{10}{text
~modifications}\global\let\K@bx\relax}\def\K@by{\K@bp
~{11}{modifications}\global\let\K@by\relax}\def\K@bz{\K@bp{12}{rotated
~labels}\global\let\K@bz\relax}\def\K@ca{\K@bp{13}{points}\global\let\K@ca
~\relax}\def\K@cb{\K@bp{14}{box and frame modifications}\global\let\K@cb
~\relax}\def\K@cc{\K@bp{15}{dots}\global\let\K@cc\relax}\def\K@bp#1#2{\gdef
~\D@kuviolevel{#1}\gdef\K@cd{\message{#2}}\relax}\def
~\kuviofmt{\xdef\K@bc{\everyjob={\the\everyjob\message{** kuvio.tex (p216)
~Anders G S Svensson **}}}\K@bc\K@bo\K@bq\K@br\K@bs\K@bt\K@bu\K@bv\K@bw
~\K@bx\K@by\K@bz\K@ca\K@cb\K@cc\let\kuviofmt\relax}\def\K@ce{\K@cd\catcode
~`\(=\K@at\catcode`\_=\K@au\catcode`\^=\K@av\catcode`\<=\K@aw\catcode
~`\>=\K@ax\catcode`\:=\K@ay\catcode`\/=\K@az\catcode`\*=\K@ba\catcode
~`\|=\K@bb\let\K@cf\let\def\let{\global\K@cf}\K@cf\def\gdef}\K@bk
~\dvipsspecial#1{\special{ps:#1}}\def\kuviospecial{\let\K@cg}\kuviospecial
~\dvipsspecial\def\kuviopatchlevel{216}\def
~\kuviodate{19 Feb 1996}\newhelp\K@ch{Expect chaos to ensue.}\def
~\kuviorequire#1{\relax\ifnum#1>\kuviopatchlevel\relax\errhelp=\K@ch
~\errmessage{Hmm. Looks like I'll need a newer patchlevel of kuvio.tex.}\fi
~}\def\K@ci{}\def\K@cj#1{}\def\K@ck#1#2{}\def\K@cl{{}}\def\K@cm{0}\def\K@cn
~{.5}\def\K@co{1}\def\K@cp{-1}\def\K@cq#1#2{#1#2}\def\K@cr#1#2{#2#1}\def
~\K@cs#1#2{{#2}{#1}}\def\K@ct#1#2{\let\K@bf#1\let#1#2\let#2\K@bf}\def\K@cu
~#1\K@cv{\def\K@cw{#1}\K@cx}\def\K@cx{\K@cw\let\K@cy\K@cx\else\let\K@cy
~\relax\fi\K@cy}\let\K@cv\fi\newif\ifkuvio@\def\K@cz{\relax\kuvio@false
~}\newwrite\K@da\newwrite\K@db\newread\K@dc\newread\K@dd\newcount\K@de
~\newcount\K@df\newcount\K@dg\newcount\K@dh\newcount\K@di\newcount\K@dj
~\newcount\K@dk\newcount\K@dl\newdimen\K@dm\newdimen\K@dn\newdimen\K@do
~\newdimen\K@dp\newtoks\K@dq\newtoks\K@dr\newbox\K@ds\newbox\K@dt
~\newbox\K@du
??V\begingroup\K@ce\def\K@dv#1{\bgroup\K@dw\K@dx{\K@dy\K@dr}{\the\K@dr}\def
??V\K@dz{\egroup\egroup\egroup\K@ea}\def\Modify{\egroup\egroup\egroup\K@by
??V\K@eb}\let\K@ec\relax\let\K@ed\relax\let\K@ee\relax\K@ef\setbox\K@du
??V=\hbox\bgroup\vbox\bgroup\halign\bgroup&\K@eg\K@eh{##}\cr}\def\endDiagram
??V{\K@dz\egroup}%
~\def\K@ei{\K@dv}\K@bk\Diagram{\bgroup\K@bt\K@ej\K@ek\K@el\everyDiagram
~\K@ei{Diagram}}\K@bl\everyDiagram\relax
??V\def\K@ea{\K@em\global\setbox\K@ds=\hbox{\hbox to \K@en sp{\hfil\vbox to
??V\K@eo sp{\vfil}}\K@ep\K@eq\K@er}\K@dx{}{{}{1}{\gdef\noexpand\K@es{\K@es
??V}\gdef\noexpand\K@et{\K@et}\gdef\noexpand\K@eu{\K@eu}\gdef\noexpand\K@ev
??V{\K@ev}}}\K@ew\K@ex\K@ey\egroup\Displaybox\K@ds}%
~\let\K@ew\relax\let\K@dw\relax\let\K@dx\K@ck\K@bg\compileto{\bgroup\K@ez
~\K@fa}\def\K@fa#1{\egroup\def\K@ej{\K@fb{#1}\global\let\K@ej\relax\let
~\K@ek\relax}}\def\K@ez{\catcode`\ =9 \catcode`\??I=9 }\K@bg\recompileto
~{\bgroup\K@ez\K@fc}\def\K@fc#1{\egroup\def\K@ej{\K@fd{#1}\global\let\K@ej
~\relax\let\K@ek\relax}}\K@bg\recompile{\let\K@fb\K@fd}\K@bg\nocompile{\let
~\K@fe\K@ck}\K@bg\forcekdg{\def\K@el{\def\K@ei{\K@ff}\def\K@fg{\K@fh}\def
~\K@fi{\K@fj}\K@bu}}\let\K@el\relax\def\K@fb#1{\K@fe0{#1}}\def\K@fd#1{\K@fe
~1{#1}}\K@bg\autocompileto{\bgroup\K@ez\K@fk}\def\K@fk#1{\egroup\K@fl
~\autocompilecounter\def\K@ek{\global\advance\autocompilecounter by 1 \K@fb
~{#1-\the\autocompilecounter}}}\K@bg\autocompile{\kuviocs\autocompileto
~{.\jobname}}\let\K@ej\relax\let\K@ek\relax\edef\K@fl#1{\noexpand\newcount
~#1\gdef\noexpand\K@fl##1{\global##1=0 }}\def\K@fe#1#2{\let\K@el\relax\gdef
~\K@fm{#2}\ifcase#1 \immediate\openin\K@dc=#2.kuv \def\K@fn{\K@fd{#2}}\ifeof
~\K@dc\immediate\closein\K@dc\def\K@bc{\K@fd{#2}}\else\def\K@dw{\K@fo
~{reading #2.kuv}}\def\K@ei{\K@fp}\def\K@fg{\K@fq}\def\K@fi{\K@fr}\let\K@bc
~\K@bv\fi\or\immediate\openout\K@da=#2.kuv \let\K@dx\K@fs\def\K@dw{\K@fo
~{compiling #2.kuv}}\def\K@ei{\K@ff}\def\K@fg{\K@fh}\def\K@fi{\K@fj}\let
~\K@bc\K@bu\let\K@ew\K@ft\fi\K@bc}\def\K@fm{\jobname}\def\K@fs
~#1#2{{#1\immediate\write\K@da{#2}}}\def\K@ft{\immediate\closeout\K@da\K@fu
~{done}}\def\K@fo#1{\message{(#1}}\def\K@fu#1{\message{#1)}}
??U\begingroup\K@ce\def\K@ff{\K@fv\K@dv}\def\K@fh{\K@fv\K@fw}\def\K@fv
??U#1#2{\let\K@fo\message\let\K@fu\message\def\K@fx{\K@fy{\relax
??U}\expandafter\def\csname#2\endcsname{#1{}}\input\K@fm.kdg \K@fy
??U{}}\expandafter\expandafter\expandafter\K@fz\expandafter\expandafter
??U\csname#2\endcsname\csname end#2\endcsname\K@fx}%
~\def\K@ga#1#2#3#4{\def\K@bc##1#3{\def\K@gb{#1#2##1#3}\gdef\K@fy####1{\K@gc
~{#1#2##1#3####1}}#4}\K@bc}\def\K@fz{\K@ga{}}\def\K@gd{\K@ga{\relax}}\def
~\K@ge{\K@ga{\relax\relax}}\def\K@gc#1{\immediate\openout\K@db=\K@fm.kdg
~\K@dr={#1}\immediate\write\K@db{\the\K@dr}\immediate\closeout\K@db}%
??U\def\K@fj#1{\let\K@fo\message\let\K@fu\message\expandafter\expandafter
??U\expandafter\K@gf\expandafter\expandafter\csname#1\endcsname\csname
??Uend#1\endcsname{\K@gg{#1}}}%
~\let\K@gf\K@fz
??U\def\K@gg#1{\expandafter\expandafter\expandafter\K@gh\expandafter
??U\expandafter\csname end#1\endcsname\K@gb\Modify\K@gi\expandafter\def
??U\csname#1\endcsname{\K@gj{}}\K@fy{\relax}\input\K@fm.kdg \K@fy{}}\endgroup
~\def\K@gk#1#2#3#4#5\Modify#6\K@gi{\def\K@gl{#6}\ifx\K@gl\K@ci\def\K@bc
~##1#1{\def\K@gm####1{####1={\vbox {\offinterlineskip\hbox{$##1$}\null
~}}}}\K@bc#5\else\def\K@gm##1{##1={\vbox {\offinterlineskip\hbox
~{$#5$}\null}}}\fi}\let\K@gn\K@gk\def\K@go{\K@gk{}}\def\K@gh{\K@gk{}{}}%
??T\begingroup\K@ce\def\K@fp{\K@gp\K@gq\K@ff\K@dv}\def\K@fq{\K@gp\K@gr\K@fh
??T\K@fw}\def\K@fr{\let\K@gs\K@gt\K@gp\K@gu\K@fj\K@gj}\def\K@gp
??T#1#2#3{\immediate\read\K@dc to\K@bf\ifx\K@bf\K@gv\def\K@bc{\K@gs
??T#1#3}\else\def\K@bc{\immediate\closein\K@dc\K@fn#2}\fi\K@bc}%
~\def\kuviorevision#1{\def\K@dy##1{##1={{kuvio}{216}{#1}}}\def\K@gv
~{{kuvio}{216}{#1} }}\kuviorevision{}
??T\def\K@gs#1#2#3{\def\K@fx{\K@gw#1#2{#3}}\expandafter\expandafter
??T\expandafter\K@fz\expandafter\expandafter\csname#3\endcsname\csname
??Tend#3\endcsname\K@fx}\def\K@gw#1#2#3{\def\K@bc{\K@fy{\relax}\expandafter
??T\def\csname#3\endcsname{\K@fn#2{}}\input\K@fm.kdg \K@fy{}}\immediate\openin
??T\K@dd=\K@fm.kdg \ifeof\K@dd\immediate\closein\K@dd\immediate\closein\K@dc
??T\let\K@fo\message\let\K@fu\message\else\immediate\read\K@dd to \K@bf
??T\immediate\closein\K@dd\ifx\K@bf\K@gb\let\K@bc#1\else\immediate\closein
??T\K@dc\let\K@fo\message\let\K@fu\message\fi\fi\K@bc}\def\K@gt#1#2#3{\def
??T\K@fx{\expandafter\expandafter\expandafter\K@gh\expandafter\expandafter
??T\csname end#3\endcsname\K@gb\Modify\K@gi\K@gw#1#2{#3}}\expandafter
??T\expandafter\expandafter\K@gf\expandafter\expandafter\csname#3\endcsname
??T\csname end#3\endcsname\K@fx}\def\K@gq{\K@dw\def\K@gx{\K@gy\K@gz\let\gr
??T\K@ha}\K@hb\K@hc\K@hd\K@he\K@hf\let\stop\relax\let\grav\relax\let\base
??T\relax\K@hg\K@hh\K@hi\catcode`\@=11 \immediate\read\K@dc to\K@bf
??T\expandafter\K@hj\K@bf\K@fu{done}\K@ex\K@ey\egroup\Displaybox\K@ds}\def
??T\K@gr{\let\K@hk\K@hl\let\K@hm\K@hj\K@gu}\def\K@hj#1#2#3#4#5 {\def\K@en
??T{#1}\def\K@hn{#5}\def\K@ho{#3}\def\K@hp{#4}\global\setbox\K@ds=\hbox{\hbox
??Tto #1sp{\hfil\vbox to #2sp{\vfil}}\K@hq}}\def\K@gu{\K@dw\def\K@gx{\let\gr
??T\K@ha}\catcode`\@=11 \immediate\read\K@dc to\K@bf\expandafter\K@hm\K@bf
??T\K@fu{done}\K@ex\K@hr\egroup\Displaybox\K@ds}\def\K@hm#1#2#3#4 {\def\K@en
??T{#1}\def\K@hp{#4}\global\setbox\K@ds=\hbox{\hbox to #1sp{\hfil\vbox to
??T#2sp{\vfil}}\K@hq}}\def\K@hq{\immediate\read\K@dc to\K@bf\ifeof\K@dc\def
??T\K@bc{\immediate\closein\K@dc}\else\def\K@bc{\expandafter\K@hs\K@bf\K@hq
??T}\fi\K@bc}\def\K@hs#1#2#3 {\def\K@bf{#1}\ifx\K@bf\K@ci\def\K@bc
??T##1{{##1\ifnum#2=0 \immediate\read\K@dc to\K@bf\expandafter\K@ht\K@bf\fi
??T}}\else\def\K@bc{\K@hu{#1}{#2}}\fi\K@bc{#3}}\def\K@ht#1#2#3 {\K@hu
??T{#1}{#2}{#3}}\def\K@hu#1#2#3{\llap{\hbox to \K@en sp{\hskip#1 \smash{\hbox
??T{\raise#2\hbox{\K@hv#3}}}\hss}}}\endgroup
~\K@bk\newDiagram{\bgroup\K@ez\K@hw}\def\K@hw#1{\egroup\expandafter\ifx
~\csname D@every#1@\endcsname\relax\expandafter\expandafter\expandafter\let
~\expandafter\expandafter\csname D@@#1@@\endcsname\csname#1\endcsname
~\expandafter\def\csname D@every#1@\endcsname{\expandafter\expandafter
~\expandafter\let\expandafter\expandafter\csname#1\endcsname\csname
~D@@#1@@\endcsname}\fi\expandafter\K@bk\csname#1\endcsname{\bgroup\K@bt
~\csname D@every#1@\endcsname\K@ej\K@ek\K@el\everyDiagram\csname
~every#1\endcsname\K@ei{#1}}\expandafter\K@bk\csname end#1\endcsname{\K@dz
~\egroup}\expandafter\let\csname every#1\endcsname\relax}
??V\def\K@ef{\K@df=-1\relax\K@de=-1\relax\let\K@hx\K@cp\everycr{\noalign
??V{\global\advance\K@df by 1\relax\ifnum\K@hx<\K@de\xdef\K@hx{\the\K@de}\fi
??V\global\K@de=-1}}\global\K@dg=0 \global\K@dh=0 \let\K@hy\K@ci\let\K@hz
??V\K@ci\let\K@ia\K@ci\let\K@ib\K@ci\let\K@ic\K@ci\let\K@id\K@ci\let\K@ie
??V\K@ci\let\K@if\K@ci\let\K@ig\K@ci\let\K@ih\K@ci\let\K@ii\K@ci\global
??V\let\K@ij\relax\global\let\K@hn\K@cp\global\let\K@ik\K@ci\K@il\let\\\cr
??V\let\K@es\K@cp\let\K@et\K@cp\let\K@eu\K@cm\let\K@ev\K@cm\let\K@im\K@cm
??V\let\K@in\K@ci\K@hg\K@hd\K@he\K@hb\K@gy\let\gr\K@ha}%
~\def\K@hb{\let\K@io(\let\K@ip\(\catcode`\(=\active\let\K@iq_\let\K@ir
~\_\catcode`\_=\active\let\K@is^\let\K@it\^\catcode`\^=\active\let\K@iu
~<\let\K@iv\<\catcode`\<=\active\let\K@iw>\let\K@ix\>\catcode`\>=\active
~\let\K@iy:\let\K@iz\:\catcode`\:=\active}{\catcode`\(=\active\catcode
~`\_=\active\catcode`\^=\active\catcode`\<=\active\catcode`\>=\active
~\catcode`\:=\active\gdef\K@ja{\let(\K@jb\let<\K@jc\let>\K@jd\let_\K@je
~\let^\K@jf\let:\K@jg\let\((\let\__\let\^^\let\<<\let\>>\let\::} \gdef
~\K@gy{\let(\K@io\let\(\K@ip\let_\K@iq\let\_\K@ir\let^\K@is\let\^\K@it
~\let<\K@iu\let\<\K@iv\let>\K@iw\let\>\K@ix\let:\K@iy\let\:\K@iz} \gdef
~\K@hc{\def(##1,##2){}\let_\K@jh\let^\K@jh\let<\K@jh\let>\K@jh\let:\K@cj
~\let\((\let\__\let\^^\let\<<\let\>>\let\::} \gdef\K@ji{\let(\K@jj\let
~_\K@jk\let^\K@jl\let<\K@jm\let>\K@jn\let:\K@jo} }\def\K@jj{(}\def\K@jk
~{_}\def\K@jl{^}\def\K@jm{<}\def\K@jn{>}\def\K@jo{:}\def\K@jh{\def\K@jp
~##1{\K@cj}\let\K@jq\K@cj\futurelet\K@jr\K@js}\def\K@hg{\let\K@jt\dh\let
~\K@ju\dt\let\K@jv\up\let\K@jw\rt\let\K@jx\mv\let\K@jy\nl\let\K@jz\ru
~\let\K@ka\rr\let\K@kb\rm\let\K@kc\ro\let\K@kd\pt\let\K@ke\fd\let\K@kf
~\rl\let\K@kg\br\let\K@kh\pp\let\K@ki\jt\let\K@kj\jh\let\K@kk\jn\let
~\K@kl\rw\let\K@km\hx\let\K@kn\hy\let\K@ko\tx\let\K@kp\ty\let\K@kq\fx
~\let\K@kr\fy\let\K@ks\ts\let\K@kt\hs\let\K@ku\fs\let\K@kv\tr\let\K@kw
~\hr\let\K@kx\fr\let\K@ky\db\let\K@kz\lb\let\K@la\rb\let\K@lb\nw\let
~\K@lc\tl\let\K@ld\hd\let\K@le\pl\let\K@lf\pd\let\K@lg\lw}\def\K@lh{\let
~\dh\K@li\let\dt\K@lj\let\up\K@lk\let\rt\K@ll\let\mv\K@lm\let\nl\K@ln
~\let\ru\K@lo\let\rr\K@lp\let\rm\K@lq\let\ro\K@lr\let\pt\K@ls\let\fd
~\K@lt\let\rl\K@lu\let\br\K@lv\let\pp\K@lw\let\jt\K@lx\let\jh\K@ly\let
~\jn\K@lz\let\rw\K@ma\let\K@mb\K@mc\let\K@md\K@me\let\hx\K@mf\let\hy
~\K@mg\let\tx\K@mh\let\ty\K@mi\let\fx\K@mj\let\fy\K@mk\let\ts\K@ml\let
~\hs\K@mm\let\fs\K@mn\let\tr\K@mo\let\hr\K@mp\let\fr\K@mq\let\db\K@mr
~\let\lb\K@ms\let\rb\K@mt\let\nw\K@mu\let\lw\K@mv}\def\K@gz{\let\dh\K@jt
~\let\dt\K@ju\let\up\K@jv\let\rt\K@jw\let\mv\K@jx\let\nl\K@jy\let\ru
~\K@jz\let\rr\K@ka\let\rm\K@kb\let\ro\K@kc\let\pt\K@kd\let\fd\K@ke\let
~\rl\K@kf\let\br\K@kg\let\pp\K@kh\let\jt\K@ki\let\jh\K@kj\let\jn\K@kk
~\let\rw\K@kl\let\K@mb\K@kl\let\K@md\K@kl\let\hx\K@km\let\hy\K@kn\let
~\tx\K@ko\let\ty\K@kp\let\fx\K@kq\let\fy\K@kr\let\ts\K@ks\let\hs\K@kt
~\let\fs\K@ku\let\tr\K@kv\let\hr\K@kw\let\fr\K@kx\let\db\K@ky\let\lb
~\K@kz\let\rb\K@la\let\nw\K@lb\let\tl\K@lc\let\hd\K@ld\let\pl\K@le\let
~\pd\K@lf\let\lw\K@lg}\def\K@hh{\let\dh\K@cj\let\dt\K@cj\let\up\K@cj\let
~\rt\K@cj\let\mv\K@cj\let\nl\relax\let\ru\K@cj\let\rr\K@cj\let\rm\K@cj
~\let\ro\relax\let\pt\K@cj\let\fd\K@cj\let\rl\relax\let\br\relax\let\pp
~\relax\let\jt\relax\let\jh\relax\let\jn\relax\let\rw\K@cj\let\K@mb
~\K@cj\let\K@md\K@cj\let\hx\K@cj\let\hy\K@cj\let\tx\K@cj\let\ty\K@cj
~\let\fx\K@cj\let\fy\K@cj\let\ts\K@cj\let\hs\K@cj\let\fs\K@cj\let\tr
~\K@cj\let\hr\K@cj\let\fr\K@cj\let\db\relax\let\lb\relax\let\rb\relax
~\let\nw\relax\let\lw\K@jh}
??V\def\K@mw#1{\xdef\K@hy{\K@hy{\the\K@de}{#1}}}\def\K@mx#1{\relax\K@di=\K@df
??V\multiply\K@di by -1 \advance\K@di by \K@ho\xdef\K@hz{{\the\K@di}{#1}\K@hz
??V}}\def\K@my#1{\xdef\K@ia{\K@ia{\the\K@de}{#1}}}\def\K@mz#1{\relax\K@di
??V=\K@df\multiply\K@di by -1 \advance\K@di by \K@ho\xdef\K@ib{{\the\K@di
??V}{#1}\K@ib}}\def\K@na#1{\xdef\K@ic{\K@ic{\the\K@de}{#1}}}\def\K@nb{\xdef
??V\K@id{\K@id{\the\K@df}{\the\K@de}}}\def\K@nc{\xdef\K@ie{\K@ie{\the\K@df
??V}{\the\K@de}}}\def\K@nd{\xdef\K@if{\K@if{\the\K@df}{\the\K@de}}}\def\K@ne
??V{\xdef\K@ig{\K@ig{\the\K@df}{\the\K@de}}}\def\K@nf{\xdef\K@ih{\K@ih{\the
??V\K@df}{\the\K@de}}}\def\K@ng{\xdef\K@ii{\K@ii{\the\K@df}{\the\K@de}}}\def
??V\K@nh{\global\let\K@ni\K@co}\def\K@nj{\xdef\K@ij{\the\K@de}}\def\K@nk{\xdef
??V\K@hn{\the\K@df}}\def\K@nl{\def\grav{\relax\K@nj}\def\base{\relax\K@nk
??V}}\def\K@nm{\let\K@nj\relax\let\K@nk\relax}\def\K@hd{\let\K@nn\stop\let
??V\K@no\grav\let\K@np\base\let\K@nq\safe\let\K@nr\nodot\let\K@ns\dx\let
??V\K@nt\dy\let\K@nu\dl\let\K@nv\dr\let\K@nw\mx\let\K@nx\my\let\K@ny\ml
??V\let\K@nz\mr\let\K@oa\ax\let\K@ob\al\let\K@oc\ar}\def\K@he{\K@nl\def
??V\stop{\relax\K@nh}\def\safe{\relax\K@od}\def\nodot{\relax\K@oe}\def\dl
??V{\relax\K@nb}\def\dr{\relax\K@nc}\def\ml{\relax\K@nd}\def\mr{\relax\K@ne
??V}\def\al{\relax\K@nf}\def\ar{\relax\K@ng}\def\mx{\relax\K@my}\def\my
??V{\relax\K@mz}\def\dx{\relax\K@mw}\def\dy{\relax\K@mx}\def\ax{\relax\K@na
??V}}\def\K@hf{\let\K@nb\relax\let\K@nc\relax\let\K@nd\relax\let\K@ne
??V\relax\let\K@nf\relax\let\K@ng\relax\let\K@my\K@cj\let\K@mz\K@cj\let
??V\K@mw\K@cj\let\K@mx\K@cj\let\K@na\K@cj}\def\K@of{\let\stop\K@nn\let\grav
??V\K@no\let\base\K@np\let\safe\K@nq\let\nodot\K@nr\let\dx\K@ns\let\dy
??V\K@nt\let\dl\K@nu\let\dr\K@nv\let\mx\K@nw\let\my\K@nx\let\ml\K@ny\let
??V\mr\K@nz\let\ax\K@oa\let\al\K@ob\let\ar\K@oc}\def\K@od#1{{\K@of#1}}%
~\def\K@jb#1,#2){\gdef\K@og{#1@#2}}\def\K@jc{\K@oh0}\def\K@jd{\K@oh1}\def
~\K@je{\K@oh2}\def\K@jf{\K@oh3}\def\K@oh#1{\def\K@jp##1{\let\K@oi\K@cl\let
~\K@oj\K@cl\let\K@ok\K@cl\K@ol##1;@\K@oi\K@oj\K@ok\edef\K@bc{\noexpand
~\K@om {#1{\K@oi}{\K@oj}{\K@ok}}}\K@bc} \def\K@jq{\K@om{#1}}\futurelet\K@jr
~\K@js}\def\K@om#1#2{\K@dr={#1@#2}\K@on}\def\K@oo#1#2{\K@dr={#1@#2}\def\K@bc
~{\K@op#1{{}}@{#2}\K@on}\ifnum\K@oq=1\relax\else\ifnum\K@oq=5\relax\else
~\let\K@bc\K@on\fi\fi\K@bc}\def\K@op#1#2#3@#4{\setbox\K@dt\hbox{$\K@gx
~\labelstyle\hskip\K@or sp\hskip\K@os{\K@ot}{}sp \hskip\K@ou sp \hskip\K@ov
~{\K@ot}{}sp \hskip\K@ow#4\hskip\K@ow\hskip\K@os{\K@ot}{}sp\hskip\K@or sp
~\hskip\K@ov{\K@ot}{}sp \hskip\K@ou sp$}\K@di=\wd\K@dt\xdef\K@ox{\K@ox
~{#2}{\the\K@di}}}\def\K@oy#1{\if\noexpand\K@jr\noexpand#1\let\K@bc\K@oz
~\else\if\noexpand\K@jr\K@bn\def\K@bc{\K@pa\futurelet\K@jr\K@js @}\else
~\let\K@bc\K@jq\fi\fi\K@bc}\def\OptDelim#1#2{\def\K@oz#1##1#2{\K@jp
~{##1}}\def\K@js{\K@oy{#1}}}\OptDelim[] \def\K@on{\edef\K@bc{\global\K@dq
~={\the\K@dq{\the\K@dr}}}\K@bc}\def\K@jg#1{\K@ol#1;@\K@pb\K@pc\K@pd}\def
~\K@ol#1;#2@#3#4#5{\def\K@bf{#2}\ifx\K@bf\K@ci\let\K@pe\K@pf\K@pg
~#1;@#3#4#5\else\let\K@ph\K@pi\let\K@pj\K@pk\K@pl#1;#2@#3#4#5\fi}\def
~\K@pg#1;#2@#3#4#5{\def\K@bf{#2}\ifx\K@bf\K@ci\def\K@bf{#1,;}\else\edef
~\K@bf{#1,#2}\fi\expandafter\K@pe\K@bf @#3#4#5}\def\K@pf
~#1,#2;@#3#4#5{\setbox\K@du=\hbox{\K@do=#1pt}\def\K@bf{#2} \ifdim\wd\K@du
~>\z@\ifx\K@bf\K@ci\K@pm#1@#5\else\setbox\K@du=\hbox{\K@do=#2pt}\ifdim\wd
~\K@du>\z@\K@pm#1@#4\K@pm\K@pn#2@#5\else\K@pm#1@#5\xdef#3{\K@pn#2}\fi\fi
~\else\gdef#3{#1}\ifx\K@bf\K@ci\else\K@pm\K@pn#2@#5\fi\fi}\def\K@pn
~#1,{#1}\def\K@po#1;{#1}\def\K@pl#1;#2@#3#4#5{\K@pp#1,@#3#4#5\def\K@bf
~{#2}\ifx\K@bf\K@ci\else\edef\K@bf{\K@po#2}\expandafter\K@pp\K@bf
~,@#3#4#5\fi}\def\K@pp#1,#2@#3#4#5{\def\K@bf{#2}\ifx\K@bf\K@ci\K@ph
~#1@#3#5\else\K@pj#1,#2@#4#5\fi}\def\K@pi#1@#2#3{\setbox\K@du=\hbox{\K@do
~=#1pt}\ifdim\wd\K@du>\z@\K@pm#1@#3\else\gdef#2{#1}\fi}\def\K@pk
~#1,#2,@#3#4{\K@pm#1@#3\K@pm#2@#4}\def\K@pm#1@#2{\K@do=#1 \K@di=\K@do\xdef
~#2{\the\K@di}}\let\K@hp\K@cm\def\K@pq#1{\bgroup\edef\endgr{\noexpand\K@pr
~{\K@hp}\egroup}\K@ps{#1}\K@hp\edef\K@bc{\noexpand\K@pr{\K@hp}}\K@bc}\def
~\K@ha{\relax\K@pq}\def\K@pt#1{\K@ps{#1}\K@pu\global\let\K@pu\K@pu}\let\gr
~\K@ha\def\K@ps#1#2{\K@do=1sp \K@do=#1\K@do\ifdim\K@do>1sp \else\ifdim
~\K@do<\z@\else\def#2{#1}\fi\fi}\def\gray{\gr\K@pv}\def
~\endgray{\endgr}
~\def\K@li#1{\let\K@ph\K@pw\let\K@pj
~\K@px\K@pl#1;@\K@py\K@pz\K@qa}\def\K@mf#1{\gdef\K@pz{#1}}\def\K@mg#1{\K@do
~=#1 \K@di=\K@do\xdef\K@qa{\the\K@di}}\def\K@lj#1{\let\K@ph\K@pw\let\K@pj
~\K@px\K@pl#1;@\K@qb\K@qc\K@qd}\def\K@mh#1{\gdef\K@qc{#1}}\def\K@mi#1{\K@do
~=#1 \K@di=\K@do\xdef\K@qd{\the\K@di}}\def\K@lt#1{\K@li{#1}\K@lj{#1}}\def
~\K@mj#1{\K@mh{#1}\K@mf{#1}}\def\K@mk#1{\K@mi{#1}\K@mg{#1}}\def\K@ml#1{\gdef
~\K@qe{#1}}\def\K@mm#1{\gdef\K@qf{#1}}\def\K@mn#1{\K@ml{#1}\K@mm{#1}}\def
~\K@mo#1{\K@do=#1 \K@di=\K@do\xdef\K@qg{\the\K@di}}\def\K@mp#1{\K@do=#1
~\K@di=\K@do\xdef\K@qh{\the\K@di}}\def\K@mq#1{\K@mo{#1}\K@mp{#1}}\def\K@px
~#1,#2@#3#4{\K@do=#2 \K@di=\K@do\xdef#4{\the\K@di}\gdef#3{#1}}\def\K@lk
~#1{\K@do=#1 \K@di=\K@do\xdef\K@qi{\the\K@di}}\def\K@ll#1{\K@do=#1 \K@di
~=\K@do\xdef\K@qj{\the\K@di}}\def\K@lm#1{\K@qk#1@}\def\K@qk#1,#2@{\K@ll
~{#1}\K@lk{#2}}\def\K@lo#1{\K@do=#1 \K@di=\K@do\xdef\K@ql{\the\K@di}}\def
~\K@lp#1{\K@do=#1 \K@di=\K@do\xdef\K@qm{\the\K@di}}\def\K@lq#1{\K@qn#1@}\def
~\K@qn#1,#2@{\K@lp{#1}\K@lo{#2}}\def\K@mv{\let\K@jp\K@qo\def\K@jq{\global
~\let\K@qp\K@ci}\futurelet\K@jr\K@qq}\let\K@qr\relax\K@bg\kuvioldlabels
~{\K@cz\let\K@qr\K@ci}\def\K@qo#1{\egroup\K@do=#1 \gdef\K@ow{#1}}\def\K@ln
~{\global\let\K@qs\K@ci}\def\K@lr{\global\let\K@qt\K@cl}\def\K@mr{\global
~\let\K@qu\K@qv}\def\K@ms{\gdef\K@qu{2}}\def\K@mt{\gdef\K@qu{3}}\let\K@qw
~\K@cm\let\K@qv\K@co\def\braced{\K@cz\let\K@qw\K@co\let\K@qv\K@cm}\def
~\loose{\K@cz\let\K@qw\K@cm\let\K@qv\K@co\columndist=0pt}\def\K@mu{\global
~\let\K@qx\K@ci}\def\K@ma#1{\K@do=#1 \K@di=\K@do\divide\K@di by 2 \xdef\K@qy
~{=\the\K@di}\def\K@qz{\egroup\K@cj}}\def\K@mc#1{\K@do=#1 \global\K@dr
~={{#1}}}\def\K@me#1{\K@do=#1 \gdef\K@ra{#1}}\def\K@qz#1{\egroup\K@do=#1
~\K@di=\K@do\xdef\K@qy{\the\K@di}}\def\K@lu{\global\let\K@rb\K@ci\global\let
~\K@rc\K@ci\let\K@jp\K@qz\let\K@jq\relax\futurelet\K@jr\K@qq}\def\K@lv
~{\global\let\K@rc\K@ci\let\K@jp\K@qz\let\K@jq\relax\futurelet\K@jr\K@qq
~}\def\K@lw{\global\let\K@rd\K@ci}\def\K@lx{\global\let\K@re\K@cl}\def\K@ly
~{\global\let\K@rf\K@cl}\def\K@lz{\K@lx\K@ly}\let\K@rg\K@cl\def\joined
~{\K@cz\let\K@rg\K@ci}\def\K@ls#1{\K@ca\K@rh#1,@}\def\K@ri#1{\K@ca\K@rj#1,@}
??O\begingroup\K@ce\def\K@rh#1,#2@{{\K@ji\K@ed\def\K@bf{#2}\ifx\K@bf\K@ci
??O\K@rk#1@,@\else\K@rk#1@#2@\fi }}\def\K@rk#1@#2,@{\xdef\K@in{{#1}\K@in
??O}\xdef\K@rl{{#1@#2}\K@rl}}\def\K@rj#1,#2@{{\K@ji\K@ed\def\K@bf{#2}\ifx
??O\K@bf\K@ci\ifx\K@in\K@ci\K@rk#1@,@\else\ifx\K@rm\K@ci\edef\K@gl
??O{#1}\K@dr={\K@rn {\csname D@pt@#1@x\endcsname}{\csname D@pt@#1@y\endcsname
??O}{\csname D@pt@#1@type\endcsname}}\expandafter\K@ro\K@in{}\ifx\K@gl\relax
??O\else\K@rk#1@,@\fi\else\K@rk#1@,@\fi\fi\else\K@rk#1@#2@\fi }}\def
??O\K@ro#1#2{\def\K@bf{#1}\ifx\K@bf\K@gl\the\K@dr\K@dr={}\let\K@gl\relax\fi
??O\def\K@bf{#2}\ifx\K@bf\K@ci\let\K@bc\relax\else\def\K@bc{\K@ro{#2}}\fi
??O\K@bc}\def\K@rn#1#2#3{\ifx\K@rp\K@ci\xdef\K@rp{#1,#2}\expandafter\ifx
??O#3\relax\else\xdef\K@rq{\K@rr{#3}{}}\fi\else\xdef\K@rm
??O{#1,#2}\expandafter\ifx#3\relax\else\xdef\K@rs{\K@rr{#3}{}}\fi\fi}%
~\def\K@pr#1{\K@cg{#1 setgray}}\def\K@pw#1@#2#3{\K@do=#1 \K@di=-\K@do\xdef
~#2{\the\K@di}}
??V\def\K@eg{\global\advance\K@de by 1\relax\global\let\K@ni\K@cm\let\K@rt
??V\K@ru}\def\K@eh#1{\K@dr={#1}\edef\K@bc{\gdef\expandafter\noexpand\csname
??VD@Entry@\the\K@dh @\endcsname{\noexpand\D@\the\K@df @\the\K@de @\the\K@dr
??V@}}\K@bc\global\advance\K@dh by 1\relax\setbox\K@du=\hbox{\K@hi\K@hf
??V\K@hc\K@hh\let\gr\K@cj\let\endgr\relax\global\setbox\K@dt=\hbox\bgroup
??V\K@rv$\vertexstyle\K@gx#1\K@rw}}\def\K@rw{\K@rt\K@di=\wd\K@dt\K@dj=\ht
??V\K@dt\K@dk=\dp\K@dt {\count8=\K@di\advance\count8 by \K@dj\advance\count8
??Vby \K@dk\relax\ifnum\count8>0 \K@nh\fi }\ifnum\K@ni=1\relax\expandafter
??V\xdef\csname D@whd@\the\K@df @\the\K@de @\endcsname{\the\K@di @\the\K@dj
??V@\the\K@dk @}\xdef\K@ik{\K@ik{\the\K@df}{\the\K@de}}\fi}\def\K@em{\advance
??V\K@df by -1 \xdef\K@ho{\the\K@df}\let\K@rx\K@cm\let\K@ry\K@ho\let\K@rz
??V\K@cm\let\K@sa\K@hx\K@sb\let\K@sc\K@sd\K@nm\let\K@nh\relax\K@se\K@sf
??V\ifx\K@sg\relax\else\xdef\K@sh{\K@hy\K@ic\K@ia\K@hz\K@ib\K@id\K@ie\K@if
??V\K@ig\K@ih\K@ii}\ifx\K@sh\K@ci {\K@dm=\K@hx\K@dm\K@di=\K@dm\xdef\K@en
??V{\the\K@di}\K@dn=\K@ho\K@dn\K@di=\K@dn\xdef\K@eo{\the\K@di}}\def\K@bs
??V\K@si{\K@sj}\fi\fi\K@bs\K@si\K@sk{.}\K@dx{}{{\K@en}{\K@eo}{\K@ho}{\K@hp
??V}{\K@hn}}\K@hf\let\st@p\relax\K@nm}%
~\let\K@sk\K@cj
??V\def\K@sb{\def\K@bc{\K@sl{0}{0}}\expandafter\K@bc\K@ik{}{}}\def\K@sl
??V#1#2#3#4{\def\K@bf{#3}\ifx\K@bf\K@ci\def\K@bc{\K@sm{#1}{#2}\K@sn\K@sl
??V{#1}{#2}{}{}}\else\def\K@bc{\K@so{#1}{#2}{#3}{#4}}\fi\ifnum#1>\K@ho
??V\relax\let\K@bc\relax\fi\K@bc}\def\K@so#1#2#3#4{\def\K@bc{\K@sm
??V{#1}{#2}\K@sn\K@sl{#1}{#2}{#3}{#4}}\ifnum#1=#3\relax\ifnum#2=#4\relax
??V\def\K@bc{\K@sn\K@sl{#1}{#2}}\fi\fi\K@bc}\def\K@sn#1#2#3{\ifnum#3<\K@hx
??V\relax\K@di=#3 \advance\K@di by 1 \edef\K@bc{\noexpand#1{#2}{\the\K@di
??V}}\else\K@di=#2 \advance\K@di by 1 \edef\K@bc{\noexpand#1{\the\K@di
??V}{0}}\fi\K@bc}\def\K@sm#1#2{\global\expandafter\let\csname
??VD@whd@#1@#2@\endcsname\relax}\def\K@se{\K@sp\K@sq0\K@sr\K@ja\K@lh\let
??V\gr\K@pt\let\endgr\relax\let\D@\K@ss\K@st0\K@dh\K@su}\def\K@sv{\global
??V\let\K@sw\K@co\global\expandafter\let\csname D@as@1@\endcsname\relax
??V\global\let\K@sx\K@co\global\expandafter\let\csname D@cs@1@\endcsname
??V\relax\global\let\K@sy\K@co\global\expandafter\let\csname
??VD@ws@1@\endcsname\relax\global\let\K@sz\relax}\def\K@ta#1{\global
??V\expandafter\let\csname D@Bindings@#1@\endcsname\K@ci\K@di=#1\relax\ifnum
??V#1<\K@hx\advance\K@di by 1 \edef\K@bc{\noexpand\K@ta{\the\K@di}}\else\let
??V\K@bc\relax\fi\K@bc}\def\K@tb#1{\K@di=#1\relax\ifnum#1<\K@hx\advance
??V\K@di by 1 \edef\K@bc{\noexpand\K@tc{#1}{\the\K@di}}\K@bc{\expandafter\K@td
??V\K@cs}\K@ck\K@td\edef\K@bc{\noexpand\K@tb{\the\K@di}}\else\let\K@bc\relax
??V\fi\K@bc}\def\K@sp{\ifx\K@ij\relax\ifx\K@te\relax\K@di=\K@hx\divide
??V\K@di by 2 \multiply\K@di by 2 \ifnum\K@di=\K@hx\multiply\K@di by 5 \else
??V\multiply\K@di by 5 \advance\K@di by 5 \fi\else\K@di=\K@te\multiply\K@di
??Vby 10 \fi\else\K@di=\K@ij\multiply\K@di by 10 \fi\xdef\K@ij{\the\K@di
??V}}\def\K@su#1{\edef\K@tf{\expandafter\noexpand\csname D@Entry@\the
??V#1@\endcsname}\csname D@Entry@\the#1@\endcsname}\def\K@ss#1@#2@#3@{\K@df
??V=#1\relax\K@de=#2\relax\K@tg#3\K@th}\def\K@eq{\K@gy\K@ec\K@gz\let\gr
??V\K@ha\K@st\K@im\K@dg\K@ti}\def\K@st#1#2#3{\edef\K@bf{#2{\the#2}}\global
??V#2=#1 \expandafter\K@tj\K@bf#3}\def\K@tj#1#2#3{\ifnum#1=#2 \let\K@bc\relax
??V\else\def\K@bc{#3#1\global\advance#1 by 1 \K@tj#1{#2}#3}\fi\K@bc}\def
??V\K@ti#1{\csname D@Cell@\the#1@\endcsname}\def\K@tk#1{\ifcase#1 \relax\or
??V\K@tl0{-1}\K@tm\K@tn{-1}0\K@tl01\K@to\K@tp{-1}\K@hx\or\K@tl{-1}{-1}\K@tm
??V\K@tn00\K@tl11\K@to\K@tp\K@ho\K@hx\or\K@tl{-1}0\K@tm\K@tn0{-1}\K@tl
??V10\K@to\K@tp\K@ho{-1}\or\K@tl{-1}1\K@tm\K@tn0\K@hx\K@tl1{-1}\K@to\K@tp
??V\K@ho0\or\K@tl01\K@tm\K@tn{-1}\K@hx\K@tl0{-1}\K@to\K@tp{-1}0\or\K@tl
??V11\K@tm\K@tn\K@ho\K@hx\K@tl{-1}{-1}\K@to\K@tp00\or\K@tl10\K@tm\K@tn\K@ho
??V{-1}\K@tl{-1}0\K@to\K@tp0{-1}\or\K@tl1{-1}\K@tm\K@tn\K@ho0\K@tl{-1}1\K@to
??V\K@tp0\K@hx\fi}\def\K@tl#1#2#3#4#5#6{\K@di=\K@df\K@dj=\K@de\let#3\relax
??V\K@cu\ifnum\K@di=#5\relax\let\K@bc\K@tq\else\ifnum\K@dj=#6\relax\let
??V\K@bc\K@tq\else\let\K@bc\K@tr\fi\fi\K@bc{#1}{#2}{#3}{#4}\ifx#3\relax
??V\K@cv}\def\K@tq#1#2#3#4{\xdef#3{\the\K@di}\xdef#4{\the\K@dj}}\def\K@tr
??V#1#2#3#4{\advance\K@di by #1\relax\advance\K@dj by #2\relax\K@ts{\the
??V\K@di}{\the\K@dj}\K@bf\ifx\K@bf\relax\else\xdef#3{\the\K@di}\xdef#4{\the
??V\K@dj}\fi}%
~\def\K@tt#1#2#3{\ifx\K@rq\relax\xdef\K@bc{\noexpand\K@sc @\K@tn,\noexpand
~\K@tu @\K@tm,\noexpand\K@tv @}\K@bc\else\let\K@tu\K@tn\let\K@tv\K@tm\fi
~\ifnum#1=9 \let\K@tw\K@tx\else\K@ty1\K@qc\K@tz\K@ty1\K@pz\K@ua\ifnum
~#1=0 \ifx\K@qe\relax\else\K@ub\K@tu by {-1}\K@qe\fi\ifx\K@qg\relax
~\else\K@ub\K@tv by 1\K@qg\fi\fi\K@di=\K@qc\advance\K@di by \K@tu\edef
~\K@tu{\the\K@di}\K@di=\K@qd\advance\K@di by \K@tv\edef\K@tv{\the\K@di}\ifx
~\K@rs\relax\edef\K@bc{\noexpand\K@sc @\K@tp,\noexpand\K@uc @\K@to
~,\noexpand\K@ud @}\K@bc\else\let\K@uc\K@tp\let\K@ud\K@to\fi\ifnum#1=0
~\ifx\K@qf\relax\else\K@ub\K@uc by {-1}\K@qf\fi\ifx\K@qh\relax\else
~\K@ub\K@ud by 1\K@qh\fi\fi\K@di=\K@pz\advance\K@di by \K@uc\edef\K@uc
~{\the\K@di}\K@di=\K@qa\advance\K@di by \K@ud\edef\K@ud{\the\K@di}\K@ue
~\ifnum\K@uf=0 \K@di=\K@ug\relax\ifnum\K@di<0 \multiply\K@di by -1 \fi\edef
~\K@tw{\the\K@di}\else\ifnum\K@ug=0 \K@di=\K@uf\relax\ifnum\K@di<0
~\multiply\K@di by -1 \fi\edef\K@tw{\the\K@di}\else\K@uh\fi\fi\fi
~\ifnum\K@tw>0 \K@ui{#1}{#2}{#3}\fi}\def\K@ue{\K@di=\K@uc\multiply\K@di by
~-1 \advance\K@di by \K@tu\multiply\K@di by -1 \edef\K@ug{\the\K@di}\K@di
~=\K@ud\multiply\K@di by -1 \advance\K@di by \K@tv\multiply\K@di by -1 \edef
~\K@uf{\the\K@di}}\def\K@ui#1#2#3{\ifnum#1=0 \K@uj{#1}\K@oq\else\def\K@oq
~{#1}\fi {\ifnum#1=0 \ifx\K@tz\K@ci\let\K@tz\K@cm\let\K@uk\K@cm\let\K@ul
~\K@cm\fi\K@um\K@tu\K@tv\K@tz\K@uk\K@ul\ifx\K@ua\K@ci\let\K@ua\K@cm\let
~\K@un\K@cm\let\K@uo\K@cm\fi\K@um\K@uc\K@ud\K@ua\K@un\K@uo\fi\K@up\K@oq
~}\K@uq#1{#2}\ifx\K@rl\K@ci\else\K@ur{#1}#2\fi\ifnum\K@us<\K@ut\relax
~\else\ifx\K@pu\K@ci\else\edef\K@pu{\noexpand\K@pr{\K@pu}}\fi\K@uu
~{#1}{#2}{#3}\fi}\def\K@uu#1#2#3{\ifnum#1=9 \K@uv{#1}{#2}{#3}\else\K@uw
~{#1}{#2}{#3}\fi\ifx\K@rb\relax\def\K@gl{#3}\ifx\K@gl\K@ci\else\ifx
~\K@qt\K@ux\def\K@uy{\K@bz\K@uz}\else\let\K@uy\K@va\fi\K@vb
~{#1}{#2}#3{}\fi\fi}
??S\begingroup\K@ce\def\K@uv#1#2#3{\K@di=\K@tu\advance\K@di by \K@qj\edef
??S\K@vc{\the\K@di}\K@di=\K@tv\advance\K@di by \K@qi\edef\K@vd{\the\K@di
??S}\K@um\K@vc\K@vd000\ifx\K@qs\relax {\dimen0=\K@tw sp \K@dx{}{{}{0}{\dimen
??S0=\K@tw sp}}\edef\K@bc##1{\noexpand\K@ve1@\K@vc sp,\K@vd sp@{\noexpand\K@hv
??S0{\noexpand\K@vf\noexpand\K@vg{\K@vh}\hskip\K@qm sp\raise\K@ql sp
??S\noexpand\K@hv2{##1}\noexpand\K@vi}}}\ifx\K@rc\K@ci\K@bc{\K@pr{1}\K@hv
??S0{\hskip\K@qb sp \lower\K@vj\hbox{$\vcenter{ \hbox{\vrule width \K@us sp
??Sheight \K@qy sp depth \K@qy sp}}$}\hskip\K@py sp}\hskip\K@vk\dimen0}\fi
??S\ifx\K@rb\relax\K@dr={\K@vl#2}\K@bc{\K@pu\K@hv0{\hskip\K@qb sp \the\K@dr
??S{\K@us sp}\hskip\K@py sp}\hskip\K@vk\dimen0}\fi }\fi}%
~\def\K@uw#1#2#3{\K@di=\K@tu\advance\K@di by \K@uc\divide\K@di by 2 \advance
~\K@di by \K@qj\edef\K@vc{\the\K@di}\K@di=\K@tv\advance\K@di by \K@ud
~\divide\K@di by 2 \advance\K@di by \K@qi\edef\K@vd{\the\K@di}\ifx\K@qs
~\relax\edef\K@bc##1{\noexpand\K@ve1@\K@vc sp,\K@vd sp@{\noexpand\K@hv
~0{\noexpand\K@vf\noexpand\K@vm{\K@uf}{\K@ug}\hskip\K@qm sp\raise\K@ql sp
~\noexpand\K@hv1{##1}\noexpand\K@vi}}}\ifx\K@rc\K@ci\K@bc{\K@pr{1}\hbox
~{\hskip\K@qb sp \lower\K@vj\hbox{$\vcenter{ \hbox{\vrule width \K@us sp
~height \K@qy sp depth \K@qy sp}}$}\hskip\K@py sp}}\fi\ifx\K@rb\relax\K@dr
~={\K@vl#2}\K@bc{\K@pu\hbox{\hskip\K@qb sp \the\K@dr{\K@us sp}\hskip\K@py
~sp}}\fi\fi}\def\K@vb#1#2#3{\def\K@bf{#3}\ifx\K@bf\K@ci\let\K@bc\relax
~\else\def\K@bc{\K@uy{#1}{#2}#3@\K@vb{#1}{#2}}\fi\K@bc}
??O\def\K@ur#1#2#3{\edef\K@gl{\K@vn{#2}}\def\K@bf{\K@vo{#1}{#2}}\expandafter
??O\K@bf\K@rl{}}\def\K@vo#1#2#3#4{\K@vp{#1}{#2}#3@\def\K@bf{#4}\ifx\K@bf\K@ci
??O\let\K@bc\relax\else\def\K@bc{\K@vo{#1}{#2}{#4}}\fi\K@bc}\def\K@vp
??O#1#2#3@#4@{{\K@ji\K@ed\ifnum#1=9 \global\expandafter\let\csname
??OD@pt@#3@x\endcsname\K@tn\global\expandafter\let\csname D@pt@#3@y\endcsname
??O\K@tm\global\expandafter\let\csname D@pt@#3@type\endcsname\relax\else
??O\def\K@bf{#4}\ifx\K@bf\K@ci\else\let\K@gl\K@bf\fi\K@di=\K@qb\advance\K@di
??Oby \K@qm\relax\K@vq\K@tu\K@ug\K@vq\K@tv\K@uf\K@di=\K@py\multiply\K@di
??Oby -1 \advance\K@di by \K@qm\relax\K@vq\K@uc\K@ug\K@vq\K@ud\K@uf\K@do
??O=\K@tu sp \multiply\K@do by -1 \advance\K@do by \K@uc sp \K@do=\K@gl\K@do
??O\K@di=\K@do\advance\K@di by \K@tu\advance\K@di by \K@qj\relax\K@vr\K@dj
??O=\K@ql\times\K@uf\over{-\K@tw}\advance\K@di by \K@dj\expandafter\xdef
??O\csname D@pt@#3@x\endcsname{\the\K@di}\K@do=\K@tv sp \multiply\K@do by -1
??O\advance\K@do by \K@ud sp \K@do=\K@gl\K@do\K@di=\K@do\advance\K@di by
??O\K@tv\advance\K@di by \K@qi\relax\K@vr\K@dj=\K@ql\times\K@ug\over\K@tw
??O\advance\K@di by \K@dj\expandafter\xdef\csname D@pt@#3@y\endcsname{\the
??O\K@di}\expandafter\gdef\csname D@pt@#3@type\endcsname{#2}\fi }}%
~\def\K@vq#1#2{\K@dk=\K@tw\relax\K@vs\K@dj=\K@di\times#2\over\K@dk\advance
~\K@dj by #1 \edef#1{\the\K@dj}}\def\K@vr#1=#2\times#3\over#4{\K@dk=#2 \K@dl
~=#4\relax\K@vs#1=\K@dk\times#3\over\K@dl}\let\K@ux\K@cl\def\rotatedlabels
~{\let\K@ux\K@ci}\def\K@va#1#2#3#4@#5@{{\def\K@bf{#4}\ifx\K@bf\K@ci\else
~\K@vt#4\fi\K@vu{#1}{#2}\let\K@vv\K@co\K@vw{#3}\multiply\K@dp by \K@vv
~\advance\K@dp by \K@ql sp \edef\K@bc##1{\noexpand\K@vx @\K@vc sp,\K@vd
~sp@{##1}{\the\K@do}{\the\K@dp}}\K@bc{\K@vy{$\labelstyle#5$}}}}\def\K@vt
~#1#2#3{\def\K@bf{#1}\ifx\K@bf\K@cl\else\def\K@pb{#1}\fi\def\K@bf{#2}\ifx
~\K@bf\K@cl\else\def\K@pc{#2}\fi\def\K@bf{#3}\ifx\K@bf\K@cl\else\def\K@pd
~{#3}\fi}
??P\begingroup\K@ce\def\K@uz#1#2#3#4@#5@{{\def\K@bf{#4}\ifx\K@bf\K@ci\else
??P\K@vt#4\fi\K@vu{#1}{#2}\let\K@vv\K@co\let\K@vz\K@wa\K@vw{#3}\multiply
??P\K@dp by \K@vv\advance\K@dp by \K@ql sp \edef\K@bc##1{\noexpand\K@vx
??P@\K@vc sp,\K@vd sp@{##1}{\the\K@do}{\the\K@dp}}\K@bc{\K@vy{$\labelstyle#5$}}}}%
~\def\K@vu#1#2{\def\K@wb{\K@wc}\edef\K@vx{\noexpand\K@wd{\K@uf}{\K@ug}}\edef
~\K@vw{\noexpand\K@vz{\K@oq}}\K@do=\K@us sp \K@do=\K@pb\K@do\advance\K@do
~by \K@pc sp \advance\K@do by \K@qb sp \advance\K@do by \K@qm sp \K@dp=\K@tw
~sp \advance\K@do by -.5\K@dp\K@dp=\K@or sp \advance\K@dp by \K@os#2sp
~\advance\K@dp by \K@pd sp}
??S\def\K@vu#1#2{\ifnum#1=9 \def\K@wb{\K@we}\edef\K@vx{\noexpand\K@wf\K@vh
??S}\edef\K@vw{\noexpand\K@wg{\K@vh}}\K@do=\K@us sp \K@do=\K@pb\K@do
??S\advance\K@do by \K@pc sp \advance\K@do by \K@qb sp \advance\K@do by \K@qm
??Ssp \K@dp=-\K@tw sp \advance\K@do by \K@vk\K@dp\K@dp=\K@or sp \advance\K@dp
??Sby \K@os#2sp \advance\K@dp by \K@pd sp \else\def\K@wb{\K@wc}\edef\K@vx
??S{\noexpand\K@wd{\K@uf}{\K@ug}}\edef\K@vw{\noexpand\K@vz{\K@oq}}\K@do=\K@us
??Ssp \K@do=\K@pb\K@do\advance\K@do by \K@pc sp \advance\K@do by \K@qb sp
??S\advance\K@do by \K@qm sp \K@dp=\K@tw sp \advance\K@do by -.5\K@dp\K@dp
??S=\K@or sp \advance\K@dp by \K@os#2sp \advance\K@dp by \K@pd sp \fi}\def
??S\K@wf#1@#2,#3@#4#5#6{\K@dr={#4}\K@dx{}{{}{0}{\let\noexpand\K@vy\expandafter
??S\noexpand\K@vy\noexpand\K@gx}}\K@ve0@#2,#3@{\K@vf\K@wh{#1}\hskip#5 \raise
??S#6 \hbox{\K@wb{#1}{}\the\K@dr}\K@vi}}%
~\def\K@gx{\K@gy\K@gz\K@ec\let\gr\K@ha}\def\K@wh#1{\edef\K@bc{\noexpand\K@vg
~{#1}}\K@bc}\def\K@we#1#2{\K@di=#1 \multiply\K@di by -1 \edef\K@bc
~{\noexpand\K@vg{\the\K@di}}\K@bc}\def\K@wd#1#2@#3,#4@#5#6#7{\K@dr={#5}\K@dx
~{}{{}{0}{\let\noexpand\K@vy\expandafter\noexpand\K@vy\noexpand\K@gx}}\K@ve
~0@#3,#4@{\K@vf\K@wi{#1}{#2}\hskip#6 \raise#7 \hbox{\K@wb{#1}{#2}\the\K@dr
~}\K@vi}}\def\K@wi#1#2{\edef\K@bc{\noexpand\K@vm{#1}{#2}}\K@bc}\def\K@wc
~#1#2{\K@di=#1 \multiply\K@di by -1 \edef\K@bc{\noexpand\K@vm{\the\K@di
~}{#2}}\K@bc}\def\K@uq#1#2{\ifnum#1=9 \relax\else\ifnum#1=0 \K@wj\K@qb
~\K@re\K@tz\K@rq\K@wk{#2}\K@wj\K@py\K@rf\K@ua\K@rs\K@wl{#2}\else\K@wm\K@qb
~\K@re\K@wk{#2}\K@wm\K@py\K@rf\K@wl{#2}\fi\fi\K@di=\K@qb\advance\K@di by
~\K@py\multiply\K@di by -1 \advance\K@di by \K@tw\edef\K@us{\the\K@di
~}}\def\K@wj#1#2#3#4#5#6{\K@di=#1\relax\ifx#2\K@rg\advance\K@di by \K@wn
~\advance\K@di by \K@wo#6\else\ifx#3\K@ci\else\advance\K@di by #5\relax
~\advance\K@di by \K@wp\advance\K@di by \K@wq#6\fi\fi\edef#1{\the\K@di}}
??O\def\K@wj#1#2#3#4#5#6{\K@di=#1\relax\ifx#2\K@rg\advance\K@di by \K@wn
??O\advance\K@di by \K@wo#6\else\ifx#3\K@ci\ifx#4\relax\else\advance\K@di
??Oby \K@wr\advance\K@di by \K@ws#6\advance\K@di by #4\relax\advance\K@di
??Oby \K@wt\fi\else\advance\K@di by #5\relax\advance\K@di by \K@wp
??O\advance\K@di by \K@wq#6\fi\fi\edef#1{\the\K@di}}%
~\def\K@wm#1#2#3#4{\K@di=#1\relax\ifx#2\K@rg\advance\K@di by \K@wn
~\advance\K@di by \K@wo#4\else\advance\K@di by #3\relax\advance\K@di by
~\K@wp\advance\K@di by \K@wq#4\fi\edef#1{\the\K@di}}
??S\def\K@wg#1#2{\K@wu\K@di=#1\mod{360}\edef\K@vh{\the\K@di}\ifnum\K@di=0 \def
??S\K@gl{{1}}\else\ifnum\K@di=90 \def\K@gl{{7}}\else\ifnum\K@di=180 \def
??S\K@gl{{5}}\else\ifnum\K@di=270 \def\K@gl{{3}}\else\divide\K@di by 90
??S\ifcase\K@di\def\K@gl{{8}}\or\def\K@gl{{6}}\or\def\K@gl{{4}}\or\def
??S\K@gl{{2}}\fi\fi\fi\fi\fi\expandafter\K@vz\K@gl{#2}}%
~\def\K@vz#1#2{\ifcase#1 \relax\or\let\K@bc\K@wv\or\let\K@bc\K@ww\or
~\let\K@bc\K@wx\or\let\K@bc\K@wy\or\let\K@bc\K@wz\or\let\K@bc\K@xa\or
~\let\K@bc\K@xb\or\let\K@bc\K@xc\fi\K@bc{#2}}\def\K@wv#1{\ifcase#1
~\K@cf\K@vy\K@xd\or\K@cf\K@vy\K@xe\let\K@vv\K@cp\or\K@cf\K@vy\K@xe\let
~\K@vv\K@cp\or\K@cf\K@vy\K@xd\fi}\def\K@ww#1{\ifcase#1 \K@cf\K@vy\K@xf
~\let\K@vv\K@cp\or\K@cf\K@vy\K@xg\or\K@cf\K@vy\K@xf\let\K@vv\K@cp\or
~\K@cf\K@vy\K@xg\fi}\def\K@wx#1{\ifcase#1 \K@cf\K@vy\K@xh\let\K@vv\K@cp
~\or\K@cf\K@vy\K@xi\or\K@cf\K@vy\K@xh\let\K@vv\K@cp\or\K@cf\K@vy\K@xi
~\fi}\def\K@wy#1{\ifcase#1 \K@cf\K@vy\K@xj\let\K@vv\K@cp\or\K@cf\K@vy
~\K@xk\or\K@cf\K@vy\K@xk\or\K@cf\K@vy\K@xj\let\K@vv\K@cp\fi}\def\K@wz
~#1{\ifcase#1 \K@cf\K@vy\K@xe\or\K@cf\K@vy\K@xd\let\K@vv\K@cp\or\K@cf
~\K@vy\K@xe\or\K@cf\K@vy\K@xd\let\K@vv\K@cp\fi}\def\K@xa#1{\ifcase#1
~\K@cf\K@vy\K@xf\or\K@cf\K@vy\K@xg\let\K@vv\K@cp\or\K@cf\K@vy\K@xf\or
~\K@cf\K@vy\K@xg\let\K@vv\K@cp\fi}\def\K@xb#1{\ifcase#1 \K@cf\K@vy\K@xh
~\or\K@cf\K@vy\K@xi\let\K@vv\K@cp\or\K@cf\K@vy\K@xi\let\K@vv\K@cp\or
~\K@cf\K@vy\K@xh\fi}\def\K@xc#1{\ifcase#1 \K@cf\K@vy\K@xj\or\K@cf\K@vy
~\K@xk\let\K@vv\K@cp\or\K@cf\K@vy\K@xk\let\K@vv\K@cp\or\K@cf\K@vy\K@xj
~\fi}\def\K@xh#1{\K@hv2{\lower\K@vj\hbox{#1}}}\def\K@xj#1{\K@hv2{\vbox
~{\offinterlineskip\hbox{#1}\null}}}\def\K@xd#1{\K@hv1{\vbox
~{\offinterlineskip\hbox{#1}\null}}}\def\K@xg#1{\K@hv0{\vbox
~{\offinterlineskip\hbox{#1}\null}}}\def\K@xi#1{\K@hv0{\lower\K@vj\hbox
~{#1}}}\def\K@xk#1{\K@hv0{\vtop{\offinterlineskip\null\hbox{#1}}}}\def\K@xe
~#1{\K@hv1{\vtop{\offinterlineskip\null\hbox{#1}}}}\def\K@xf#1{\K@hv2{\vtop
~{\offinterlineskip\null\hbox{#1}}}}\def\K@xl#1#2{\K@vg{180}}
??P\def\K@wa#1#2{\ifcase#1 \relax\or\let\K@bc\K@xm\or\let\K@bc\K@xn\or
??P\let\K@bc\K@xo\or\let\K@bc\K@xp\or\let\K@bc\K@xq\or\let\K@bc\K@xr\or
??P\let\K@bc\K@xs\or\let\K@bc\K@xt\fi\K@bc{#2}}\def\K@xm#1{\let\K@wb\K@ck
??P\ifcase#1 \K@cf\K@vy\K@xd\or\K@cf\K@vy\K@xe\let\K@vv\K@cp\or\K@cf
??P\K@vy\K@xe\let\K@vv\K@cp\or\K@cf\K@vy\K@xd\fi}\def\K@xn#1{\let\K@wb
??P\K@ck\ifcase#1 \K@cf\K@vy\K@xe\let\K@vv\K@cp\or\K@cf\K@vy\K@xd\or
??P\K@cf\K@vy\K@xe\let\K@vv\K@cp\or\K@cf\K@vy\K@xd\fi}\def\K@xo#1{\K@cf
??P\K@vy\K@xd\ifcase#1 \let\K@vv\K@cp\let\K@wb\K@xl\or\let\K@wb\K@ck\or
??P\let\K@vv\K@cp\let\K@wb\K@xl\or\let\K@wb\K@ck\fi}\def\K@xp#1{\let\K@wb
??P\K@xl\ifcase#1 \K@cf\K@vy\K@xd\let\K@vv\K@cp\or\K@cf\K@vy\K@xe\or
??P\K@cf\K@vy\K@xe\or\K@cf\K@vy\K@xd\let\K@vv\K@cp\fi}\def\K@xq#1{\let
??P\K@wb\K@xl\ifcase#1 \K@cf\K@vy\K@xe\or\K@cf\K@vy\K@xd\let\K@vv\K@cp\or
??P\K@cf\K@vy\K@xe\or\K@cf\K@vy\K@xd\let\K@vv\K@cp\fi}\def\K@xr#1{\let
??P\K@wb\K@xl\ifcase#1 \K@cf\K@vy\K@xe\or\K@cf\K@vy\K@xd\let\K@vv\K@cp\or
??P\K@cf\K@vy\K@xe\or\K@cf\K@vy\K@xd\let\K@vv\K@cp\fi}\def\K@xs#1{\K@cf
??P\K@vy\K@xd\ifcase#1 \let\K@wb\K@ck\or\let\K@vv\K@cp\let\K@wb\K@xl\or
??P\let\K@vv\K@cp\let\K@wb\K@xl\or\let\K@wb\K@ck\fi}\def\K@xt#1{\let\K@wb
??P\K@ck\ifcase#1 \K@cf\K@vy\K@xd\or\K@cf\K@vy\K@xe\let\K@vv\K@cp\or
??P\K@cf\K@vy\K@xe\let\K@vv\K@cp\or\K@cf\K@vy\K@xd\fi}\endgroup
~\def\K@cf#1#2{\def#1{#2}}\def\K@uj#1#2{\ifnum\K@uf>0 \ifnum\K@ug>0 \def
~#2{8}\else\ifnum\K@ug=0 \def#2{7}\else\def#2{6}\fi\fi\else\ifnum\K@uf
~=0 \ifnum\K@ug>0 \def#2{1}\else\ifnum\K@ug=0 \relax\else\def#2{5}\fi\fi
~\else\ifnum\K@ug>0 \def#2{2}\else\ifnum\K@ug=0 \def#2{3}\else\def
~#2{4}\fi\fi\fi\fi}\def\K@xu#1#2#3#4{\ifx#1\K@rg\let#2\K@ck\fi\ifx
~#3\K@rg\let#4\K@ck\fi}\def\K@up#1{\global\let\K@wk\K@cm\global\let\K@wl
~\K@cm\K@di=\K@uf\relax\ifnum\K@di<0 \multiply\K@di by -1 \fi\K@dj=\K@ug
~\relax\ifnum\K@dj<0 \multiply\K@dj by -1 \fi\ifcase#1 \let\K@bc\relax\or
~\K@xv\K@bc\or\K@xw\K@bc\or\K@xx\K@bc\or\K@xy\K@bc\or\K@xz\K@bc\or
~\K@ya\K@bc\or\K@yb\K@bc\or\K@yc\K@bc\fi\K@bc}\def\K@xv#1{\K@xu\K@re
~\K@yd\K@rf\K@ye\def#1{\K@yd1\K@wk\K@ye1\K@wl}}\def\K@xw#1{\K@xu\K@re\K@yd
~\K@rf\K@ye\def#1{\K@yd2\K@wk\K@ye2\K@wl}}\def\K@xx#1{\K@xu\K@re\K@yd\K@rf
~\K@ye\def#1{\K@yd3\K@wk\K@ye3\K@wl}}\def\K@xy#1{\K@xu\K@re\K@yd\K@rf\K@ye
~\def#1{\K@yd4\K@wk\K@ye4\K@wl}}\def\K@xz#1{\K@xu\K@re\K@ye\K@rf\K@yd
~\K@ct\K@pz\K@qc\K@ct\K@qa\K@qd\K@ct\K@tz\K@ua\K@ct\K@uk\K@un\K@ct\K@ul
~\K@uo\def#1{\K@yd1\K@wl\K@ye1\K@wk}}\def\K@ya#1{\K@xu\K@re\K@ye\K@rf\K@yd
~\K@ct\K@pz\K@qc\K@ct\K@qa\K@qd\K@ct\K@tz\K@ua\K@ct\K@uk\K@un\K@ct\K@ul
~\K@uo\def#1{\K@yd2\K@wl\K@ye2\K@wk}}\def\K@yb#1{\K@xu\K@re\K@ye\K@rf\K@yd
~\K@ct\K@pz\K@qc\K@ct\K@qa\K@qd\K@ct\K@tz\K@ua\K@ct\K@uk\K@un\K@ct\K@ul
~\K@uo\def#1{\K@yd3\K@wl\K@ye3\K@wk}}\def\K@yc#1{\K@xu\K@re\K@ye\K@rf\K@yd
~\K@ct\K@pz\K@qc\K@ct\K@qa\K@qd\K@ct\K@tz\K@ua\K@ct\K@uk\K@un\K@ct\K@ul
~\K@uo\def#1{\K@yd4\K@wl\K@ye4\K@wk}}\def\K@yd#1#2{\ifcase#1 \relax\or
~\K@yf{#2}\or\K@yg{#2}\or\K@yh{#2}\or\K@yi{#2}\fi}\def\K@yf#1{\K@dk=\K@tz
~\advance\K@dk by \K@uk\advance\K@dk by \K@ul\relax\ifnum\K@dk>0 \K@dk
~=\K@tz\divide\K@dk by -2 \advance\K@dk by \K@qc\multiply\K@dk by -1 \K@vs
~\K@dl=\K@dk\times\K@tw\over\K@dj\xdef#1{\the\K@dl}\fi}\def\K@yg#1{\K@dk
~=\K@tz\advance\K@dk by \K@uk\advance\K@dk by \K@ul\relax\ifnum\K@dk>0
~\K@dk=\K@ul\advance\K@dk by \K@vj\advance\K@dk by \K@qd\K@dl=\K@tz\divide
~\K@dl by -2 \advance\K@dl by \K@qc\multiply\K@dl by -1 \K@yj#1\fi}\def\K@yh
~#1{\K@dk=\K@tz\advance\K@dk by \K@uk\advance\K@dk by \K@ul\relax\ifnum
~\K@dk>0 \K@dk=\K@vj\advance\K@dk by \K@ul\relax\K@vs\K@dl=\K@dk\times\K@tw
~\over\K@di\xdef#1{\the\K@dl}\fi}\def\K@yi#1{\K@dk=\K@tz\advance\K@dk by
~\K@uk\advance\K@dk by \K@ul\relax\ifnum\K@dk>0 \K@dk=\K@ul\advance\K@dk
~by \K@vj\advance\K@dk by \K@qd\K@dl=\K@tz\divide\K@dl by 2 \advance\K@dl
~by \K@qc\relax\K@yj#1\fi}\def\K@ye#1#2{\ifcase#1 \relax\or\K@yk{#2}\or
~\K@yl{#2}\or\K@ym{#2}\or\K@yn{#2}\fi}\def\K@yk#1{\K@dk=\K@ua\advance
~\K@dk by \K@un\advance\K@dk by \K@uo\relax\ifnum\K@dk>0 \K@dk=\K@ua
~\divide\K@dk by 2 \advance\K@dk by \K@pz\relax\K@vs\K@dl=\K@dk\times\K@tw
~\over\K@dj\xdef#1{\the\K@dl}\fi}\def\K@yl#1{\K@dk=\K@ua\advance\K@dk by
~\K@un\advance\K@dk by \K@uo\relax\ifnum\K@dk>0 \K@dk=\K@vj\advance\K@dk
~by -\K@un\advance\K@dk by \K@qa\multiply\K@dk by -1 \K@dl=\K@ua\divide
~\K@dl by 2 \advance\K@dl by \K@pz\relax\K@yj#1\fi}\def\K@ym#1{\K@dk=\K@ua
~\advance\K@dk by \K@un\advance\K@dk by \K@uo\relax\ifnum\K@dk>0 \K@dk
~=-\K@vj\advance\K@dk by \K@un\relax\K@vs\K@dl=\K@dk\times\K@tw\over\K@di
~\xdef#1{\the\K@dl}\fi}\def\K@yn#1{\K@dk=\K@ua\advance\K@dk by \K@un
~\advance\K@dk by \K@uo\relax\ifnum\K@dk>0 \K@dk=\K@vj\advance\K@dk by
~-\K@un\advance\K@dk by \K@qa\multiply\K@dk by -1 \K@dl=\K@ua\divide\K@dl
~by -2 \advance\K@dl by \K@pz\multiply\K@dl by -1 \K@yj#1\fi}\def\K@yj
~#1{\ifnum\K@dk=0 \ifnum\K@dl<0 \K@vs\K@dk=\K@dl\times\K@tw\over\K@dj\xdef
~#1{\the\K@dk}\else\global\let#1\K@cm\fi\else\ifnum\K@dl=0 \ifnum\K@dk
~<0 \K@vs\K@dl=\K@dk\times\K@tw\over\K@di\xdef#1{\the\K@dl}\else\global
~\let#1\K@cm\fi\else\ifnum\K@dk<0 \ifnum\K@dl<0 \K@yo#1\else\K@vs\K@dl
~=\K@dk\times\K@tw\over\K@di\xdef#1{\the\K@dl}\fi\else\ifnum\K@dl<0 \K@vs
~\K@dk=\K@dl\times\K@tw\over\K@dj\xdef#1{\the\K@dk}\else\K@yo#1\fi\fi\fi
~\fi}\def\K@yo#1{\K@do=\K@di sp \divide\K@do by \K@dk\K@dp=\K@dj sp
~\divide\K@dp by \K@dl\ifdim\K@do=\K@dp\K@vs\K@dh=\K@dk\times\K@tw\over
~\K@di\K@vs\K@dg=\K@dl\times\K@tw\over\K@dj\ifnum\K@dh<\K@dg\xdef#1{\the
~\K@dh}\else\xdef#1{\the\K@dg}\fi\else\ifdim\K@dp<\K@do\K@vs\K@dl=\K@dk
~\times\K@tw\over\K@di\xdef#1{\the\K@dl}\else\K@vs\K@dk=\K@dl\times\K@tw
~\over\K@dj\xdef#1{\the\K@dk}\fi\fi}
??V\def\K@er{\K@hc\K@hh\K@ee\let\gr\K@cj\let\endgr\relax\K@hi\let\D@\K@yp
??V\K@st0\K@dh\K@yq\K@yr}%
~\def\K@hi{\def\K@ys##1##2{\K@rt\let\K@rt\relax}\def\K@yt##1##2{\K@rt\let
~\K@rt\relax}\def\K@yu##1##2{\K@rt\let\K@rt\relax}\def\K@yv##1{\K@rt\let
~\K@rt\relax}}
??V\def\K@yq#1{\csname D@Entry@\the#1@\endcsname}\def\K@eb{\K@em\let\K@ec
??V\K@yw\let\K@ed\K@yx\let\K@ee\K@yy\def\K@um{\K@yz}\let\K@dz\K@za\let
??V\K@rt\egroup\global\setbox\K@ds=\hbox\bgroup\hbox to \K@en sp{\hfil\vbox
??Vto \K@eo sp{\vfil}}\K@ep\K@eq\K@er\let\gr\K@pt\let\endgr\relax\K@lh
??V\K@ja\K@zb\K@zc\K@zd\let\\\relax\setbox\K@du=\hbox\bgroup}%
??Q\begingroup\K@ce\def\K@zd{\let\hd\K@ze\let\pd\K@zf\let\tl\K@zg\let\pl
??Q\K@zh\let\pt\K@ri\def\Text{\K@bx\K@zi}\def\Txt{\K@bx\K@zj}\def\Math{\K@bx
??Q\K@zk}\def\Label{\K@bx\K@zl}\def\Vertex{\K@bx\K@zm}\def\Pt{\K@ca\K@zn}\def
??Q\Dot{\K@bx\K@zo}\def\Box{\K@cb\K@zp}\def\Frame{\K@cb\K@zq}}%
~\def\K@ze#1{\gdef\K@zr{\K@dr={#1}}}\def\K@zg#1{\gdef\K@zs{\K@dr={#1}}}\def
~\K@zf#1{\hd{\phantom{#1}}}\def\K@zh#1{\tl{\phantom{#1}}}
??V\def\K@za{\K@rt\egroup\K@dx{}{{}{1}{\gdef\noexpand\K@es{\K@es}\gdef
??V\noexpand\K@et{\K@et}\gdef\noexpand\K@eu{\K@eu}\gdef\noexpand\K@ev{\K@ev
??V}}}\K@ew\K@ex\K@ey\egroup\Displaybox\K@ds}%
??X\begingroup\K@ce\def\K@zt{$\egroup\null\egroup\K@di=\wd\K@ds\xdef\K@en
??X{\the\K@di}\K@dj=\ht\K@ds\xdef\K@eo{\the\K@dj}\K@dx{}{{\K@en}{\K@eo
??X}{}{\K@hp}}\let\K@gi\K@zu\K@zv}\def\K@zw{\K@di=\K@en\relax\K@dj=\K@eo
??X\relax\K@zv}\def\K@zv{\ifx\K@zx\K@ci\K@dm=\K@di sp \divide\K@dm by \K@zy
??X\let\K@hx\K@zy\else\K@dm=\K@zx\divide\K@di by \K@dm\edef\K@hx{\the\K@di
??X}\fi\let\K@rz\K@cm\let\K@sa\K@hx\ifx\K@zz\K@ci\K@dn=\K@dj sp \divide
??X\K@dn by \K@Aa\let\K@ho\K@Aa\else\K@dn=\K@zz\divide\K@dj by \K@dm\edef
??X\K@ho{\the\K@dj}\fi\let\K@rx\K@cm\let\K@ry\K@ho\let\K@sc\K@Ab\let\K@in
??X\K@ci\let\K@ec\K@yw\let\K@ed\K@yx\let\K@ee\K@yy\let\K@rt\egroup\global
??X\setbox\K@ds=\hbox\bgroup\hbox to \K@en sp{\vbox to \K@eo sp{\vss}\hss
??X}\K@Ac\K@ep{\llap{\box\K@ds}\K@Ad{\K@gm\K@dr}{{0pt}{0pt}{0{\the\K@dr
??X}}}}\K@Ae\let\gr\K@pt\let\endgr\relax\K@hg\K@lh\K@hb\K@ja\K@zb\K@zc
??X\K@zd\K@Af\setbox\K@du=\hbox\bgroup}\def\K@zu{\K@rt\egroup\K@ew\K@ex
??X\K@hr\egroup\Displaybox\K@ds}%
??Q\def\K@zb{\let\K@Ag/\let\K@Ah\/\catcode`\/=\active\let\K@Ai|\let\K@Aj
??Q\|\catcode`\|=\active\let\K@Ak*\let\K@Al\*\catcode`\*=\active}{\catcode
??Q`\/=\active\catcode`\|=\active\catcode`\*=\active\catcode`\(=\active
??Q\gdef\K@zc{\let(\K@Am\let/\K@An\let|\K@Ao\let*\K@Ap\let\((\let\//\let
??Q\||\let\**}\gdef\K@yw{\let/\K@Ag\let\/\K@Ah\let|\K@Ai\let\|\K@Aj\let
??Q*\K@Ak\let\*\K@Al}\gdef\K@yy{\let/\K@cj\let|\K@cj\let*\K@cj\let\//\let
??Q\||\let\**}\gdef\K@yx{\let/\K@Aq\let|\K@Ar\let*\K@As}}\def\K@Aq{/}\def
??Q\K@Ar{|}\def\K@As{*}\def\K@Am#1,#2){\ifx\K@rp\K@ci\xdef\K@rp{#1,#2}\else
??Q\ifx\K@rm\K@ci\xdef\K@rm{#1,#2}\fi\fi}\def\K@An#1{\edef\K@bf{\K@At
??Q#1.@}\ifx\K@bf\K@ci\else\global\let\K@vh\K@bf\fi}\def\K@Ao#1{\K@do=#1
??Q\K@di=\K@do\xdef\K@tx{\the\K@di}}\def\K@Ap#1{\gdef\K@vk{#1}}%
~\def\K@At#1.#2@{#1}\def\K@ex#1{#1\K@ds\K@Au\K@ds\K@Av\K@ds\K@Aw\K@ds
~\K@Ax\K@ds}
??V\def\K@ey#1{\ifmmode\ifnum\K@ho=0 \let\K@Av\K@cj\fi\fi\K@do=\K@eu sp
??V\advance\K@do by \K@Ay\K@dp=\K@ev sp \advance\K@dp by \K@Ay\relax\ifx
??V\K@hn\K@cp\else\let\K@Az\K@Ba\let\K@Av\K@cj\fi\global\setbox#1=\hbox
??V{\hskip\K@do\vtop{\offinterlineskip\vbox\bgroup\vskip\K@Ay\K@Az
??V#1\vskip\K@Ay}\hskip\K@dp}}\def\K@Az#1{\ifx\K@Bb\relax\vskip-\K@vj\vskip
??V\ifnum\K@es=-1 \K@vj\else\K@es sp\fi\box#1\vskip\K@vj\egroup\else\vskip
??V\ifnum\K@es=-1 \K@vj\else\K@es sp\fi\egroup\vskip-\K@vj\box#1\vskip\K@vj
??V\fi\vskip\ifnum\K@et=-1 -\K@vj\else\K@et sp\fi}\def\K@Ba#1{\K@di=-\K@ho
??V\advance\K@di by \K@hn\relax\multiply\K@di by -1 \K@di=\csname D@Y\the
??V\K@di @\endcsname\relax\vskip-\K@vj\vskip\ifnum\K@es=-1 \K@vj\else\K@es
??Vsp\fi\hbox{\raise\K@vj\hbox{\lower\K@di sp\box#1}}\egroup\vskip\K@vj
??V\vskip\ifnum\K@et=-1 -\K@vj\else\K@et sp\fi}%
??X\def\K@hr#1{\global\setbox#1=\hbox{\hskip\K@hk\vtop{\offinterlineskip
??X\vbox\bgroup\vskip\K@hk\ifx\K@Bc\relax\box#1\egroup\else\egroup\box#1
??X\fi\vskip\K@hk}\hskip\K@hk}}%
~\def\K@Aw#1{\global\setbox#1=\hbox{\hskip\K@Bd\vtop{\vbox{\vskip\K@Be\box
~#1} \vskip\K@Bf}\hskip\K@Bg}}\def\K@Bh#1{\ifmmode\K@Bi#1\fi}\let\K@Av\K@Bh
~\def\K@Bi#1{\global\setbox#1=\hbox{$\vcenter{\box#1}$}}\K@bg\center{\K@cz
~\let\K@Av\K@Bi}\K@bm\centre\center\def\K@Bj{\K@cz\let\K@Av\K@Bh}\K@bh
~\uncenter\K@Bj\K@bh\uncentre\K@Bj\K@bh\centermath\K@Bj\K@bh\centremath
~\K@Bj\K@bg\deep{\K@cz\let\K@Av\K@cj\let\K@Bb\K@ci\let\K@Bc\K@ci}\K@bg
~\tall{\K@cz\let\K@Av\K@cj\let\K@Bb\relax\let\K@Bc\relax}\let\K@Bb\K@ci\let
~\K@Bc\relax\let\K@Au\K@cj\K@bg\landscape{\K@cz\def\K@Au{\K@bo\K@Bk
~}}\K@bg\oplandscape{\K@cz\def\K@Au{\K@bo\K@Bl}}
??Z\begingroup\K@ce\def\K@Bk#1{\global\setbox#1=\hbox{\K@Bm\box#1\K@Bn}\K@Bo
??Z\K@Bg\K@Bd}\def\K@Bl#1{\global\setbox#1=\hbox{\K@Bp\box#1\K@Bq}\K@Bo\K@Bd
??Z\K@Bg}\def\K@Bo#1#2{\let\K@bf\K@Be\let\K@Be#1\let#1\K@Bf\let\K@Bf#2\let
??Z#2\K@bf}\def\K@Bm{\K@Br\bgroup\setbox\K@du=\hbox\bgroup\K@Bs}\def\K@Bn
??Z{\K@Bt\egroup\K@do=\ht\K@du\advance\K@do by \dp\K@du\hbox to \K@do{\hfil
??Z\vbox to \wd\K@du{\vfill\K@vf\K@vg{90}\K@xg{\box\K@du}\K@vi }}\egroup
??Z}\K@bh\endland\K@Bn\def\K@Bp{\K@Br\bgroup\setbox\K@du=\hbox\bgroup\K@Bs
??Z}\def\K@Bq{\K@Bt\egroup\K@do=\ht\K@du\advance\K@do by \dp\K@du\hbox to
??Z\K@do{\hfil\vbox to \wd\K@du{\vfill\K@vf\K@vg{-90}\K@xf{\box\K@du}\K@vi
??Z}}\egroup}\K@bh\endopland\K@Bq\def\K@Bu{\K@Br\bgroup\setbox\K@du=\hbox
??Z\bgroup\K@Bs}\K@bg\endflip{\K@Bt\egroup\K@do=\ht\K@du\ht\K@du=\dp\K@du\dp
??Z\K@du=\K@do\hbox to \wd\K@du{\hfil\K@vf\K@vg{180}\hbox to \z@{\box\K@du
??Z\hss}\K@vi }\egroup}\def\K@Bv{\K@cg{currentpoint currentpoint translate 1
??Z-1 scale neg exch neg exch translate}}\def\K@Bw{\K@cg{currentpoint
??Zcurrentpoint translate -1 1 scale neg exch neg exch translate}}\def\K@Bx
??Z{\K@Br\bgroup\setbox\K@du=\hbox\bgroup\K@Bs}\K@bg\endhmir{\K@Bt\egroup
??Z\K@do=\ht\K@du\ht\K@du=\dp\K@du\dp\K@du=\K@do\hbox{\K@vf\K@Bv\box\K@du
??Z\K@vi}\egroup}\def\K@By{\K@Br\bgroup\setbox\K@du=\hbox\bgroup\K@Bs}\K@bg
??Z\endvmir{\K@Bt\egroup\hbox to \wd\K@du{\hfil\K@vf\K@Bw\hbox to \z@{\box
??Z\K@du\hss}\K@vi}\egroup}\def\K@Bz#1{\K@Br\bgroup\edef\K@bc{\noexpand\K@vg
??Z{\K@At#1.@}}\setbox\K@du=\hbox\bgroup\K@Bs}\K@bg\endrot{\K@Bt\egroup\K@do
??Z=\ht\K@du\advance\K@do by -\dp\K@du\hbox{\phantom{\copy\K@du}\hbox to \z@
??Z{\hskip-.5\wd\K@du\raise.5\K@do\hbox{\K@vf\K@bc\hbox to \z@{\hss\vbox to
??Z\z@{\vss\box\K@du\vss}\hss}\K@vi}\hss}}\egroup}\endgroup
~\K@bk\land{\relax\K@bo\K@Bm}\K@bg\opland{\relax\K@bo\K@Bp}\K@bg\flip{\relax
~\K@bo\K@Bu}\K@bg\hmir{\relax\K@bo\K@Bx}\K@bg\hmirror{\relax\K@bo\K@Bv}\K@bg
~\vmir{\relax\K@bo\K@By}\K@bg\rot{\relax\K@bo\K@Bz}\def\K@Br#1{\ifmmode
~\ifinner\def\K@Bs{$}\else\def\K@Bs{$\displaystyle}\fi\def\K@Bt{$}\else
~\let\K@Bs\bgroup\let\K@Bt\egroup\fi #1\let\K@Br\relax}\K@bg\squish{\relax
~\ifmmode\def\next{\mathpalette\K@Ca}\else\let\next\K@Cb\fi\next}\def
~\K@Cb#1{\hbox to \z@{\hss#1\hss}}\def\K@Ca#1#2{\hbox to \z@{\hss$\m@th
~#1{#2}$\hss}}\K@bg\squash{\relax\ifmmode\def\next{\mathpalette\K@Cc}\else
~\let\next\K@Cd\fi\next}\def\K@Cd#1{\setbox\z@\hbox to \z@{\hss#1\hss
~}\K@Ce}\def\K@Cc#1#2{\setbox\z@\hbox to \z@{\hss$\m@th#1{#2}$\hss}\K@Ce
~}\def\K@Ce{\ht\z@\z@\dp\z@\z@\box\z@}
??Y\begingroup\K@ce\def\K@Cf{\K@Br\bgroup\hbox\bgroup\K@Cg\vtop\bgroup
??Y\vbox\bgroup\K@Ch\vskip\K@Be\vskip\K@Ci\hbox\bgroup\hskip\K@Bd\hskip
??Y\K@Ci\K@Bs}\K@bg\endframe{\K@Bt\hskip\K@Bg\hskip\K@Ci\egroup\egroup
??Y\vskip\K@Bf\vskip\K@Ci\K@Ch\egroup\K@Cg\egroup\egroup}%
~\K@bg\frame{\relax\K@bq\K@Cf}\K@bg\framed{\let\K@Ax\K@Cj}\let\K@Ax\K@cj
~\def\K@Cj#1{\global\setbox#1=\hbox{\K@Cg\vtop{\vbox{\K@Ch\box#1}\K@Ch
~}\K@Cg}}\def\K@Cg{\K@Ck\vrule width \K@Cl\K@Cm}\def\K@Ch{\K@Ck\hrule
~height \K@Cl\K@Cm}
??Y\def\K@Cn#1#2{\K@Br\bgroup\K@Co#1@\K@qc\K@qd\K@Co#2@\K@pz\K@qa\setbox
??Y8=\hbox\bgroup\K@Cg\vtop\bgroup\vbox\bgroup\K@Ch\vskip\K@Be\vskip\K@Ci
??Y\hbox\bgroup\hskip\K@Bd\hskip\K@Ci\K@Bs}\def\K@Cp{\K@Bt\hskip\K@Bg\hskip
??Y\K@Ci\egroup\egroup\vskip\K@Bf\vskip\K@Ci\K@Ch\egroup\K@Cg\egroup
??Y\dimen4=-\wd8 \advance\dimen4 by \K@qc\relax\multiply\dimen4 by -1
??Y\advance\dimen4 by \K@pz\relax\ifdim\dimen4>\z@\dimen8=\ht8 \dimen6=-\dp
??Y8 \advance\dimen8 by \K@qa\advance\dimen6 by \K@qd\multiply\dimen6 by -1
??Y\K@do=\K@qc\K@dp=\K@pz\multiply\K@dp by -1 \setbox\K@du=\hbox{\ifdim\K@do
??Y>\z@\hskip\K@do\fi\vrule width\dimen4 height \dimen8 depth \dimen6 \ifdim
??Y\K@dp>\z@\hskip\K@dp\fi}\multiply\K@dp by -1 \hbox{\K@Cq\box\K@du\K@Cr
??Y\llap{\K@ha1\vrule width\wd8 height \ht8 depth \dp8 \endgr\ifdim\K@dp>\z@
??Y\hskip\K@dp\fi}\llap{\box8 \ifdim\K@dp>\z@\hskip\K@dp\fi}}\else\box8 \fi
??Y\egroup}\def\K@Cs{\K@Bt\hskip\K@Bg\hskip\K@Ci\egroup\egroup\vskip\K@Bf
??Y\vskip\K@Ci\K@Ch\egroup\K@Cg\egroup\K@do=-\wd8 \advance\K@do by \K@qc
??Y\multiply\K@do by -1 \advance\K@do by \K@pz\relax\ifdim\K@do>\z@\dimen
??Y8=\ht8 \dimen6=-\dp8 \setbox\K@du=\hbox{\phantom{\copy8}}\ifdim\K@qa>\z@
??Y\setbox\K@du=\vbox{\offinterlineskip\hbox to \wd8{\hskip\K@qc\vrule
??Ywidth \K@do height \K@qa\hss} \box\K@du}\setbox8=\vbox{\vskip\K@qa\box8}\fi
??Y\ifdim\K@qd<\z@\K@dp=\K@qd\multiply\K@dp by -1 \setbox\K@du=\vtop
??Y{\offinterlineskip\box\K@du\hbox to \wd8{\hskip\K@qc\vrule width \K@do
??Yheight \K@dp\hss}}\setbox8=\vtop{\box8\vskip\K@dp}\fi\ifdim\K@qa<\z@
??Y\advance\dimen8 by \K@qa\fi\ifdim\K@qd>\z@\advance\dimen6 by \K@qd\fi
??Y\multiply\dimen6 by -1 \K@do=\K@qc\multiply\K@do by -1 \K@dp=\K@pz\setbox
??Y\K@du=\hbox{\ifdim\K@do>\z@\vrule width \K@do height \dimen8 depth \dimen
??Y6 \fi\box\K@du\ifdim\K@dp>\z@\vrule width \K@dp height \dimen8 depth
??Y\dimen6 \fi}\hbox{\K@Cq\box\K@du\K@Cr\llap{\hbox{\ifdim\K@do>\z@\hskip
??Y\K@do\fi\box8 \ifdim\K@dp>\z@\hskip\K@dp\fi}}}\else\box8 \fi\egroup
??Y}\K@bg\endshade{\K@Bt\hskip\K@Bg\hskip\K@Ci\egroup\egroup\vskip\K@Bf
??Y\vskip\K@Ci\K@Ch\egroup\K@Cg\egroup\dimen4=-\wd8 \advance\dimen4 by
??Y\K@qc\relax\multiply\dimen4 by -1 \advance\dimen4 by \K@pz\relax\ifdim
??Y\dimen4>\z@\dimen8=\ht8 \dimen6=-\dp8 \advance\dimen8 by \K@qa\advance
??Y\dimen6 by \K@qd\multiply\dimen6 by -1 \K@do=\K@qc\K@dp=\K@pz\multiply
??Y\K@dp by -1 \setbox\K@du=\hbox{\ifdim\K@do>\z@\hskip\K@do\fi\vrule
??Ywidth\dimen4 height \dimen8 depth \dimen6 \ifdim\K@dp>\z@\hskip\K@dp\fi
??Y}\multiply\K@dp by -1 \hbox{\K@Cq\box\K@du\K@Cr\llap{\K@ha1\dimen6=\K@Cl
??Y\dimen2=\wd8 \K@do=\ht8 \K@dp=\dp8 \ifdim\K@qd<\z@\advance\K@dp by -\dimen
??Y6 \fi\ifdim\K@qa>\z@\advance\K@do by -\dimen6 \fi\ifdim\K@qc<\z@
??Y\advance\dimen2 by -\dimen6 \fi\ifdim\K@pz>\z@\advance\dimen2 by -\dimen
??Y6 \else\dimen6=\z@\fi\vrule width\dimen2 height \K@do depth \K@dp\hskip
??Y\dimen6 \endgr\ifdim\K@dp>\z@\hskip\K@dp\fi}\llap{\box8 \ifdim\K@dp>\z@
??Y\hskip\K@dp\fi}}\else\box8 \fi\egroup}\def\K@Co#1,#2@#3#4{\edef
??Y#3{#1}\edef#4{#2}}\endgroup
~\K@bg\shade{\relax\K@bq\K@Cn}\K@bg\NoTeXviewHack{\def\endshade{\K@Cp}}\K@bg
~\TeXshades{\def\endshade{\K@Cs}}
??N\begingroup\K@ce\def\K@zp{\K@rt\let\K@pu\K@ci\let\K@rp\K@ci\let\K@rm
??N\K@ci\let\K@qj\K@cm\let\K@qi\K@cm\let\K@qm\K@cm\let\K@ql\K@cm\def\K@rt
??N{\egroup\K@Ct}\setbox\K@du=\hbox\bgroup}\def\K@Ct{\ifx\K@rm\K@ci\else
??N\K@Cu\expandafter\K@Cv\K@rp @\K@tn\K@tm\edef\K@bc{\noexpand\K@sc @\K@tn
??N,\noexpand\K@tu @\K@tm,\noexpand\K@tv @}\K@bc\expandafter\K@Cv\K@rm @\K@tp
??N\K@to\edef\K@bc{\noexpand\K@sc @\K@tp,\noexpand\K@uc @\K@to,\noexpand\K@ud
??N@}\K@bc\ifnum\K@tu>\K@uc\relax\K@ct\K@tu\K@uc\fi\ifnum\K@tv>\K@ud\relax
??N\K@ct\K@tv\K@ud\fi\K@di=\K@tu\multiply\K@di by -1 \advance\K@di by \K@uc
??N\relax\K@dj=\K@tv\multiply\K@dj by -1 \advance\K@dj by \K@ud\relax\K@do
??N=\K@tu sp \advance\K@do by \K@qj sp \advance\K@do by \K@qm sp \K@dp=\K@tv
??Nsp \advance\K@dp by \K@qi sp \advance\K@do by \K@ql sp \edef\K@bc{\noexpand
??N\K@ve0@\the\K@do,\the\K@dp @{\noexpand\K@Bs\vrule width \the\K@di sp
??Nheight \the\K@dj sp \noexpand\K@Bt}}\K@bc\fi}%
??Q\def\K@Cu{\ifx\K@pu\K@ci\def\K@Bs{\relax}\def\K@Bt{\relax}\else\edef\K@Bs
??Q{\noexpand\K@ha{\K@pu}}\def\K@Bt{\endgr}\fi}\endgroup
??N\def\K@zq{\K@rt\let\K@pu\K@ci\let\K@rp\K@ci\let\K@rm\K@ci\let\K@qj\K@cm
??N\let\K@qi\K@cm\let\K@qm\K@cm\let\K@ql\K@cm\let\K@ra\K@Cw\def\K@rt
??N{\egroup\K@Cx}\setbox\K@du=\hbox\bgroup\def\rw{\K@md}}\def\K@Cx{\ifx\K@rm
??N\K@ci\else\bgroup\K@Cu\expandafter\K@Cv\K@rp @\K@tn\K@tm\edef\K@bc
??N{\noexpand\K@sc @\K@tn,\noexpand\K@tu @\K@tm,\noexpand\K@tv @}\K@bc
??N\expandafter\K@Cv\K@rm @\K@tp\K@to\edef\K@bc{\noexpand\K@sc @\K@tp
??N,\noexpand\K@uc @\K@to,\noexpand\K@ud @}\K@bc\ifnum\K@tu>\K@uc\relax\K@ct
??N\K@tu\K@uc\fi\ifnum\K@tv>\K@ud\relax\K@ct\K@tv\K@ud\fi\K@di=\K@tu
??N\multiply\K@di by -1 \advance\K@di by \K@uc\relax\K@dj=\K@tv\multiply\K@dj
??Nby -1 \advance\K@dj by \K@ud\relax\K@do=\K@tu sp \advance\K@do by \K@qj
??Nsp \advance\K@do by \K@qm sp \K@dp=\K@tv sp \advance\K@dp by \K@qi sp
??N\advance\K@do by \K@ql sp \edef\K@bc{\noexpand\K@ve0@\the\K@do,\the\K@dp
??N@{\noexpand\K@Bs\noexpand\K@Cy{\the\K@di sp}{\the\K@dj sp}\noexpand\K@Bt
??N}}\K@bc\egroup\fi}\def\K@Cy#1#2{\hbox to #1{\vbox to #2{\vfil}\hfil}\llap
??N{\vbox to #2{ \vfil\hbox to #1{\vrule width \K@ra height #2 \hss\vrule
??Nwidth \K@ra height #2} \vfil}}\llap{\hbox to #1{\hfil\vbox to #2{ \hrule
??Nheight \K@ra width #1 \vss\hrule height \K@ra width #1}\hfil}}}\endgroup
??X\def\K@gj#1{\bgroup\K@dw\K@dx{\K@dy\K@dr}{\the\K@dr}\def\K@gi{\K@Cz}\def
??X\Modify{\K@by\K@Af\K@zt}\global\setbox\K@ds=\vbox\bgroup\offinterlineskip
??X\hbox\bgroup$}\K@bk\endFigure{\K@gi\egroup}%
~\def\K@fi{\K@gj}\K@bk\Figure{\bgroup\K@br\K@ej\K@ek\K@el\everyFigure\K@fi
~{Figure}}\K@bl\everyFigure\relax
??X\def\K@Cz{$\egroup\null\egroup\K@dx{\K@di=\wd\K@ds\K@dj=\ht\K@ds}{{\the
??X\K@di}{\the\K@dj}{}{\K@hp}}\K@ew\K@ex\K@hr\egroup\Displaybox\K@ds}%
~\K@bk\newFigure{\bgroup\K@ez\K@Da}\def\K@Da#1{\egroup\expandafter\ifx
~\csname D@every#1@\endcsname\relax\expandafter\expandafter\expandafter\let
~\expandafter\expandafter\csname D@@#1@@\endcsname\csname#1\endcsname
~\expandafter\def\csname D@every#1@\endcsname{\expandafter\expandafter
~\expandafter\let\expandafter\expandafter\csname#1\endcsname\csname
~D@@#1@@\endcsname}\fi\expandafter\K@bk\csname#1\endcsname{\bgroup\K@br
~\csname D@every#1@\endcsname\K@ej\K@ek\K@el\everyFigure\csname
~every#1\endcsname\K@fi{#1}}\expandafter\K@bk\csname end#1\endcsname{\K@gi
~\egroup}\expandafter\let\csname every#1\endcsname\relax}
??X\def\K@fw#1#2#3{\bgroup\K@dw\K@dx{\K@dy\K@dr}{\the\K@dr}\let\K@pe\K@Db
??X\K@pg#2;@\K@en\K@zx\K@zy\K@pg#3;@\K@eo\K@zz\K@Aa\K@dx{}{{\K@en}{\K@eo
??X}{}{\K@hp}}\let\K@Dc\K@zu\let\K@hk\K@hl\let\K@Ad\K@ck\global\setbox\K@ds
??X=\hbox to \K@en sp {\vbox to \K@eo sp{\vss}\hss}\K@by\K@Af\K@zw}\K@bk
??X\endGraph{\K@Dc\egroup}%
~\def\K@fg{\K@fw}\K@bk\Graph{\bgroup\K@br\K@ej\K@ek\K@el\everyGraph\K@fg
~{Graph}}\K@bl\everyGraph\relax
??X\def\K@Ad{\K@dx}%
~\def\K@Af{\def\K@um##1##2##3##4##5{}}\K@bk\newGraph{\bgroup\K@ez\K@Dd}\def
~\K@Dd#1{\egroup\K@br\expandafter\ifx\csname D@every#1@\endcsname\relax
~\expandafter\expandafter\expandafter\let\expandafter\expandafter\csname
~D@@#1@@\endcsname\csname#1\endcsname\expandafter\def\csname
~D@every#1@\endcsname{\expandafter\expandafter\expandafter\let\expandafter
~\expandafter\csname#1\endcsname\csname D@@#1@@\endcsname}\fi\expandafter
~\K@bk\csname#1\endcsname{\bgroup\K@br\csname D@every#1@\endcsname\K@ej
~\K@ek\K@el\everyGraph\csname every#1\endcsname\K@fg{#1}}\expandafter
~\K@bk\csname end#1\endcsname{\K@Dc\egroup}\expandafter\let\csname
~every#1\endcsname\relax}
??X\def\K@Db#1,#2;@#3#4#5{\K@do=#1\relax\ifdim\K@do<\z@\multiply\K@do by -1
??X\fi\K@di=\K@do\edef#3{\the\K@di}\def\K@bf{#2}\ifx\K@bf\K@ci\else\setbox
??X\K@du=\hbox{\K@do=\K@pn#2pt}\ifdim\wd\K@du>\z@\K@do=\K@pn#2\relax\ifdim
??X\K@do=\z@\else\edef#4{\the\K@do}\fi\else\K@di=\K@pn#2\relax\ifnum\K@di
??X>\z@\edef#5{\the\K@di}\fi\fi\fi}\endgroup
~\K@bg\gridlines{\K@cz\K@De\let\K@gf\K@gd\let\K@gh\K@go}\K@bg\overgrid
~{\K@cz\K@De\let\K@gf\K@ge\let\K@gh\K@gn\let\K@Ac\K@cr}\let\K@Ac\K@cq\def
~\K@De{\let\K@fz\K@gd\let\K@ep\K@Df\let\K@ep\K@Df\def\K@Cz{\K@zt\K@zu
~}}\let\K@ep\relax\def\K@Dg#1#2{\K@do=1sp \K@do=#1\K@do\ifdim\K@do>1sp
~\else\ifdim\K@do<\z@\else\K@do=1000sp \K@do=#1\K@do\ifdim\K@do=\z@\let
~#2\K@ci\else\def#2{#1}\fi\fi\fi}\def\K@Df{\let\K@tz\K@ci\let\K@ua\K@ci
~\let\K@rp\K@ci\let\K@rm\K@ci\let\K@zr\K@ci\let\K@zs\K@ci\let\K@qd\K@cm
~\let\K@qc\K@ci\let\K@qa\K@cm\let\K@pz\K@ci\let\K@qj\K@cm\let\K@qi\K@cm
~\let\K@qm\K@cm\let\K@ql\K@cm\def\K@qb{-65536}\let\K@py\K@qb\let\K@pu\K@Dh
~\let\K@qs\relax\let\K@rb\relax\let\K@rc\relax\let\K@rl\K@ci\let\K@qe
~\relax\let\K@qf\relax\let\K@qg\relax\let\K@qh\relax\let\K@rq\relax\let
~\K@rs\relax\ifnum\K@ho>0 \K@di=0 \K@cu\let\K@tm\K@rx\edef\K@tn{\the\K@di
~}\let\K@to\K@ry\let\K@tp\K@tn {\K@Af\K@tt0{{Line}{}}{}}\ifnum\K@di<\K@hx
~\advance\K@di by 1 \K@cv\fi\ifnum\K@hx>0 \K@di=0 \K@cu\edef\K@tm{\the
~\K@di}\let\K@tn\K@rz\let\K@to\K@tm\let\K@tp\K@sa {\K@Af\K@tt
~0{{Line}{}}{}}\ifnum\K@di<\K@ho\advance\K@di by 1 \K@cv\fi}\K@bk\newcell
~{\bgroup\K@ez\K@Di}\def\K@Di#1{\egroup\expandafter\K@bk\csname#1\endcsname
~{\K@Dj{#1}} \expandafter\K@bk\csname a#1\endcsname{\null\K@yu0{#1}}
~\expandafter\K@bk\csname r#1\endcsname{\null\K@yu1{#1}} \expandafter\K@bk
~\csname rd#1\endcsname{\null\K@yu2{#1}} \expandafter\K@bk\csname
~d#1\endcsname{\null\K@yu3{#1}} \expandafter\K@bk\csname ld#1\endcsname
~{\null\K@yu4{#1}} \expandafter\K@bk\csname l#1\endcsname{\null\K@yu5{#1}}
~\expandafter\K@bk\csname lu#1\endcsname{\null\K@yu6{#1}} \expandafter\K@bk
~\csname u#1\endcsname{\null\K@yu7{#1}} \expandafter\K@bk\csname
~ru#1\endcsname{\null\K@yu8{#1}} \expandafter\K@bk\csname b#1\endcsname
~{\null\K@yu{10}{#1}} \expandafter\edef\csname D@#1b@x\endcsname
~##1##2{\lower\K@vj\hbox{$\expandafter\noexpand\csname#1box\endcsname
~{##2}$}} \expandafter\ifx\csname#1box\endcsname\relax\expandafter\K@bi
~\expandafter\edef\csname#1box\endcsname##1{\hbox to ##1{$\expandafter
~\noexpand\csname#1fill\endcsname$}} \fi\expandafter\let\csname
~D@#1@lwp@\endcsname\K@cm\expandafter\let\csname D@#1@lp@\endcsname\K@cm
~\expandafter\let\csname D@#1@cp@\endcsname\K@cm\expandafter\let\csname
~D@#1@jp@\endcsname\K@cm\expandafter\let\csname D@#1@bp@\endcsname\K@cm
~\expandafter\let\csname D@#1@lpt@\endcsname\K@ci\expandafter\let\csname
~D@#1@ppt@\endcsname\K@ci\expandafter\let\csname D@#1@pp@\endcsname\K@cm
~\expandafter\let\csname D@#1@ap@\endcsname\K@cm}\def\K@Dj#1{\K@rt\K@bw
~\K@Dk\edef\K@pb{\K@Dl{#1}{}}\let\K@oi\K@pb\global\K@dr={{#1}{}}\def\K@rt
~{\egroup\K@Dm}\setbox\K@du=\hbox\bgroup}
??S\def\K@Dn#1{\K@rt\K@Dk\edef\K@pb{\K@Dl{Fillcell}{}}\let\K@oi\K@pb\global
??S\K@dr={{Fillcell}{#1}}\def\K@rt{\egroup\K@Dm}\setbox\K@du=\hbox\bgroup}\def
??S\K@Do#1{\K@rt\K@Dk\edef\K@pb{\K@Dl{Boxcell}{}}\let\K@oi\K@pb\global\K@dr
??S={{Boxcell}{#1}}\def\K@rt{\egroup\K@Dm}\setbox\K@du=\hbox\bgroup}\def\K@Dp
??S{\K@rt\K@Dk\edef\K@pb{\K@Dl{Rule}{}}\let\K@oi\K@pb\edef\K@bc{\global
??S\K@dr={{\K@Dq}}}\K@bc\def\K@rt{\egroup\expandafter\K@Dr\the\K@dr\K@Dm
??S}\setbox\K@du=\hbox\bgroup\def\rw{\K@mb}}\def\K@Dr#1{\K@do=#1 \divide
??S\K@do by 2 \K@di=\K@pd\advance\K@di by \K@do\xdef\K@pd{\the\K@di}\K@di
??S=\K@qy\advance\K@di by \K@do\xdef\K@qy{\the\K@di}\global\K@dr
??S={{Rule}{#1}}}\def\K@Dk{\let\K@rp\K@ci\let\K@rm\K@ci\let\K@vh\K@cm\let
??S\K@vk\K@cn\let\K@tx\K@Ds\let\K@zr\K@ci\let\K@zs\K@ci\let\K@pu\K@ci\let
??S\K@qs\relax\let\K@rc\relax\let\K@rb\relax\let\K@qy\K@cm\let\K@rl\K@ci
??S\let\K@pc\K@cm\let\K@oj\K@pc\let\K@pd\K@cm\let\K@ok\K@pd\let\K@qt\K@ci
??S\let\K@re\K@ci\let\K@rf\K@ci\let\K@rq\relax\let\K@rs\relax\let\K@qe\relax
??S\let\K@qf\relax\let\K@qg\relax\let\K@qh\relax\let\K@qj\K@cm\let\K@qi\K@cm
??S\let\K@qm\K@cm\let\K@ql\K@cm\let\K@qb\K@cm\let\K@qd\K@cm\let\K@qc\K@ci
??S\let\K@py\K@cm\let\K@qa\K@cm\let\K@pz\K@ci\K@dq={}}\def\K@Dm{\ifx\K@rc
??S\K@ci\expandafter\K@Dt\K@qy @\K@di\K@Du {\expandafter\K@Dv\the\K@dr}\edef
??S\K@qy{\the\K@di}\fi\ifx\K@rp\K@ci\else\expandafter\K@Dw\fi}\def\K@Dw{\ifx
??S\K@rm\K@ci\ifnum\K@tx>0 \expandafter\K@Cv\K@rp @\K@tn\K@tm\edef\K@bc
??S{\noexpand\K@tt9{\the\K@dr}{\the\K@dq}}{\K@gx\let\K@up\K@cj\K@bc }\fi
??S\else\let\K@Dx\relax\let\K@Dy\relax\def\K@Dz{\let\K@up\K@cj\K@bc
??S}\expandafter\K@Cv\K@rp @\K@tn\K@tm\expandafter\K@Cv\K@rm @\K@tp\K@to
??S\let\K@tz\K@ci\let\K@ua\K@ci\K@Ea\K@zs\K@qe\K@qg\K@rq\K@re\K@rp\K@tz
??S\K@uk\K@ul\K@Eb\K@Ec\K@tn\K@tm\K@Dy\K@Ea\K@zr\K@qf\K@qh\K@rs\K@rf\K@rm
??S\K@ua\K@un\K@uo\K@Ed\K@Ee\K@tp\K@to\K@Dx\edef\K@bc{\noexpand\K@tt0{\the
??S\K@dr}{\the\K@dq}}{\K@gx\K@Dz\K@Dy\K@Dx }\fi}\def\K@Ea
??S#1#2#3#4#5#6#7#8#9{\ifx#1\K@ci\def\K@fx{\K@Ef#2#3#4#5#6#7#8#9}\else\def
??S\K@fx{\K@Eg#1#7#8#9#2#3#4}\fi\K@fx}\def\K@Ef#1#2#3#4#5#6#7#8{\let#1\relax
??S\let#2\relax\ifx\K@sc\K@Ab\else\ifx#3\relax\ifx#4\K@ci\expandafter
??S\K@Eh#5@#6#7#8\fi\fi\fi\def\K@fx##1##2##3##4##5{}\K@fx}\def\K@Eg
??S#1#2#3#4#5#6#7#8#9{\def\K@Dz{\K@bc}\setbox\K@dt=\hbox{$\vertexstyle\K@gx
??S#1\the\K@dr$}\K@do=\wd\K@dt\K@di=\K@do\edef#2{\the\K@di}\K@do=\ht\K@dt\K@di
??S=\K@do\edef#3{\the\K@di}\K@do=\dp\K@dt\K@di=\K@do\edef#4{\the\K@di}\ifx
??S#5\relax\def#8{\relax}\else\K@ty{-1}#5#2\K@di=#5\relax\multiply\K@di by
??S-2 \edef#8{\hskip\the\K@di sp}\fi\ifx#6\relax\def#9{\lower\noexpand\K@vj
??S}\else\K@di=-\K@vj\advance\K@di by #6\relax\edef#9{\raise\the\K@di
??Ssp}\fi\def\K@fx{\K@Ei#1#7#8#9}\K@fx}\def\K@Ei#1#2#3#4#5#6#7{\edef
??S#7{\noexpand#1\ifx#2\relax\noexpand\K@Ej1@#5,#6@\else\noexpand\K@ve
??S1@#5sp,#6sp@\fi {#4\hbox{#3\noexpand\noexpand\noexpand\K@gx $\noexpand
??S\noexpand\noexpand\vertexstyle\noexpand\the\noexpand\K@dr$}}}}\def\K@Eh
??S#1,#2@#3#4#5{\K@Ek#1.,#2.@#3#4#5}\def\K@Ek#1.#2,#3.#4@#5#6#7{\def\K@bf
??S{#2#4}\ifx\K@bf\K@ci\ifnum#1>-1 \ifnum#1>\K@hx\relax\else\ifnum#3>-1
??S\ifnum#3>\K@ho\relax\else\K@di=-\K@ho\advance\K@di by #3 \multiply\K@di
??Sby -1 \K@ts{\the\K@di}{#1}\K@bf\ifx\K@bf\relax\else\expandafter\K@El
??S\K@bf#5#6#7\def\K@Dz{\K@bc}\fi\fi\fi\fi\fi\fi}%
~\def\K@Cv#1,#2@#3#4{\gdef#3{#1}\gdef#4{#2}}
??V\def\K@yu#1#2{\K@rt\K@Em\def\K@oq{#1}\def\K@ot{#2}\global\K@dr
??V={{#2}{}}\def\K@rt{\egroup\K@En{#1}}\setbox\K@du=\hbox\bgroup}\def\K@ys
??V#1#2{\K@rt\K@Em\def\K@oq{#1}\def\CellName{Fillcell}\global\K@dr
??V={{Fillcell}{#2}}\def\K@rt{\egroup\K@En{#1}}\setbox\K@du=\hbox\bgroup}\def
??V\K@yt#1#2{\K@rt\K@Em\def\K@oq{#1}\def\CellName{Boxcell}\global\K@dr
??V={{Boxcell}{#2}}\def\K@rt{\egroup\K@En{#1}}\setbox\K@du=\hbox\bgroup}\def
??V\K@yv#1{\K@rt\K@Em\def\K@oq{#1}\def\CellName{Rule}\edef\K@bc{\global\K@dr
??V={{\K@Dq}}}\K@bc\def\K@rt{\egroup\expandafter\K@Eo\the\K@dr\K@En
??V{#1}}\setbox\K@du=\hbox\bgroup\def\rw{\K@mb}}\def\K@Eo#1{\K@do=#1 \divide
??V\K@do by 2 \K@di=\K@pd\advance\K@di by \K@do\xdef\K@pd{\the\K@di}\K@di
??V=\K@qy\advance\K@di by \K@do\xdef\K@qy{\the\K@di}\global\K@dr
??V={{Rule}{#1}}}\def\K@Em{\let\K@qu\K@qw\let\K@qx\relax\let\K@pu\K@ci\let
??V\K@qs\relax\let\K@qp\K@qr\def\K@ow{0pt}\let\K@rc\relax\let\K@rb\relax
??V\let\K@qy\K@cm\let\K@rd\relax\let\K@py\K@cm\let\K@qb\K@cm\let\K@pb\K@ci
??V\let\K@rl\K@ci\let\K@pc\K@cm\let\K@pd\K@cm\let\K@qt\K@ci\let\K@re\K@ci
??V\let\K@rf\K@ci\let\K@qj\K@cm\let\K@qi\K@cm\let\K@qm\K@cm\let\K@ql\K@cm
??V\let\K@qd\K@cm\let\K@qc\K@ci\let\K@qa\K@cm\let\K@pz\K@ci\let\K@og\K@ci
??V\let\K@ox\K@ci\K@dq={}}\def\K@Ep#1#2#3#4#5#6#7#8#9{\ifcase#1 #2\or#3\or
??V#4\or#5\or#6\or #7\or#8\or#9\fi}\def\K@En#1{\def\K@oq{#1}\let\K@fx\K@ci
??V\ifx\K@og\K@ci\ifnum#1=0 \let\K@fx\relax\else {\K@tk{#1}}\fi\else\ifnum
??V#1=10 \edef\K@to{\the\K@df}\edef\K@tp{\the\K@de}\expandafter\K@Eq\K@og
??V@\K@tm\K@tn\K@Er{\let\K@fx\relax}\K@Es\K@oq\else\ifnum#1=0 \edef\K@tm
??V{\the\K@df}\edef\K@tn{\the\K@de}\expandafter\K@Eq\K@og @\K@to\K@tp\K@Er
??V{\let\K@fx\relax}\K@Es\K@oq\else {\K@tk{#1}}\fi\fi\fi\let\K@Et\K@cm
??V\ifx\K@qp\relax\expandafter\K@Eu\K@ox{}{}\fi\K@Ev\K@oq\ifx\K@fx\K@ci\ifx
??V\K@qs\K@ci\let\K@pu\K@cl\fi\ifx\K@rc\K@ci\expandafter\K@Dt\K@qy @\K@dk
??V\K@Du {\expandafter\K@Dv\the\K@dr}\edef\K@qy{\ifx\K@rb\relax1\else2\fi\the
??V\K@dk}\fi\ifx\K@rd\relax\K@Ew\K@oq{\the\K@dg}\K@fx\global\advance\K@dg
??Vby 1 \else\K@dk=\K@im\advance\K@dk by -1 \xdef\K@im{\the\K@dk}\K@Ew\K@oq
??V\K@im\K@fx\fi\fi\K@fx}\def\K@Eu#1#2{\def\K@bf{#1}\ifx\K@bf\K@ci\let
??V\K@bc\relax\else\ifx\K@bf\K@cl\ifx\K@pb\K@ci\def\K@bf{\K@Dl{\K@ot
??V}{}}\else\let\K@bf\K@pb\fi\fi\ifdim\K@bf pt=.5pt \ifnum#2>\K@Et\relax
??V\def\K@Et{#2}\fi\fi\let\K@bc\K@Eu\fi\K@bc}\def\K@Es#1{\ifnum\K@to>\K@tm
??V\ifnum\K@tp>\K@tn\def#1{2}\else\ifnum\K@tp=\K@tn\def#1{3}\else\def
??V#1{4}\fi\fi\else\ifnum\K@to=\K@tm\ifnum\K@tp>\K@tn\def#1{1}\else
??V\ifnum\K@tp=\K@tn\relax\else\def#1{5}\fi\fi\else\ifnum\K@tp>\K@tn\def
??V#1{8}\else\ifnum\K@tp=\K@tn\def#1{7}\else\def#1{6}\fi\fi\fi\fi}\def
??V\K@Er#1{\ifnum\K@to<0\relax #1\else\ifnum\K@to>\K@ho\relax #1\else\ifnum
??V\K@tp<0\relax #1\else\ifnum\K@tp>\K@hx\relax #1\fi\fi\fi\fi}\def\K@Dt
??V#1#2@#3#4#5{\if#1=#3=#2 \else #3=#1#2 \advance#3 by #4 \advance#3 by #5
??V\fi}\def\K@Ew#1#2#3{\edef#3{\gdef\expandafter\noexpand\csname
??VD@Cell@#2@\endcsname{\noexpand\K@Ex#1@\the\K@dr @\K@tm @\K@tn @\K@to @\K@tp
??V@{\the\K@dq}@{{\K@pb}{\K@rl}}{{\K@pc}{\K@pd}}{{\K@qb}{\K@py}}{{\K@qc
??V}{\K@qd}}{{\K@pz}{\K@qa}}{{\K@pu}{\K@qt}}{{\K@qj}{\K@qm}}{{\K@qi}{\K@ql
??V}}@{\K@qy}{\K@re}{\K@rf}@}}}\def\K@Eq#1@#2@#3#4{\K@di=\K@At#2.@\relax
??V\multiply\K@di by -1 \advance\K@di by \K@df\edef#3{\the\K@di}\K@di=\K@At
??V#1.@\relax\advance\K@di by \K@de\edef#4{\the\K@di}}\def\K@Ey#1{\ifnum
??V#1=1\relax\ifx\K@qx\relax\K@Ez\K@tm\K@tn\K@tp\K@Et\fi\else\ifnum
??V#1=5\relax\ifx\K@qx\relax\K@Ez\K@tm\K@tp\K@tn\K@Et\fi\else\ifnum\K@tn
??V=\K@tp\else\K@Fa\fi\fi\fi}\def\K@Ez#1#2#3#4{{\K@di=\K@sw\relax\advance
??V\K@di by 1 \K@Fb{A:r#1,c#2-#3}\expandafter\xdef\csname D@as@\K@sw
??V@\endcsname{\noexpand\K@Fc{#1}{#2}{#3}{#4}\expandafter\noexpand\csname
??VD@as@\the\K@di @\endcsname}\xdef\K@sw{\the\K@di}}\global\expandafter\let
??V\csname D@as@\K@sw @\endcsname\relax}\def\K@Fa{{\ifnum\K@tn>\K@tp\relax
??V\K@ct\K@tn\K@tp\K@ct\K@tm\K@to\fi\K@Fb{W:r\K@tm-\K@to,c\K@tn-\K@tp}\K@di
??V=\K@sy\relax\advance\K@di by 1 \expandafter\xdef\csname D@ws@\K@sy
??V@\endcsname{\noexpand\K@Fd {\K@tm}{\K@tn}{\K@to}{\K@tp}{\K@qu}\expandafter
??V\noexpand\csname D@ws@\the\K@di @\endcsname}\xdef\K@sy{\the\K@di}}\global
??V\expandafter\let\csname D@ws@\K@sy @\endcsname\relax}%
~\K@bk\displaystretches{\K@cz\let\K@Fb\K@Fe\let\K@sk\K@Fe}\let\K@Fb\K@cj
~\def\K@vl#1#2{\csname D@#1b@x\endcsname{#2}}\def\K@ov#1#2{\csname
~D@#1@lwp@\endcsname}\def\K@os#1#2{\csname D@#1@lp@\endcsname}\def\K@Dv
~#1#2{\csname D@#1@bp@\endcsname}\def\K@wq#1#2{\csname D@#1@cp@\endcsname
~}\def\K@wo#1#2{\csname D@#1@jp@\endcsname}
??O\def\K@ws#1#2{\csname D@#1@pp@\endcsname}\def\K@rr#1#2{\csname
??OD@#1@ap@\endcsname}%
~\def\K@Dl#1#2{\expandafter\ifx\csname D@#1@lpt@\endcsname\K@ci\K@Ff\else
~\csname D@#1@lpt@\endcsname\fi}
??O\def\K@vn#1{\expandafter\ifx\csname D@#1@ppt@\endcsname\K@ci\K@Fg\else
??O\csname D@#1@ppt@\endcsname\fi}%
~\def\D@Fillcellb@x#1#2{\lower\K@vj\hbox to #2{$#1$}}\K@bk\aFillcell#1{\null
~\K@ys0{#1}}\K@bk\rFillcell#1{\null\K@ys1{#1}}\K@bk\rdFillcell#1{\null\K@ys
~2{#1}}\K@bk\dFillcell#1{\null\K@ys3{#1}}\K@bk\ldFillcell#1{\null\K@ys
~4{#1}}\K@bk\lFillcell#1{\null\K@ys5{#1}}\K@bk\luFillcell#1{\null\K@ys
~6{#1}}\K@bk\uFillcell#1{\null\K@ys7{#1}}\K@bk\ruFillcell#1{\null\K@ys
~8{#1}}\K@bk\bFillcell#1{\null\K@ys{10}{#1}}\K@bk\Fillcell#1{\K@bw\K@Dn
~{#1}}\def\D@Boxcellb@x#1#2{\lower\K@vj\hbox{$#1{#2}$}}\K@bk\aBoxcell
~#1{\null\K@yt0{#1}}\K@bk\rBoxcell#1{\null\K@yt1{#1}}\K@bk\rdBoxcell#1{\null
~\K@yt2{#1}}\K@bk\dBoxcell#1{\null\K@yt3{#1}}\K@bk\ldBoxcell#1{\null\K@yt
~4{#1}}\K@bk\lBoxcell#1{\null\K@yt5{#1}}\K@bk\luBoxcell#1{\null\K@yt
~6{#1}}\K@bk\uBoxcell#1{\null\K@yt7{#1}}\K@bk\ruBoxcell#1{\null\K@yt
~8{#1}}\K@bk\bBoxcell#1{\null\K@yt{10}{#1}}\K@bk\Boxcell#1{\K@bw\K@Do
~{#1}}\def\D@Ruleb@x#1#2{\lower\K@vj\hbox{$\vcenter{ \hbox{\vrule width#2
~height #1}}$}}\K@bk\aRule{\null\K@yv0}\K@bk\rRule{\null\K@yv1}\K@bk\rdRule
~{\null\K@yv2}\K@bk\dRule{\null\K@yv3}\K@bk\ldRule{\null\K@yv4}\K@bk\lRule
~{\null\K@yv5}\K@bk\luRule{\null\K@yv6}\K@bk\uRule{\null\K@yv7}\K@bk\ruRule
~{\null\K@yv8}\K@bk\bRule{\null\K@yv{10}}\K@bk\Rule{\K@bw\K@Dp}\def\K@hv
~#1#2{\hbox to \z@{\ifnum#1=0 \else\hss\fi\smash{#2}\ifnum#1=2 \else\hss\fi
~}}\def\K@ve#1@#2,#3@#4{\llap{\hbox to \K@en sp{\hskip#2 \smash{\hbox{\raise
~#3\hbox{\K@hv#1{#4}}}}\hss}}\K@dx{\K@Fh\K@hv\K@Fh\K@wi\K@Fh\K@wh\K@Fh
~\K@vm\K@Fh\K@vg\K@Fh\K@wc\K@Fh\K@we\K@Fh\K@ha\K@Fh\K@pt\K@Fh\endgr\K@Fh
~\K@vj\K@Fh\K@cg}{{#2}{#3}{#1{#4}}}}\def\K@Fh#1{\def#1{\noexpand#1}}\def
~\K@Ej#1@#2,#3@#4{\K@sc @#2,\K@Fi @#3,\K@Fj @\K@ve#1@\K@Fi sp,\K@Fj
~sp@{#4}}\def\K@Fk @#1,#2@#3,#4@{\K@do=#1\K@dm\ifdim\K@do<\z@\K@di=\K@do
~\edef#2{\the\K@di}\else\K@Fl#1.@\K@Fm\K@Fn\ifnum\K@Fm<\K@hx\K@di=\csname
~D@X\K@Fm @\endcsname\ifnum\K@Fn>0 \K@do=\csname D@dx\K@Fm @\endcsname sp
~\K@do=.\K@Fn\K@do\advance\K@di by \K@do\fi\edef#2{\the\K@di}\else\K@di
~=\K@Fm\advance\K@di by -\K@hx\K@do=\csname D@X\K@hx @\endcsname sp
~\advance\K@do by \the\K@di\K@dm\advance\K@do by .\K@Fn\K@dm\K@di=\K@do
~\edef#2{\the\K@di}\fi\fi\K@do=#3\K@dn\ifdim\K@do<\z@\K@di=\K@do\edef
~#4{\the\K@di}\else\K@Fl#3.@\K@Fo\K@Fp\ifnum\K@Fo<\K@ho\K@di=\csname
~D@Y\K@Fo @\endcsname\ifnum\K@Fp>0 \K@do=\csname D@dy\K@Fo @\endcsname sp
~\K@do=.\K@Fp\K@do\advance\K@di by \K@do\fi\edef#4{\the\K@di}\else\K@di
~=\K@Fo\advance\K@di by -\K@ho\K@do=\K@eo sp \advance\K@do by \the\K@di
~\K@dn\advance\K@do by .\K@Fp\K@dn\K@di=\K@do\edef#4{\the\K@di}\fi\fi
~}\def\K@sd @#1,#2@#3,#4@{\K@do=#1\K@dm\K@di=\K@do\edef#2{\the\K@di}\K@do
~=#3\K@dn\K@di=\K@do\edef#4{\the\K@di}}\def\K@Ab @#1,#2@#3,#4@{\setbox
~\K@du=\hbox{\global\K@do=#1\K@dm=0pt}\K@di=\K@do\edef#2{\the\K@di}\setbox
~\K@du=\hbox{\global\K@do=#3\K@dn=0pt}\K@di=\K@do\edef#4{\the\K@di}}\def
~\K@Fl#1.#2@#3#4{\def\K@gl{#1}\ifx\K@gl\K@ci\let#3\K@cm\else\def#3{#1}\fi
~\def\K@gl{#2}\ifx\K@gl\K@ci\let#4\K@cm\else\edef#4{\K@At#2@}\fi}\def
~\K@vm#1#2{\K@cg{currentpoint currentpoint translate #1 neg #2 atan rotate
~neg exch neg exch translate}}\def\K@vg#1{\K@cg{currentpoint currentpoint
~translate #1 neg rotate neg exch neg exch translate}}\def\K@vf{\K@cg
~{gsave}}\def\K@vi{\K@cg{grestore}}\def\K@Fq#1,#2@#3,#4@#5{\K@Ej
~#2@#3,#4@{\hbox{\K@vf\K@vg{#1}\K@hv{#2}{#5}\K@vi}}}
??R\begingroup\K@ce\def\K@zl#1{\K@Fr{\lower\K@vj\hbox{$\labelstyle#1$}}}\def
??R\K@zm#1{\K@Fr{\lower\K@vj\hbox{$\vertexstyle#1$}}}\def\K@zi#1{\K@Fr
??R{#1}}\def\K@zj#1{\K@Fr{\lower\K@vj\hbox{$\vcenter{\hbox{#1}}$}}}\def\K@zk
??R#1{\K@Fr{\lower\K@vj\hbox{$#1$}}}\def\K@zo{\K@zm{\diagramdot}}%
??O\def\K@zn#1{\K@rt\let\K@rp\K@ci\def\K@rt{\egroup\K@Fs{#1}}\setbox\K@du
??O=\hbox\bgroup}\def\K@Fs#1{\ifx\K@rp\K@ci\else\def\K@bf{\K@Ft
??O{#1}}\expandafter\K@bf\K@rp @\fi}\def\K@Ft#1#2,#3@{{\K@ji\K@ed
??O\expandafter\xdef\csname D@pt@#1@x\endcsname{#2}\expandafter\xdef\csname
??OD@pt@#1@y\endcsname{#3}\global\expandafter\let\csname
??OD@pt@#1@type\endcsname\relax\xdef\K@in{{#1}\K@in}}}\endgroup
??R\def\K@Fr#1{\K@rt\def\K@rt{\egroup\K@Fu}\let\K@pu\K@ci\let\K@rp\K@ci\let
??R\K@rm\K@ci\let\K@rq\relax\let\K@rs\relax\let\K@vh\K@cm\let\K@vk\K@cn\let
??R\K@pb\K@cn\let\K@qj\K@cm\let\K@qi\K@cm\let\K@qm\K@cm\let\K@ql\K@cm
??R\setbox\K@dt=\hbox{\K@gx#1}\K@dx{\K@dr={\K@gx#1}\edef\K@bf{\the\wd\K@dt
??R}}{{}{0}{\K@do=\K@bf\setbox\K@dt=\hbox{\the\K@dr}}}\setbox\K@du=\hbox
??R\bgroup}\def\K@Fu{\ifx\K@rp\K@ci\else\K@Cu\relax\ifx\K@rm\K@ci\K@Fv
??R\else\K@Fw\fi\K@bc{\hskip\K@qm sp\raise\K@ql sp \K@hv2{\K@hv0{\K@Bs\box
??R\K@dt\K@Bt}\hskip\K@vk\K@do}}\fi}\def\K@Fv{\expandafter\K@Cv\K@rp @\K@tn
??R\K@tm\ifx\K@rq\relax\edef\K@bc{\noexpand\K@sc @\K@tn,\noexpand\K@tu
??R@\K@tm,\noexpand\K@tv @}\else\def\K@bc{\let\K@tu\K@tn\let\K@tv\K@tm}\fi
??R\K@bc\K@di=\K@tu\advance\K@di by \K@qj\relax\K@dj=\K@tv\advance\K@dj by
??R\K@qi\relax\K@do=\wd\K@dt\edef\K@bc##1{\noexpand\K@ve0@\the\K@di sp,\the
??R\K@dj sp@{\noexpand\K@vf\noexpand\K@vg{\K@vh}##1\noexpand\K@vi}}}\def\K@Fw
??R{\expandafter\K@Cv\K@rp @\K@tn\K@tm\ifx\K@rq\relax\edef\K@bc{\noexpand
??R\K@sc @\K@tn,\noexpand\K@tu @\K@tm,\noexpand\K@tv @}\else\def\K@bc{\let
??R\K@tu\K@tn\let\K@tv\K@tm}\fi\K@bc\expandafter\K@Cv\K@rm @\K@tp\K@to\ifx
??R\K@rs\relax\edef\K@bc{\noexpand\K@sc @\K@tp,\noexpand\K@uc @\K@to
??R,\noexpand\K@ud @}\else\def\K@bc{\let\K@uc\K@tp\let\K@ud\K@to}\fi\K@bc
??R\K@ue\K@uh\K@di=\K@tu\advance\K@di by \K@qj\relax\K@dj=\K@tv\advance\K@dj
??Rby \K@qi\relax\K@do=\wd\K@dt\K@dp=\K@tw sp \edef\K@bc##1{\noexpand\K@ve
??R0@\the\K@di sp,\the\K@dj sp@{\noexpand\K@vf\noexpand\K@vm{\K@uf}{\K@ug
??R}\K@dp=\the\K@dp\hskip\K@pb\K@dp\noexpand\K@wc{\K@uf}{\K@ug}\noexpand
??R\K@vg{\K@vh}##1\noexpand\K@vi}}}\endgroup
??W\begingroup\K@ce\def\K@si{\K@sg\expandafter\K@Fx\expandafter\K@Fy\K@hy
??W{}{}\expandafter\K@Fx\expandafter\K@Fz\K@ic{}{}\expandafter\K@Fx
??W\expandafter\K@Ga\K@ia{}{}\expandafter\K@Fx\expandafter\K@Gb\K@id
??W{}{}\expandafter\K@Gc\expandafter\K@ie\K@ie{}{}\expandafter\K@Fx
??W\expandafter\K@Gd\K@ie{}{}\expandafter\K@Fx\expandafter\K@Ge\K@ih
??W{}{}\expandafter\K@Gc\expandafter\K@ii\K@ii{}{}\expandafter\K@Fx
??W\expandafter\K@Gf\K@ii{}{}\expandafter\K@Fx\expandafter\K@Gg\K@if
??W{}{}\expandafter\K@Gc\expandafter\K@ig\K@ig{}{}\expandafter\K@Fx
??W\expandafter\K@Gh\K@ig{}{}\K@Gi\global\expandafter\let\expandafter\K@en
??W\csname D@X\K@sa @\endcsname\expandafter\K@Fx\expandafter\K@Gj\K@hz
??W{}{}\expandafter\K@Fx\expandafter\K@Gk\K@ib{}{}\K@Gl\global\expandafter
??W\let\expandafter\K@eo\csname D@Y\K@ry @\endcsname\K@sj}\def\K@Fx
??W#1#2#3{\def\K@bf{#3}\ifx\K@bf\K@ci\let\K@bc\relax\else\def\K@bc
??W{#1{#2}{#3}\K@Fx#1}\fi\K@bc}\def\K@Gc#1{\let#1\K@ci\let\K@bf\K@ci\K@di
??W=0 \K@Gm#1}\def\K@Gm#1#2#3{\def\K@gl{#2}\ifx\K@gl\K@ci\xdef#1{#1\K@bf}\let
??W\K@bc\relax\else\def\K@bc{\K@Gm#1}\ifnum#2>\K@di\xdef#1{#1\K@bf}\def
??W\K@bf{{#2}{#3}}\K@di=#2 \else\xdef\K@bf{{#2}{#3}\K@bf}\fi\fi\K@bc}\def
??W\K@Gb#1#2{\K@Gn{#2}{#1}{#2}{-1}{<0}{\K@Go d\K@Gp}}\def\K@Gg#1#2{\K@Gn
??W{#2}{#1}{#2}{-1}{<0}{\K@Go m\K@Gq}}\def\K@Ge#1#2{\K@Gn
??W{#2}{#1}{#2}{-1}{<0}{\K@Go a\K@Gr}}\def\K@Go#1#2#3#4#5{\K@Gs{#1l}\K@Gt
??W{#4}{#5}{#3}{#5}\multiply\K@dj by -1 #2{#5}{\K@dj}}\def\K@Gd#1#2{\K@Gn
??W{#2}{#1}{#2}1{>\K@hx}{\K@Gu d\K@Gv}}\def\K@Gh#1#2{\K@Gn{#2}{#1}{#2}1{>\K@hx
??W}{\K@Gu m\K@Gq}}\def\K@Gf#1#2{\K@Gn{#2}{#1}{#2}1{>\K@hx}{\K@Gu a\K@Gr}}\def
??W\K@Gu#1#2#3#4#5{\K@Gs{#1r}\K@Gt{#4}{#5}{#5}{#3}#2{#5}{\K@dj}}\def\K@Gt
??W#1#2#3#4{\K@sg\K@ts{#1}{#2}\K@bf\expandafter\K@El\K@bf\K@ua\K@un\K@uo
??W\K@dj=\K@tz\advance\K@dj by \K@ua\divide\K@dj by 2 \advance\K@dj by
??W\csname D@X#3@\endcsname\multiply\K@dj by -1 \advance\K@dj by \csname
??WD@X#4@\endcsname\relax}\def\K@Fy#1#2{\K@Gs{dx}\K@Gw X{#1}{#2}\K@Gp\K@hx
??W1}\def\K@Gj#1#2{\K@Gs{dy}\K@Gw Y{#1}{#2}\K@Gx\K@ho1}\def\K@Ga#1#2{\K@Gs
??W{mx}\K@Gw X{#1}{#2}\K@Gq\K@hx0}\def\K@Gk#1#2{\K@Gs{my}\K@Gw Y{#1}{#2}\K@Gy
??W\K@ho0}\def\K@Fz#1#2{\K@Gs{ax}\K@Gw{}{#1}{#2}\K@Gr0{}}\def\K@Gw
??W#1#2#3#4#5#6{\setbox\K@du=\hbox{$\K@dp=#3pt$}\def\K@bc{#4{#2}}\ifdim\wd
??W\K@du=\z@\K@di=#2 \advance\K@di by 1 \ifnum\K@di>#5 \K@dp=\z@\else\K@dp
??W=\csname D@#1\the\K@di @\endcsname sp \multiply\K@dp by -1 \advance\K@dp
??Wby \csname D@#1#2@\endcsname sp \K@dp=#3\K@dp\ifcase#6 \multiply\K@dp by
??W-1 \or\edef\K@bc{\noexpand#4{\the\K@di}}\fi\fi\else\K@dp=#3 \fi\K@dk
??W=\K@dp\K@bc{\K@dk}}\endgroup
~\K@bg\flexible{\K@cz\def\K@Gz{\K@Ha}\def\K@Ev{\K@Ey}\def\K@Hb{\K@Hc}\def
~\K@sq{\K@ta}\def\K@sr{\K@sv}\let\K@om\K@oo\def\K@sf{\K@Hd}}\def\K@He{\def
~\K@Gz{\K@Hf}\let\K@Ev\K@cj\let\K@Hb\relax\def\K@sq{\K@ta0\K@tb}\let\K@sr
~\relax\let\K@sf\relax}\let\K@te\relax
??V\def\K@Hd{\K@Gs{A}\csname D@as@1@\endcsname\K@Gs{W}\csname
??VD@ws@1@\endcsname\K@Gs{C}\csname D@cs@1@\endcsname}%
~\K@bg\gravitateleft{\K@cz\let\K@te\K@cm}\K@bg\gravitateright{\K@cz\def\K@te
~{\K@hx}}
??V\def\K@Hg#1#2{\K@Gn{#2}{#1}{#2}{-1}{<0}\K@Hh}\def\K@Gn#1#2#3#4#5#6{\K@di
??V=#1 \advance\K@di by #4\relax\ifnum\K@di#5\relax\let\K@bc\K@ck\else
??V\K@ts{#2}{\the\K@di}\K@bf\ifx\K@bf\relax\edef\K@bc{\noexpand\K@Gn{\the
??V\K@di}{#2}{#3}{#4}}\else\expandafter\K@El\K@bf\K@tz\K@uk\K@ul\edef\K@bc
??V{\noexpand\K@Hi{\the\K@di}{#2}{#3}{#4}}\fi\fi\K@bc{#5}{#6}}\def\K@Hi
??V#1#2#3#4#5#6{#6{#1}{#2}{#3}}\def\K@Hh#1#2#3{\K@sg\K@ts{#2}{#3}\K@bf
??V\expandafter\K@El\K@bf\K@ua\K@un\K@uo\K@dk=\K@tz\advance\K@dk by \K@ua
??V\divide\K@dk by 2 \advance\K@dk by \csname D@X#1@\endcsname\multiply\K@dk
??Vby -1 \advance\K@dk by \csname D@X#3@\endcsname\K@dj=\csname
??VD@X#1@\endcsname\advance\K@dj by \K@dm\multiply\K@dj by -1 \advance\K@dj
??Vby \csname D@X#3@\endcsname\relax\ifnum\K@dk<\K@dj\K@dj=\K@dk\fi\ifnum
??V\K@dj<0\relax\def\K@bc{\global\let\K@sc\K@Fk\K@tc{#1}{#3}\K@Hj\K@Hk\K@Hl
??V}\else\let\K@bc\relax\fi\K@bc}\def\K@Hj#1#2{\K@Gr{#1}{\K@dj}\K@td
??V{#2}{#1}}\def\K@Hl#1#2{\multiply\K@dj by -1 \K@Gr{#2}{\K@dj}\K@td
??V{#1}{#2}}\def\K@Hk#1#2{\K@dj=\K@ua\divide\K@dj by 2 \advance\K@dj by
??V\csname D@X#1@\endcsname\multiply\K@dj by -1 \K@Gr{#1}{\K@dj}\K@dj=\K@tz
??V\divide\K@dj by 2 \multiply\K@dj by -1 \advance\K@dj by \csname
??VD@X#2@\endcsname\multiply\K@dj by -1 \K@Gr{#2}{\K@dj}}\def\K@tc
??V#1#2#3#4#5{{\K@di=#1 \multiply\K@di by 10 \advance\K@di by -\K@ij\K@dj=#2
??V\multiply\K@dj by 10 \advance\K@dj by -\K@ij\K@dk=\K@di\multiply\K@dk by
??V\K@dj\ifnum\K@dk<0 \gdef\K@bc{#4}\else\ifnum\K@di<0 \gdef\K@bc{#3}\else
??V\gdef\K@bc{#5}\fi\fi }\K@bc{#1}{#2}}\def\K@td#1#2{\K@Hm
??V{B:c#1-#2}\expandafter\xdef\csname D@Bindings@#1@\endcsname{{#2}\csname
??VD@Bindings@#1@\endcsname}}%
~\K@bk\displaybindings{\K@cz\let\K@Hm\K@Fe\let\K@sk\K@Fe}\let\K@Hm\K@cj
~\K@bg\displayall{\K@cz\let\K@Hm\K@Fe\let\K@Gs\K@Fe\let\K@Fb\K@Fe\let
~\K@sk\K@Fe}
??V\def\K@Hn#1#2#3{\def\K@bf{#3}\ifx\K@bf\K@ci\let\K@bc\K@ck\else
??V\expandafter\ifx\csname D@Bound@#2@#3@\endcsname\relax {\advance#1 by
??V\csname D@X#3@\endcsname\expandafter\xdef\csname D@X#3@\endcsname{\the
??V#1}}\K@Gs{X#3}\expandafter\let\csname D@Bound@#2@#3@\endcsname\K@ci\def
??V\K@bc{\def\K@bc{\K@Hn{#1}{#3}}\expandafter\expandafter\expandafter\K@bc
??V\csname D@Bindings@#3@\endcsname{}\K@Hn}\else\let\K@bc\K@Hn\fi\fi\K@bc
??V{#1}{#2}}\def\K@Fc#1#2#3#4{\let\K@tz\K@cm\K@Ho{#1}{#2}\K@tz\K@uk\K@ul
??V\let\K@ua\K@cm\K@Ho{#1}{#3}\K@ua\K@un\K@uo\K@sg\K@dj=\K@tz\divide\K@dj
??Vby 2 \advance\K@dj by \csname D@X#2@\endcsname\relax\ifnum#4>\K@Hp\relax
??V\advance\K@dj by #4\else\advance\K@dj by \K@Hp\fi\multiply\K@dj by -1
??V\advance\K@dj by \csname D@X#3@\endcsname\K@di=\K@ua\divide\K@di by -2
??V\advance\K@dj by \K@di\ifnum\K@dj<0 \K@tc{#2}{#3}\K@Hq{\K@Hr{#4}}\K@Hs\fi
??V}\def\K@Hq#1#2{\K@Gr{#1}{\K@dj}\K@td{#2}{#1}}\def\K@Hs#1#2{\multiply\K@dj
??Vby -1 \K@Gr{#2}{\K@dj}\K@td{#1}{#2}}\def\K@Hr#1#2#3{\K@dj=\K@ua\advance
??V\K@dj by \K@tz\divide\K@dj by 2 \ifnum#1>\K@Hp\relax\advance\K@dj by
??V#1\else\advance\K@dj by \K@Hp\fi\divide\K@dj by 2 \K@di=\csname
??VD@X#3@\endcsname\multiply\K@di by -1 \advance\K@di by \K@dj\K@Gr{#3}{\K@di
??V}\K@di=\csname D@X#2@\endcsname\advance\K@di by \K@dj\multiply\K@di by
??V-1 \K@Gr{#2}{\K@di}}\def\K@Gr#1#2{\K@Gq{#1}{#2}\def\K@bc{\K@Hn
??V{#2}{#1}}{\expandafter\expandafter\expandafter\K@bc\csname
??VD@Bindings@#1@\endcsname{}}}%
~\def\displaymovements{\K@cz\let\K@Gs\K@Fe\let\K@sk\K@Fe}\let\K@Gs\K@cj
~\def\K@Fe#1{\message{(#1)}}
??V\def\K@Gp#1#2{\K@Ht X{#1}{#2}1{<\K@hx}}\def\K@Gv#1#2{\K@Ht
??VX{#1}{#2}{-1}{>0}}\def\K@Gx#1#2{\K@Ht Y{#1}{#2}1{<\K@ho}}\def\K@Ht
??V#1#2#3#4#5{\K@di=#2 \K@cu\K@Hu{#1}{\the\K@di}{#3}\ifnum\K@di#5 \advance
??V\K@di by #4 \K@cv}\def\K@Gq#1#2{\K@Hu X{#1}{#2}}\def\K@Gy#1#2{\K@Hu
??VY{#1}{#2}}\def\K@Hu#1#2#3{\global\let\K@sc\K@Fk\K@Gs{#1#2}{\advance#3 by
??V\csname D@#1#2@\endcsname\expandafter\xdef\csname D@#1#2@\endcsname{\the
??V#3}}}\def\K@Fd#1#2#3#4#5{\ifcase#5\relax\K@Hv{#2}{#4}\or\K@Hw
??V{#1}{#2}{#3}{#4}\or\K@Hx{#1}{#2}{#3}{#4}\or\K@Hy{#1}{#2}{#3}{#4}\fi}\def
??V\K@Hv#1#2{\K@sg\K@dj=\csname D@X#1@\endcsname\advance\K@dj by \K@Hz\relax
??V\multiply\K@dj by -1 \advance\K@dj by \csname D@X#2@\endcsname\relax
??V\ifnum\K@dj<0 \K@tc{#1}{#2}\K@Hq\K@Ia\K@Hs\fi}\def\K@Hw#1#2#3#4{\K@sg\let
??V\K@tz\K@cm\let\K@gl\K@cm\K@Ho{#1}{#2}\K@tz\K@uk\K@ul\K@Ho{#1}{#4}\K@gl
??V\K@un\K@uo\K@Ib{#2}{#4}\K@dj\K@tz\K@gl\let\K@gl\K@cm\let\K@ua\K@cm\K@Ho
??V{#3}{#2}\K@gl\K@uk\K@ul\K@Ho{#3}{#4}\K@ua\K@un\K@uo\K@Ib{#2}{#4}\K@dk
??V\K@gl\K@ua\relax\ifnum\K@dk<\K@dj\K@dj=\K@dk\fi\ifnum\K@dj<0 \K@tc
??V{#2}{#4}\K@Hq\K@Ic\K@Hs\fi}\def\K@Hx#1#2#3#4{\K@sg\let\K@tz\K@cm\let
??V\K@ua\K@cm\let\K@gl\K@cm\K@Ho{#1}{#2}\K@tz\K@uk\K@ul\K@Ho{#3}{#4}\K@ua
??V\K@un\K@uo\K@Ho{#3}{#2}\K@gl\K@uk\K@ul\K@Ib{#2}{#4}\K@dj\K@gl\K@ua\relax
??V\ifnum\K@dj<0 \K@tc{#2}{#4}\K@Hq\K@Ic\K@Hs\fi}\def\K@Hy#1#2#3#4{\K@sg\let
??V\K@tz\K@cm\let\K@ua\K@cm\let\K@gl\K@cm\K@Ho{#1}{#2}\K@tz\K@uk\K@ul\K@Ho
??V{#3}{#4}\K@ua\K@un\K@uo\K@Ho{#1}{#4}\K@gl\K@un\K@uo\K@Ib{#2}{#4}\K@dj
??V\K@tz\K@gl\relax\ifnum\K@dj<0 \K@tc{#2}{#4}\K@Hq\K@Ic\K@Hs\fi}\def\K@Ib
??V#1#2#3#4#5{#3=#4 \advance#3 by #5 \divide#3 by 2 \advance#3 by \csname
??VD@X#1@\endcsname\advance#3 by \K@Id\multiply#3 by -1 \advance#3 by
??V\csname D@X#2@\endcsname}\def\K@Ic#1#2{\K@dk=\K@tz\multiply\K@dk by -1
??V\advance\K@dk by \K@ua\divide\K@dk by 2 \advance\K@dk by \csname
??VD@X#1@\endcsname\advance\K@dk by \csname D@X#2@\endcsname {\advance\K@dj
??Vby \K@dk\divide\K@dj by -2 \K@Gr{#2}{\K@dj}}{\advance\K@dj by -\K@dk\divide
??V\K@dj by 2 \K@Gr{#1}{\K@dj}}}\def\K@Ia#1#2{\K@di=\K@Hz\divide\K@di by 2
??V\K@dj=\csname D@X#2@\endcsname\multiply\K@dj by -1 \advance\K@dj by \K@di
??V\K@Gr{#2}{\K@dj}\K@dj=\csname D@X#1@\endcsname\advance\K@dj by \K@di
??V\multiply\K@dj by -1 \K@Gr{#1}{\K@dj}}\def\K@Ho#1#2#3#4#5{\K@ts
??V{#1}{#2}\K@bf\ifx\K@bf\relax\else\expandafter\K@El\K@bf#3#4#5\fi}\def
??V\K@Ie#1#2#3{{\K@di=0 \K@dk=0 \K@cu\expandafter\xdef\csname D@#1\the\K@di
??V@\endcsname{\the\K@dk}\ifnum\K@di<#3\relax\advance\K@di by 1 \advance\K@dk
??Vby #2\relax\K@cv }}\def\K@Hf{\K@Ie X\K@dm\K@hx}\def\K@Ha{\K@Ie X\z@\K@hx
??V}\def\K@If{\K@Ie Y\K@dn\K@ho}\def\K@sg{\K@Gz\K@If\let\K@sg\relax}\def\K@Ig
??V#1#2#3{\K@di=0 \K@cu\ifnum\K@di<#3\relax\K@dj=\K@di\K@dk=\csname
??VD@#1\the\K@di @\endcsname\multiply\K@dk by -1 \advance\K@di by 1 \advance
??V\K@dk by \csname D@#1\the\K@di @\endcsname\expandafter\xdef\csname
??VD@d#2\the\K@dj @\endcsname{\the\K@dk}\K@cv}\def\K@sj{\K@sg\K@Ig Xx\K@hx
??V\K@Ig Yy\K@ho}\def\K@Ih#1#2#3#4{\K@di=\csname D@#10@\endcsname\K@dj=0
??V\K@dk=\K@di\global\let#3\K@cm\global\let#4\K@cm\K@cu\ifnum\csname
??VD@#1\the\K@dj @\endcsname<\K@di\xdef#3{\the\K@dj}\K@di=\csname D@#1\the
??V\K@dj @\endcsname\fi\ifnum\csname D@#1\the\K@dj @\endcsname>\K@dk\xdef
??V#4{\the\K@dj}\K@dk=\csname D@#1\the\K@dj @\endcsname\fi\ifnum\K@dj<#2
??V\advance\K@dj by 1 \K@cv\ifnum\K@di=0 \else\multiply\K@di by -1 \K@dj=0
??V\K@cu\K@dk=\csname D@#1\the\K@dj @\endcsname\advance\K@dk by \K@di
??V\expandafter\xdef\csname D@#1\the\K@dj @\endcsname{\the\K@dk}\ifnum\K@dj
??V<#2\relax\advance\K@dj by 1 \K@cv\fi}\def\K@Gi{\K@Ih X\K@hx\K@rz\K@sa
??V}\def\K@Gl{\K@Ih Y\K@ho\K@rx\K@ry}\def\K@tg{\let\K@rt\K@ru\let\K@ni\K@cm
??V\K@Ii\global\setbox\K@dt=\hbox\bgroup$\vertexstyle\K@gy\K@gz\K@rv}%
~\def\K@rv{\let\box\copy\let\unhbox\unhcopy\let\unvbox\unvcopy}\def\K@ru
~{$\egroup}
??A\begingroup\K@ce\def\K@Ij{\ifx\K@Ik\relax\expandafter\global\expandafter
??A\edef\K@tf{\let\expandafter\noexpand\csname D@\the\K@df @\the\K@de
??A@\endcsname\noexpand\K@ci}\else\expandafter\global\expandafter\let\K@tf
??A\relax\fi}%
~\def\K@Il{\expandafter\global\expandafter\let\K@tf\relax}\K@bg\dotted{\K@cz
~\K@cc\def\K@yr{\K@Im}\def\K@Ii{\K@In}\def\K@Il{\K@Ij}\def\K@Io{\K@Ip}}\let
~\K@yr\relax\let\K@Ii\relax\let\K@Io\K@ck\def\K@oe{\global\let\K@Ik\relax}
??A\def\K@In{\let\K@Ik\K@ci}\def\K@Ip#1#2{\expandafter\let\csname
??AD@#1@#2@\endcsname\K@ci}%
??V\def\K@th{\K@rt\K@ts{\the\K@df}{\the\K@de}\K@bf\ifx\K@bf\relax\K@Il
??V\else\K@Hb\fi}%
??A\def\K@Im{\K@di=0 \K@dj=0 \K@cu\K@cu\expandafter\let \expandafter\K@bf
??A\csname D@\the\K@di @\the\K@dj @\endcsname\ifx\K@bf\K@ci\else\K@dk=\K@ho
??A\advance\K@dk by -\K@di\edef\K@bc{\noexpand\K@Ej1@\the\K@dj,\the\K@dk
??A@}{\K@bc{\lower\K@vj\hbox{$\vertexstyle\diagramdot$}}}\fi\ifnum\K@dj<\K@hx
??A\advance\K@dj by 1 \K@cv\ifnum\K@di<\K@ho\advance\K@di by 1 \K@dj=0
??A\K@cv}\endgroup
??V\def\K@Hc{{\K@di=\K@sx\relax\advance\K@di by 1 \K@Fb{C:r\the\K@df,c\the
??V\K@de}\K@dj=\K@de\multiply\K@dj by 10 \ifnum\K@dj<\K@ij\relax\let\K@bc
??V\K@cr\else\let\K@bc\K@cq\fi\expandafter\xdef\csname D@cs@\K@sx
??V@\endcsname{\K@bc{\noexpand\K@Hg{\the\K@df}{\the\K@de}}{\expandafter
??V\noexpand\csname D@cs@\the\K@di @\endcsname}}\xdef\K@sx{\the\K@di}}\global
??V\expandafter\let\csname D@cs@\K@sx @\endcsname\relax}\def\K@Ex
??V#1@#2@#3@#4@#5@#6@#7@#8@#9@{\def\K@tm{#3}\def\K@tn{#4}\def\K@to{#5}\def
??V\K@tp{#6}\K@Iq\K@tm\K@tn\K@tz\K@uk\K@ul\K@Iq\K@to\K@tp\K@ua\K@un\K@uo
??V\edef\K@pb{\K@Ep0#8}\expandafter\K@Ir\K@pb\K@pb\K@rl\edef\K@pc{\K@Ep
??V1#8}\expandafter\K@Ir\K@pc\K@pc\K@pd\edef\K@qb{\K@Ep2#8}\expandafter\K@Ir
??V\K@qb\K@qb\K@py\edef\K@qc{\K@Ep3#8}\expandafter\K@Ir\K@qc\K@qc\K@qd\edef
??V\K@pz{\K@Ep4#8}\expandafter\K@Ir\K@pz\K@pz\K@qa\edef\K@pu{\K@Ep
??V5#8}\expandafter\K@Ir\K@pu\K@pu\K@qt\ifx\K@pu\K@cl\let\K@qs\K@ci\else
??V\let\K@qs\relax\fi\edef\K@qj{\K@Ep6#8}\expandafter\K@Ir\K@qj\K@qj\K@qm
??V\edef\K@qi{\K@Ep7#8}\expandafter\K@Ir\K@qi\K@qi\K@ql\let\K@rq\relax\let
??V\K@rs\relax\K@di=\K@ho\advance\K@di by -\K@tm\edef\K@tm{\the\K@di}\K@di
??V=\K@ho\advance\K@di by -\K@to\edef\K@to{\the\K@di}\edef\K@qy{\K@Ep
??V0#9{}{}{}{}{}}\edef\K@re{\K@Ep1#9{}{}{}{}{}}\edef\K@rf{\K@Ep
??V2#9{}{}{}{}{}}\expandafter\K@Is\K@qy @\ifx\K@rb\relax\ifx\K@pb\K@ci\edef
??V\K@pb{\K@Dl#2}\fi\fi\expandafter\K@tt{#1}{#2}{#7}}\def\K@Is#1#2@{\ifcase
??V#1 \let\K@rc\relax\let\K@rb\relax\or\let\K@rc\K@ci\let\K@rb\relax\def
??V\K@qy{#2}\or\let\K@rc\K@ci\let\K@rb\K@ci\def\K@qy{#2}\fi}\def\K@Ir
??V#1#2#3#4{\def#3{#1}\def#4{#2}}\def\K@Iq#1#2#3#4#5{\K@ts{#1}{#2}\K@bf\ifx
??V\K@bf\relax\let#3\K@cm\let#4\K@cm\let#5\K@cm\else\expandafter\K@El\K@bf
??V#3#4#5\fi}\def\K@El#1@#2@#3@#4#5#6{\def#4{#1}\def#5{#2}\def#6{#3}}\def
??V\K@ts#1#2#3{\expandafter\let \expandafter#3\csname D@whd@#1@#2@\endcsname}%
~\def\K@ub#1by #2#3{\K@di=#3 \multiply\K@di by #2 \advance\K@di by #1 \edef
~#1{\the\K@di}}\def\K@ty#1#2#3{\ifx#2\K@ci\let#2\K@cm\else\setbox\K@du
~=\hbox{\K@do=#2pt}\ifdim\wd\K@du>\z@\K@do=#2 \K@di=\K@do\multiply\K@di by
~#1 \edef#2{\the\K@di}\else\ifx#3\K@ci\let#2\K@cm\else\K@do=#3sp \K@do
~=#2\K@do\K@di=\K@do\multiply\K@di by -2 \advance\K@di by #3 \divide\K@di
~by -2 \edef#2{\the\K@di}\fi\fi\fi}\def\K@vs#1=#2\times#3\over#4{\ifnum
~#3<10737418 #1=#3 \multiply#1 by 100 \divide#1 by #4\relax\multiply#1 by
~#2\relax\divide#1 by 100 \else {\ifnum#2<0 \multiply#2 by -1 \multiply#4
~by -1 \fi\K@It#1=#2\times#3\over#4}\fi}\def\K@It#1=#2\times#3\over#4{\K@de
~=1 \K@cu\multiply\K@de by 2 \ifnum#2>\K@de\K@cv\K@df=#3 \divide\K@df by
~\K@de\divide#4 by \K@de\multiply\K@df by #2\relax\divide\K@df by
~#4\relax\xdef\K@bc{#1=\the\K@df}\aftergroup\K@bc}\def\K@uh{\K@di=\K@ug
~\relax\ifnum\K@di<0 \multiply\K@di by -1 \fi\K@dj=\K@uf\relax\ifnum\K@dj
~<0 \multiply\K@dj by -1 \fi\K@dk=1 \let\K@bf\K@ci\K@cu\ifnum\K@di<23171
~\ifnum\K@dj<23171 \let\K@bf\relax\fi\fi\ifx\K@bf\K@ci\divide\K@di by 2
~\divide\K@dj by 2 \multiply\K@dk by 2 \K@cv\multiply\K@di by \K@di
~\multiply\K@dj by \K@dj\advance\K@di by \K@dj\expandafter\K@Iu\the\K@di
~=\K@dj\multiply\K@dj by \K@dk\edef\K@tw{\the\K@dj}}\def\K@Iu#1=#2{{\K@dh
~=#1\relax\K@df=32768 \K@de=0 \K@cu\divide\K@df by 2 \K@dg=\K@de\advance
~\K@dg by \K@df\multiply\K@dg by \K@dg\ifnum\K@dg>\K@dh\else\advance
~\K@de by \K@df\fi\ifnum\K@df>1 \K@cv\global#2=\K@de}}\def\K@wu#1=#2\mod
~#3{#1=#2\relax\ifnum#1<0 \K@cu\advance#1 by #3\relax\ifnum#1<0 \K@cv
~\else\K@cu\ifnum#1>#3\relax\advance#1 by -#3\relax\K@cv\ifnum
~#1=#3\relax #1=0 \fi\fi}
??V\def\K@yp#1@#2@#3@{\K@Io{#1}{#2}\setbox\K@du=\hbox{\K@Iv#3\K@rt}\K@dx{\K@dr
??V={\K@Iv#3\K@rt}}{{}{0}{\setbox\K@du=\hbox{\the\K@dr}}}\K@Iw{#1}{#2}\K@di
??V=\K@ho\advance\K@di by -#1 \edef\K@bc{\noexpand\K@Ej1@#2,\the\K@di @}\K@bc
??V{\lower\K@vj\hbox{\box\K@dt}}}\def\K@Iw#1#2{\K@di=\wd\K@dt\divide\K@di by
??V2 \K@dj=\K@di\advance\K@dj by \csname D@X#2@\endcsname\advance\K@dj by
??V-\K@en\advance\K@dj by -\K@ev\relax\ifnum\K@dj>0 \advance\K@dj by \K@ev
??V\xdef\K@ev{\the\K@dj}\fi\K@dj=\K@di\advance\K@dj by -\csname
??VD@X#2@\endcsname\advance\K@dj by -\K@eu\relax\ifnum\K@dj>0 \advance\K@dj
??Vby \K@eu\xdef\K@eu{\the\K@dj}\fi\K@di=\wd\K@dt\advance\K@di by \ht\K@dt
??V\advance\K@di by \dp\K@dt\ifnum\K@di>0 \ifnum#1=\K@rx\K@dj=\ht\K@dt
??V\ifnum\K@dj>\K@es\xdef\K@es{\the\K@dj}\fi\fi\ifnum#1=\K@ry\K@dj=\dp
??V\K@dt\relax\ifnum\K@dj>\K@et\xdef\K@et{\the\K@dj}\fi\fi\fi}%
??S\def\K@yz#1#2#3#4#5{\K@di=#3 \divide\K@di by 2 \K@dj=\K@di\advance\K@dj
??Sby #1 \advance\K@dj by -\K@en\advance\K@dj by -\K@ev\relax\ifnum\K@dj>0
??S\advance\K@dj by \K@ev\xdef\K@ev{\the\K@dj}\fi\K@dj=#1\advance\K@dj by
??S\K@eu\multiply\K@dj by -1 \advance\K@dj by \K@di\relax\ifnum\K@dj>0
??S\advance\K@dj by \K@eu\xdef\K@eu{\the\K@dj}\fi\K@di=#4 \advance\K@di by
??S#2\relax\ifnum\K@di>\K@eo\relax\advance\K@di by -\K@eo\relax\ifnum\K@di
??S>\K@es\relax\xdef\K@es{\the\K@di}\fi\fi\K@di=#2 \multiply\K@di by -1
??S\advance\K@di by #5\relax\ifnum\K@di>0\relax\ifnum\K@di>\K@et\relax\xdef
??S\K@et{\the\K@di}\fi\fi}\endgroup
??V\def\K@Iv{\let\K@rt\K@ru\global\setbox\K@dt=\hbox\bgroup$\vertexstyle\K@gx
??V}\endgroup
~\def\K@vj{\fontdimen22\textfont2}\def\K@Ix#1#2#3#4{\let\K@gl\K@ci\K@Iy
~#1#2{}\ifx\K@gl\relax#3\else#4\fi}\def\K@Iy#1#2#3{\if\noexpand#1#2\let
~\K@gl\relax\fi\def\K@bf{#3}\ifx\K@bf\K@ci\let\K@bc\relax\else\def\K@bc
~{\K@Iy#1#3}\fi\K@bc}\def\K@Iz{\let\K@Ja\relax\K@dr={}\futurelet\K@jr\K@Jb
~}\def\K@Jb{\if=\noexpand\K@jr\def\K@bc{\K@Jc\expandafter\futurelet
~\expandafter\K@jr\expandafter\K@Jd @}\else\if\noexpand\K@jr\K@bn\def
~\K@bc{\K@pa\expandafter\futurelet\expandafter\K@jr\expandafter\K@Jb
~@}\else\def\K@bc{\expandafter\futurelet\expandafter\K@jr\expandafter
~\K@Je}\fi\fi\K@bc}\def\K@Jd{\if\noexpand\K@jr\K@bn\def\K@bc{\K@pa
~\expandafter\futurelet\expandafter\K@jr\expandafter\K@Je @}\else\let\K@bc
~\K@Je\fi\K@bc}\def\K@Jc#1@#2{#1}\def\K@pa#1@ {#1}\def\K@Je{\if-\noexpand
~\K@jr\def\K@bc{\K@dr={-}\K@Jc\expandafter\futurelet\expandafter\K@jr
~\expandafter\K@Jf @}\else\let\K@bc\K@Jg\fi\K@bc}\def\K@Jf{\if\noexpand
~\K@jr\K@bn\def\K@bc{\K@pa\expandafter\futurelet\expandafter\K@jr
~\expandafter\K@Jg @}\else\let\K@bc\K@Jg\fi\K@bc}\def\K@Jg{\if.\noexpand
~\K@jr\ifx\K@Ja\relax\let\K@Ja\K@ci\let\K@bc\K@Jh\else\if\noexpand\K@jr
~\K@bn\def\K@bc{\K@pa\K@Ji @}\else\let\K@bc\K@Ji\fi\fi\else\K@Ix\K@jr
~{0123456789}{\let\K@Ji\K@Jj\let\K@bc\K@Jh}{\if\noexpand\K@jr\K@bn\def\K@bc
~{\K@pa\K@Ji @}\else\let\K@bc\K@Ji\fi}\fi\K@bc}\def\K@Jh#1{\edef\K@bc
~{\K@dr={\the\K@dr#1}}\K@bc\futurelet\K@jr\K@Jg}\def\K@qq{\ifcat\noexpand
~\K@jr\bgroup\def\K@bc{\bgroup\K@ez\K@jp}\else\let\K@bc\K@jq\fi\K@bc
~}\def\K@Jk#1#2{\K@cz\def\K@jp##1{\egroup\def\K@Ji{\expandafter\let\csname
~D@##1@#1@\endcsname\K@ci}\def\K@Jj{\expandafter\edef\csname
~D@##1@#1@\endcsname{\the\K@dr}}\K@Iz}\def\K@jq{\let\K@Ji\relax\def\K@Jj
~{\edef#2{\the\K@dr}}\K@Iz}\futurelet\K@jr\K@qq}\K@bk\labelpoint{\K@Jk
~{lpt}\K@Ff}\K@bk\ptpoint{\K@Jk{ppt}\K@Fg}\def\K@Jl{\if\noexpand\K@jr\K@bn
~\def\K@bc{\K@pa\futurelet\K@jr\K@Jl @}\else\if+\noexpand\K@jr\def\K@bc
~{\K@Jc\futurelet\K@jr\K@Jm @}\else\def\K@bc{\afterassignment\K@Jn\K@do}\fi
~\fi\K@bc}\def\K@Jm{\if\noexpand\K@jr\K@bn\def\K@bc{\K@pa\futurelet\K@jr
~\K@Jm @}\else\if=\noexpand\K@jr\def\K@bc{\afterassignment\K@Jo\K@do}\else
~\def\K@bc{\afterassignment\K@Jn\K@do}\fi\fi\K@bc}\def\K@Jp#1{\edef
~#1{\the\K@do}}\def\K@Jq#1{\edef#1{\the\K@di}}\def\K@Jr#1{\K@di=\K@do\K@Jq
~#1}\def\K@Js#1{\advance\K@do by #1\edef#1{\the\K@do}}\def\K@Jt#1{\advance
~\K@di by #1\edef#1{\the\K@di}}\def\K@Ju#1{\K@di=\K@do\K@Jt#1}\def\K@Jv
~#1#2{\K@cz\def\K@jp##1{\egroup\def\K@Jo{\expandafter\K@Ju\csname
~D@##1@#1@\endcsname}\def\K@Jn{\expandafter\K@Jr\csname D@##1@#1@\endcsname
~}\futurelet\K@jr\K@Jl}\def\K@jq{\def\K@Jo{\K@Ju#2}\def\K@Jn{\K@Jr
~#2}\futurelet\K@jr\K@Jl}\futurelet\K@jr\K@qq}\K@bk\labelwidthpad{\K@Jv
~{lwp}\K@ou}\K@bk\labelpad{\K@Jv{lp}\K@or}\K@bk\breakpad{\K@Jv{bp}\K@Du
~}\K@bk\cellpush{\K@Jv{cp}\K@wp}\K@bk\ptpush{\K@Jv{pp}\K@wr}\K@bk\atpush
~{\K@Jv{ap}\K@wt}\K@bk\joinpush{\K@Jv{jp}\K@wn}\def\K@Jw#1#2{\K@cz\def\K@Jo
~{#1}\def\K@Jn{#2}\futurelet\K@jr\K@Jl}\K@bg\MinimumCellLength{\K@Jw {\K@Ju
~\K@ut}{\K@Jr\K@ut}}\K@bg\celllength{\K@Jw {\K@Ju\K@Ds}{\K@Jr\K@Ds}}\K@bg
~\cellwidth{\K@Jw {\K@Ju\K@Hp}{\K@Jr\K@Hp}}\K@bg\columndist{\K@Jw {\K@Ju
~\K@Hz}{\K@Jr\K@Hz}}\K@bg\bracewidth{\K@Jw {\K@Ju\K@Id}{\K@Jr\K@Id}}\K@bk
~\Diagrampad{\K@Jw {\K@Js\K@Ay}{\K@Jp\K@Ay}}\K@bk\Figurepad{\K@Jw {\K@Js
~\K@hk}{\K@Jp\K@hk}}\K@bk\Graphpad{\K@Jw {\K@Js\K@hl}{\K@Jp\K@hl}}\K@bg\tpad
~{\K@Jw {\K@Js\K@Be}{\K@Jp\K@Be}}\K@bg\bpad{\K@Jw {\K@Js\K@Bf}{\K@Jp\K@Bf
~}}\K@bg\lpad {\K@Jw{\K@Js\K@Bd}{\K@Jp\K@Bd}}\K@bg\rpad {\K@Jw{\K@Js\K@Bg
~}{\K@Jp\K@Bg}}\K@bg\vpad{\K@Jw {\K@dp=\K@do\K@Js\K@Be\K@do=\K@dp\K@Js\K@Bf
~}{\K@Jp\K@Be\let\K@Bf\K@Be}}\K@bg\hpad{\K@Jw {\K@dp=\K@do\K@Js\K@Bd\K@do
~=\K@dp\K@Js\K@Bg}{\K@Jp\K@Bd\let\K@Bg\K@Bd}}\K@bg\framepad{\K@Jw {\K@Js
~\K@Ci}{\K@Jp\K@Ci}}\K@bg\framerulewidth{\K@Jw {\K@Js\K@Cl}{\K@Jp\K@Cl
~}}\K@bg\Rulewidth{\K@Jw {\K@Js\K@Dq}{\K@Jp\K@Dq}}\K@bg\Framerulewidth{\K@Jw
~{\K@Js\K@Cw}{\K@Jp\K@Cw}}\def\K@il{\let\xgrid\K@dm\let\ygrid\K@dn}\def
~\K@Ae{\let\xunit\K@dm\let\yunit\K@dn}\def\K@Jx#1#2{\ifdim#1<\z@\else#2\fi
~}\K@bg\xgrid{\K@Jw {\advance\K@do by \K@dm\K@Jx\K@do{\K@dm=\K@do\let\K@Jy
~\K@co}}{\K@Jx\K@do{\K@dm=\K@do\let\K@Jy\K@co}}}\K@bg\ygrid{\K@Jw {\advance
~\K@do by \K@dn\K@Jx\K@do{\K@dn=\K@do\let\K@Jz\K@co}}{\K@Jx\K@do{\K@dn
~=\K@do\let\K@Jz\K@co}}}\K@bg\grid{\K@Jw {\K@dp=\K@do\advance\K@do by \K@dm
~\K@Jx\K@do{\K@dm=\K@do\let\K@Jy\K@co}\advance\K@dp by \K@dn\K@Jx\K@dp
~{\K@dn=\K@dp\let\K@Jz\K@co}}{\K@Jx\K@do{\K@dm=\K@do\let\K@Jy\K@co\K@dn
~=\K@do\let\K@Jz\K@co}}}\def\K@Ka#1#2{\ifdim#1>\z@#2\fi}\K@bg\xunit{\K@Jw
~{\ifx\K@zx\K@ci\else\advance\K@do by \K@zx\fi\K@Ka\K@do{\K@Jp\K@zx}}{\K@Ka
~\K@do{\K@Jp\K@zx}}}\K@bg\yunit{\K@Jw {\ifx\K@zz\K@ci\else\advance\K@do by
~\K@zz\fi\K@Ka\K@do{\K@Jp\K@zz}}{\K@Ka\K@do{\K@Jp\K@zz}}}\K@bg\unit{\K@Jw
~{\K@dp=\K@do\ifx\K@zx\K@ci\else\advance\K@do by \K@zx\fi\K@Ka\K@do{\K@Jp
~\K@zx}\ifx\K@zz\K@ci\else\advance\K@dp by \K@zz\fi\K@Ka\K@dp{\edef\K@zz
~{\the\K@dp}}}{\K@Ka\K@do{\K@Jp\K@zx\let\K@zz\K@zx}}}\K@bg\xscale{\K@cz\let
~\K@Ji\relax\def\K@Jj{\K@do=\the\K@dr\K@dm\ifdim\K@do<\z@\else\K@dm=\K@do
~\edef\K@Jy{\the\K@dr}\fi}\K@Iz}\K@bg\yscale{\K@cz\let\K@Ji\relax\def\K@Jj
~{\K@do=\the\K@dr\K@dn\ifdim\K@do<\z@\else\K@dn=\K@do\edef\K@Jz{\the\K@dr
~}\fi}\K@Iz}\K@bg\scale{\K@cz\let\K@Ji\relax\def\K@Jj{\K@do=\the\K@dr\K@dm
~\ifdim\K@do<\z@\else\K@dm=\K@do\K@dn=\the\K@dr\K@dn\edef\K@Jy{\the\K@dr
~}\let\K@Jz\K@Jy\fi}\K@Iz}\def\K@Kb{\ifnum\K@di>0 \K@Ji\fi}\K@bg\xrange
~{\K@cz\def\K@Ji{\K@Jq\K@zy\let\K@zx\K@ci}\afterassignment\K@Kb\K@di}\K@bg
~\yrange{\K@cz\def\K@Ji{\K@Jq\K@Aa\let\K@zz\K@ci}\afterassignment\K@Kb\K@di
~}\K@bg\range{\K@cz\def\K@Ji{\K@Jq\K@zy\let\K@Aa\K@zy\let\K@zx\K@ci\let
~\K@zz\K@ci}\afterassignment\K@Kb\K@di}\K@bg\gridgray{\K@cz\let\K@Ji\relax
~\def\K@Jj{\edef\K@bc{\noexpand\K@Dg{\the\K@dr}}\K@bc\K@Dh}\K@Iz}\K@bm
~\gridgrey\gridgray\K@bg\graygray{\K@cz\let\K@Ji\relax\def\K@Jj{\edef
~\K@bc{\noexpand\K@Dg{\the\K@dr}}\K@bc\K@pv}\K@Iz}\K@bm\greygrey\graygray
~\K@bg\framegray{\K@cz\let\K@Ji\relax\def\K@Jj{\edef\K@bc{\noexpand\K@Dg
~{\the\K@dr}}\K@bc\K@Kc\ifx\K@Kc\K@ci\let\K@Ck\relax\let\K@Cm\relax\else
~\edef\K@Ck{\noexpand\K@ha{\K@Kc}}\def\K@Cm{\endgr}\fi}\K@Iz}\K@bm
~\framegrey\framegray\K@bg\shadegray{\K@cz\let\K@Ji\relax\def\K@Jj{\edef
~\K@bc{\noexpand\K@Dg{\the\K@dr}}\K@bc\K@Kd\ifx\K@Kd\K@ci\let\K@Cq\relax
~\let\K@Cr\relax\else\edef\K@Cq{\noexpand\K@ha{\K@Kd}}\def\K@Cr{\endgr}\fi
~}\K@Iz}\K@bm\shadegrey\shadegray\kuviocs\MinimumCellLength=0pt \K@He\ifx
~\@@agss@arrsy@@\relax\else\let\A@rightarrow\rightarrow\let\A@leftarrow
~\leftarrow\let\A@leftrightarrow\leftrightarrow\let\A@leftharpoonup
~\leftharpoonup\let\A@leftharpoondown\leftharpoondown\let\A@rightharpoonup
~\rightharpoonup\let\A@rightharpoondown\rightharpoondown\fi\K@bg\latexTo
~{\font\K@Ke=line10 \def\K@Kf{\smash{\hbox{\kern-4pt\raise\K@vj\hbox{\K@Ke
~\char"2D}\hskip.5pt}}}\newfill{srightarrow}--\K@Kf\newfill
~{sleftarrow}\K@Kf--\let\Tofill\srightarrowfill}\K@bk\newfill{\bgroup\K@ez
~\K@Kg}\def\K@Kg#1{\egroup\K@Kh{#1}}\def\K@Kh#1#2#3#4{\expandafter\def
~\csname D@@#1fill\endcsname{\m@th\mathord{#2}\mkern-8mu \cleaders\hbox
~{$\mkern-2mu\mathord{#3}\mkern-2mu$}\hfill\mkern-8mu\mathord
~{#4}}\expandafter\K@bk\csname#1fill\endcsname{\ifmmode\csname
~D@@#1fill\endcsname\else $\csname D@@#1fill\endcsname$\fi}}
\K@ab
\rekuviodef\into{\lhook\joinrel\A@rightarrow}
\rekuviodef\cdotfill{\cleaders
   \hbox{$\m@th\mkern1.5mu\mathord\cdot\mkern1.5mu$}\hfill}
\rekuviolet\cmleftarrowfill\leftarrowfill

\newfill{leftarrow}\A@leftarrow--
\newfill{rleftarrow}{-\mkern-12mu\A@leftarrow\mkern-5mu}--
\newfill{Leftarrow}\Leftarrow==
\newfill{leftharpoonup}\A@leftharpoonup--
\newfill{leftharpoondown}\A@leftharpoondown--
\newfill{hookleftarrow}\A@leftarrow-{-\mkern-8mu\rhook}
\rekuviolet\leftintofill\hookleftarrowfill
\newfill{lleftarrow}{\A@leftarrow\mkern-13mu\A@leftarrow\mkern-5mu}--
\newfill{leftepi}{\A@leftarrow\mkern-15mu\A@leftarrow\mkern-3mu}--
\newfill{rlleftarrow}{-\mkern-12mu\A@leftarrow\mkern-13mu\A@leftarrow
   \mkern-5mu}--
\newfill{llleftarrow}{\A@leftarrow\mkern-13mu\A@leftarrow
   \mkern-13mu\A@leftarrow\mkern-10mu}--
\newfill{rllleftarrow}{-\mkern-12mu\A@leftarrow\mkern-13mu\A@leftarrow
   \mkern-13mu\A@leftarrow\mkern-10mu}--

\rekuviolet\cmrightarrowfill\rightarrowfill
\newfill{rightarrow}--\A@rightarrow
\newfill{lrightarrow}--{\mkern-5mu\A@rightarrow\mkern-12mu-}
\newfill{Rightarrow}==\Rightarrow
\newfill{rightharpoonup}--\A@rightharpoonup
\newfill{rightharpoondown}--\A@rightharpoondown
\newfill{hookrightarrow}{\lhook\mkern-8mu-}-\A@rightarrow
\rekuviolet\rightintofill\hookrightarrowfill
\newfill{rrightarrow}--{\mkern-5mu\A@rightarrow\mkern-13mu\A@rightarrow}
\newfill{rightepi}--{\mkern-3mu\A@rightarrow\mkern-15mu\A@rightarrow}
\newfill{lrrightarrow}--{\mkern-5mu\A@rightarrow\mkern-13mu\A@rightarrow
   \mkern-12mu-}
\newfill{rrrightarrow}--{\mkern-10mu\A@rightarrow\mkern-13mu\A@rightarrow
   \mkern-13mu\A@rightarrow}
\newfill{lrrrightarrow}--{\mkern-10mu\A@rightarrow\mkern-13mu\A@rightarrow
   \mkern-13mu\A@rightarrow\mkern-12mu-}

\newfill{relbar}---
\rekuviolet\linefill\relbarfill
\newfill{Relbar}===
\rekuviolet\barfill\Relbarfill
\newfill{mapsto}{\mapstochar\mkern-5mu-}-\A@rightarrow
\newfill{leftrightarrow}\A@leftarrow-\A@rightarrow
\newfill{Leftrightarrow}\Leftarrow=\Rightarrow

\def\kuviojoin{\hbox{$\mkern-8mu$}}

\kuviodef\leftarrowxfill{\leftarrowfill\kuviojoin\relbarfill}
\kuviodef\leftarrowxxfill{\leftarrowfill\kuviojoin\relbarfill
   \kuviojoin\relbarfill}
\kuviodef\xleftarrowxfill{\relbarfill\kuviojoin\leftarrowfill
   \kuviojoin\relbarfill}
\kuviodef\leftarrowxxxfill{\leftarrowfill\kuviojoin\relbarfill
   \kuviojoin\relbarfill\kuviojoin\relbarfill}
\kuviodef\xxleftarrowxfill{\relbarfill\kuviojoin\relbarfill
   \kuviojoin\leftarrowfill\kuviojoin\relbarfill}

\kuviodef\rightarrowxfill{\rightarrowfill\kuviojoin\relbarfill}
\kuviodef\rightarrowxxfill{\rightarrowfill\kuviojoin\relbarfill
   \kuviojoin\relbarfill}
\kuviodef\xrightarrowxfill{\relbarfill\kuviojoin\rightarrowfill
   \kuviojoin\relbarfill}
\kuviodef\rightarrowxxxfill{\rightarrowfill\kuviojoin\relbarfill
   \kuviojoin\relbarfill\kuviojoin\relbarfill}
\kuviodef\xxrightarrowxfill{\relbarfill\kuviojoin\relbarfill
   \kuviojoin\rightarrowfill\kuviojoin\relbarfill}

\kuviodef\lleftarrowxfill{\lleftarrowfill\kuviojoin\relbarfill}
\kuviodef\lleftarrowxxfill{\lleftarrowfill\kuviojoin\relbarfill
   \kuviojoin\relbarfill}
\kuviodef\xlleftarrowxfill{\relbarfill\kuviojoin\lleftarrowfill
   \kuviojoin\relbarfill}
\kuviodef\lleftarrowxxxfill{\lleftarrowfill\kuviojoin\relbarfill
   \kuviojoin\relbarfill\kuviojoin\relbarfill}
\kuviodef\xxlleftarrowxfill{\relbarfill\kuviojoin\relbarfill
   \kuviojoin\lleftarrowfill\kuviojoin\relbarfill}

\kuviodef\rrightarrowxfill{\rrightarrowfill\kuviojoin\relbarfill}
\kuviodef\rrightarrowxxfill{\rrightarrowfill\kuviojoin\relbarfill
   \kuviojoin\relbarfill}
\kuviodef\xrrightarrowxfill{\relbarfill\kuviojoin\rrightarrowfill
   \kuviojoin\relbarfill}
\kuviodef\rrightarrowxxxfill{\rrightarrowfill\kuviojoin\relbarfill
   \kuviojoin\relbarfill\kuviojoin\relbarfill}
\kuviodef\xxrrightarrowxfill{\relbarfill\kuviojoin\relbarfill
   \kuviojoin\rrightarrowfill\kuviojoin\relbarfill}

\kuviodef\llleftarrowxfill{\llleftarrowfill\kuviojoin\relbarfill}
\kuviodef\llleftarrowxxfill{\llleftarrowfill\kuviojoin\relbarfill
   \kuviojoin\relbarfill}
\kuviodef\xllleftarrowxfill{\relbarfill\kuviojoin\llleftarrowfill
   \kuviojoin\relbarfill}

\kuviodef\llleftarrowxxxfill{\llleftarrowfill\kuviojoin\relbarfill
   \kuviojoin\relbarfill\kuviojoin\relbarfill}
\kuviodef\xxllleftarrowxfill{\relbarfill\kuviojoin\relbarfill
   \kuviojoin\llleftarrowfill\kuviojoin\relbarfill}

\kuviodef\rrrightarrowxfill{\rrrightarrowfill\kuviojoin\relbarfill}
\kuviodef\rrrightarrowxxfill{\rrrightarrowfill\kuviojoin\relbarfill
   \kuviojoin\relbarfill}
\kuviodef\xrrrightarrowxfill{\relbarfill\kuviojoin\rrrightarrowfill
   \kuviojoin\relbarfill}
\kuviodef\rrrightarrowxxxfill{\rrrightarrowfill\kuviojoin\relbarfill
   \kuviojoin\relbarfill\kuviojoin\relbarfill}
\kuviodef\xxrrrightarrowxfill{\relbarfill\kuviojoin\relbarfill
   \kuviojoin\rrrightarrowfill\kuviojoin\relbarfill}

\rekuviolet\labelstyle\scriptstyle
\rekuviolet\vertexstyle\displaystyle
\rekuviolet\diagramdot\bullet

\rekuviodef\Displaybox#1{%
   \ifvmode
      \ifinner
         \box#1%
      \else
         \vskip-\lastskip
         \vskip\baselineskip
         \box#1%
         {\advance\parskip by -\baselineskip\vskip-\parskip}%
      \fi
   \else
      \box#1%
   \fi}

\kuviocs\grid=1cm
\kuviocs\range=1

\kuviocs\vpad=0pt
\kuviocs\hpad=0pt

\kuviocs\gridgray=.5
\kuviocs\framegray=0
\kuviocs\shadegray=0
\kuviocs\graygray=.5

\kuviocs\framepad=5pt
\kuviocs\framerulewidth=.4pt
\kuviocs\Framerulewidth=.4pt
\kuviocs\Rulewidth=5pt

\kuviocs\celllength=1cm
\kuviocs\cellwidth=1cm
\kuviocs\columndist=15mm
\kuviocs\bracewidth=1cm

\labelpoint{Fillcell}={}
\labelpoint{Boxcell}={}
\labelpoint{Rule}={}

\ptpoint{Fillcell}={}
\ptpoint{Boxcell}={}
\ptpoint{Rule}={}

\labelwidthpad{Fillcell}=0pt
\labelwidthpad{Boxcell}=0pt
\labelwidthpad{Rule}=0pt

\labelpad{Fillcell}=0pt
\labelpad{Boxcell}=0pt
\labelpad{Rule}=0pt

\breakpad{Fillcell}=0pt
\breakpad{Boxcell}=0pt
\breakpad{Rule}=0pt

\cellpush{Fillcell}=0pt
\cellpush{Boxcell}=0pt
\cellpush{Rule}=1pt

\ptpush{Fillcell}=0pt
\ptpush{Boxcell}=0pt
\ptpush{Rule}=1pt

\atpush{Fillcell}=0pt
\atpush{Boxcell}=0pt
\atpush{Rule}=0pt

\joinpush{Fillcell}=0pt
\joinpush{Boxcell}=0pt
\joinpush{Rule}=1pt

\newcell{To}
\rekuviolet\Tofill\rightarrowfill

\newcell{One}
\rekuviolet\Onefill\rightarrowfill

\newcell{Bij}
\rekuviolet\Bijfill\leftrightarrowfill

\newcell{Mapsto}
\rekuviolet\Mapstofill\mapstofill

\newcell{Into}
\rekuviolet\Intofill\rightintofill

\newcell{Epi}
\rekuviolet\Epifill\rightepifill

\newcell{Line}
\rekuviolet\Linefill\linefill

\newcell{Nul}
\rekuviolet\Nulfill\hfil

\newcell{Dots}
\rekuviolet\Dotsfill\cdotfill

\newcell{Two}
\rekuviolet\Twofill\Rightarrowfill
\labelpad{Two}=.8pt
\atpush{Two}=.8pt
\breakpad{Two}=.8pt

\newcell{Impl}
\rekuviolet\Implfill\Rightarrowfill
\labelpad{Impl}=.8pt
\atpush{Impl}=.8pt
\breakpad{Impl}=.8pt

\newcell{Bar}
\rekuviolet\Barfill\barfill
\labelpad{Bar}=.8pt
\atpush{Bar}=.8pt
\breakpad{Bar}=.8pt

\newcell{Null}
\rekuviolet\Nullfill\hfil
\labelpad{Null}=.8pt
\atpush{Null}=.8pt
\breakpad{Null}=.8pt

\newcell{Eq}
\rekuviodef\Eqfill{\hfil=\hfil}
\labelpad{Eq}=.8pt
\atpush{Eq}=.8pt
\breakpad{Eq}=.8pt

\newDiagram{Diag}
\rekuviodef\everyDiag{\kuviocs\flexible\kuviocs\xgrid=0pt}

\newDiagram{Dg}
\rekuviodef\everyDg{\kuviocs\flexible\kuviocs\xgrid=0pt
   \kuviocs\ygrid+=-2mm\kuviocs\cellwidth+=3mm
   \kuviocs\bracewidth+=-2.5mm}

\newDiagram{Long}
\rekuviodef\everyLong{\kuviocs\flexible\kuviocs\xgrid=0pt
   \kuviocs\ygrid+=-5mm\kuviocs\bracewidth+=-2.5mm}

\rekuviodefine{\long\def}\arrsy{
   \input arrsy\relax

   \rekuviolet\epi\rightepi
   \rekuviolet\mono\rightmono

   \newfill{Leftharpoonup}\A@Leftharpoonup==
   \rekuviolet\Leftparafill\Leftharpoonupfill
   \newfill{Leftharpoondown}\A@Leftharpoondown==
   \rekuviolet\Leftallofill\Leftharpoondownfill

   \newfill{Rightharpoonup}==\A@Rightharpoonup
   \rekuviolet\Rightallofill\Rightharpoonupfill
   \newfill{Rightharpoondown}==\A@Rightharpoondown
   \rekuviolet\Rightparafill\Rightharpoondownfill

   \newfill{Lleftarrow}\A@Lleftarrow\A@Rrelbar\A@Rrelbar
   \newfill{Rrightarrow}\A@Rrelbar\A@Rrelbar\A@Rrightarrow
   \newfill{Rrelbar}\A@Rrelbar\A@Rrelbar\A@Rrelbar
   \rekuviolet\equivfill\Rrelbarfill
   \newfill{Lleftrightarrow}\A@Lleftarrow\A@Rrelbar\A@Rrightarrow

   \newfill{dashbar}{\mkern8mu}\A@dashbar{\mkern8mu}
   \rekuviolet\dashfill\dashbarfill
   \newfill{rightdash}{\mkern8mu}\A@dashbar\A@shortrightarrow
   \newfill{leftdash}\A@shortleftarrow\A@dashbar{\mkern8mu}

   \newfill{rightepi}--\A@rightepi
   \newfill{leftepi}\A@leftepi--
   \newfill{rightmono}{\A@rightmonotail\mkern2mu}-\A@rightarrow
   \newfill{leftmono}\A@leftarrow-{\mkern2mu\A@leftmonotail}
   \newfill{rightmonotail}{\A@rightmonotail\mkern2mu}--
   \newfill{leftmonotail}--{\mkern2mu\A@leftmonotail}
   \newfill{rightiso}{\A@rightmonotail\mkern2mu}-\A@rightepi
   \newfill{leftiso}\A@leftepi-{\mkern2mu\A@leftmonotail}

   \newfill{rightdashmono}{\A@rightmonotail\mkern2mu}\A@dashbar
      \A@shortrightarrow
   \newfill{leftdashmono}\A@shortleftarrow\A@dashbar{\mkern2mu\A@leftmonotail}
   \newfill{rightdashmonotail}{\A@rightmonotail\mkern2mu}\A@dashbar{\mkern8mu}
   \newfill{leftdashmonotail}{\mkern8mu}\A@dashbar{\mkern2mu\A@leftmonotail}
   \newfill{rightdashepi}{\mkern8mu}\A@dashbar
      {\mkern-2.5mu\A@shortrightarrow\mkern-7.5mu\A@shortrightarrow}
   \newfill{leftdashepi}{\A@shortleftarrow\mkern-7.5mu\A@shortleftarrow
      \mkern-2.5mu}\A@dashbar{\mkern8mu}
   \newfill{rightdashiso}{\A@rightmonotail\mkern4mu}\A@dashbar
      {\mkern-2.5mu\A@shortrightarrow\mkern-7.5mu\A@shortrightarrow}
   \newfill{leftdashiso}{\A@shortleftarrow\mkern-7.5mu\A@shortleftarrow
      \mkern-2.5mu}\A@dashbar{\mkern4mu\A@leftmonotail}

   \newfill{squiggle}{\mkern8mu}\A@squiggle{\mkern8mu}

   \newcell{Allo}
   \rekuviolet\Allofill\Rightallofill
   \labelpad{Allo}=.8pt
   \atpush{Allo}=.8pt
   \breakpad{Allo}=.8pt

   \newcell{Para}
   \rekuviolet\Parafill\Rightparafill
   \labelpad{Para}=.8pt
   \atpush{Para}=.8pt
   \breakpad{Para}=.8pt

   \newcell{Three}
   \rekuviolet\Threefill\Rrightarrowfill
   \labelpad{Three}=1.5pt
   \atpush{Three}=1.5pt
   \breakpad{Three}=1.5pt

   \newcell{Equiv}
   \rekuviolet\Equivfill\Rrelbarfill
   \labelpad{Equiv}=1.5pt
   \atpush{Equiv}=1.5pt
   \breakpad{Equiv}=1.5pt

   \newcell{Nulll}
   \rekuviolet\Nulllfill\hfil
   \labelpad{Nulll}=1.5pt
   \atpush{Nulll}=1.5pt
   \breakpad{Nulll}=1.5pt

   \newcell{Dash}
   \rekuviolet\Dashfill\dashfill

   \newcell{Dashto}
   \rekuviolet\Dashtofill\rightdashfill

   \newcell{Mono}
   \rekuviolet\Monofill\rightmonofill

   \newcell{Tail}
   \rekuviolet\Tailfill\rightmonotailfill

   \newcell{Iso}
   \rekuviolet\Isofill\rightisofill
}

\kuviobye

  \def\pp{{\mathchoice
          {%displaystyle
              \kern 1pt%
              \raise 1pt
              \vbox{\hrule width5pt height0.4pt depth0pt
                    \kern -2pt
                    \hbox{\kern 2.3pt
                          \vrule width0.4pt height6pt depth0pt
                          }
                    \kern -2pt
                    \hrule width5pt height0.4pt depth0pt}%
                    \kern 1pt
           }
            {%textstyle
              \kern 1pt%
              \raise 1pt
              \vbox{\hrule width4.3pt height0.4pt depth0pt
                    \kern -1.8pt
                    \hbox{\kern 1.95pt
                          \vrule width0.4pt height5.4pt depth0pt
                          }
                    \kern -1.8pt
                    \hrule width4.3pt height0.4pt depth0pt}%
                    \kern 1pt
            }
            {%scriptstyle
              \kern 0.5pt%
              \raise 1pt
              \vbox{\hrule width4.0pt height0.3pt depth0pt
                    \kern -1.9pt  %[e]=0.15pt
                    \hbox{\kern 1.85pt
                          \vrule width0.3pt height5.7pt depth0pt
                          }
                    \kern -1.9pt
                    \hrule width4.0pt height0.3pt depth0pt}%
                    \kern 0.5pt
            }
            {%scriptscriptstyle
              \kern 0.5pt%
              \raise 1pt
              \vbox{\hrule width3.6pt height0.3pt depth0pt
                    \kern -1.5pt
                    \hbox{\kern 1.65pt
                          \vrule width0.3pt height4.5pt depth0pt
                          }
                    \kern -1.5pt
                    \hrule width3.6pt height0.3pt depth0pt}%
                    \kern 0.5pt%}
            }
        }}

  \def\mm{{\mathchoice
                       {%displaystyle
                             \kern 1pt
               \raise 1pt    \vbox{\hrule width5pt height0.4pt depth0pt
                                  \kern 2pt
                                  \hrule width5pt height0.4pt depth0pt}
                             \kern 1pt}
                       {%textstyle
                            \kern 1pt
               \raise 1pt \vbox{\hrule width4.3pt height0.4pt depth0pt
                                  \kern 1.8pt
                                  \hrule width4.3pt height0.4pt depth0pt}
                             \kern 1pt}
                       {%scriptstyle
                            \kern 0.5pt
               \raise 1pt
                            \vbox{\hrule width4.0pt height0.3pt depth0pt
                                  \kern 1.9pt
                                  \hrule width4.0pt height0.3pt depth0pt}
                            \kern 1pt}
                       {%scriptscriptstyle
                           \kern 0.5pt
             \raise 1pt  \vbox{\hrule width3.6pt height0.3pt depth0pt
                                  \kern 1.5pt
                                  \hrule width3.6pt height0.3pt depth0pt}
                           \kern 0.5pt}
                       }}

\catcode`@=11
\def\un#1{\relax\ifmmode\@@underline#1\else
        $\@@underline{\hbox{#1}}$\relax\fi}
\catcode`@=12

\let\du=\du                     % dot-under
\let\br=\ub                     % breve
\def\c{\chi}
\def\d{\delta}

\def\o{\omega}
\def\p{\pi}

\def\D{\Delta}

\def\ve{\varepsilon}

\def\cf{{\cal F}}

\def\cl{{\cal L}}
\def\cm{{\cal M}}

\def\cp{{\cal P}}

\def\bo{{\raise-.5ex\hbox{\large$\Box$}}}               % D'Alembertian
\def\pa{\partial}                                       % curly d
\def\pr{\prod}                                          % product
\def\TH{{\raise.2ex\hbox{$\displaystyle \bigodot$}\mskip-4.7mu \llap H \;}}
\def\face{{\raise.2ex\hbox{$\displaystyle \bigodot$}\mskip-2.2mu \llap {$\ddot
        \smile$}}}                                      % happy face
\def\Bar#1{\overline{#1}}                       % big bar
\def\leftrightarrowfill{$\mathsurround=0pt \mathord\leftarrow \mkern-6mu
        \cleaders\hbox{$\mkern-2mu \mathord- \mkern-2mu$}\hfill
        \mkern-6mu \mathord\rightarrow$}
\def\dvec#1{\vbox{\ialign{##\crcr
        \leftrightarrowfill\crcr\noalign{\kern-1pt\nointerlineskip}
        $\hfil\displaystyle{#1}\hfil$\crcr}}}           % <--> accent
\def\dt#1{{\buildrel {\hbox{\LARGE .}} \over {#1}}}     % dot-over for sp/sb
\def\frac#1#2{{\textstyle{#1\over\vphantom2\smash{\raise.20ex
        \hbox{$\scriptstyle{#2}$}}}}}                   % fraction
\def\sfrac#1#2{{\vphantom1\smash{\lower.5ex\hbox{\small$#1$}}\over
        \vphantom1\smash{\raise.4ex\hbox{\small$#2$}}}} % alternate fraction
\def\bfrac#1#2{{\vphantom1\smash{\lower.5ex\hbox{$#1$}}\over
        \vphantom1\smash{\raise.3ex\hbox{$#2$}}}}       % "
\def\afrac#1#2{{\vphantom1\smash{\lower.5ex\hbox{$#1$}}\over#2}}    % "
\def\[{\lfloor{\hskip 0.35pt}\!\!\!\lceil}
\def\]{\rfloor{\hskip 0.35pt}\!\!\!\rceil}

\def\du#1#2{_{#1}{}^{#2}}

\def\fracm#1#2{\hbox{\large{${\frac{{#1}}{{#2}}}$}}}

\def\tr{{\rm tr}}

\def\un{\underline}
\def\fracmm#1#2{{{#1}\over{#2}}}

\def\low#1{{\raise -3pt\hbox{${\hskip 0.75pt}\!_{#1}$}}}

\def\Dot#1{\buildrel{_{_{\hskip 0.01in}\bullet}}\over{#1}}
\def\dt#1{\Dot{#1}}

\newskip\humongous \humongous=0pt plus 1000pt minus 1000pt
\def\caja{\mathsurround=0pt}
\def\eqalign#1{\,\vcenter{\openup2\jot \caja
        \ialign{\strut \hfil$\displaystyle{##}$&$
        \displaystyle{{}##}$\hfil\crcr#1\crcr}}\,}
\newif\ifdtup

\def\pl#1#2#3{Phys.~Lett.~{\bf {#1}B} (19{#2}) #3}
\def\np#1#2#3{Nucl.~Phys.~{\bf B{#1}} (19{#2}) #3}

\def\pr#1#2#3{Phys.~Rev.~{\bf D{#1}} (19{#2}) #3}
\def\cqg#1#2#3{Class.~and Quantum Grav.~{\bf {#1}} (19{#2}) #3}
\def\cmp#1#2#3{Commun.~Math.~Phys.~{\bf {#1}} (19{#2}) #3}

\def\ibid#1#2#3{{\it ibid.}~{\bf {#1}} (19{#2}) #3}

\parskip=\medskipamount                 % space between paragraphs (LaTeX)
\thicklines                         % thick straight lines for pictures (LaTeX)

\begin{document}
\thispagestyle{empty}

{\hbox to\hsize{
\vbox{\noindent KL~--~TH 02/04   \hfill September 2002 \\
                hep-th/0209063   \hfill  }}}

\noindent
\vskip1.3cm
\begin{center}

{\Large\bf IIA String Instanton Corrections \vglue.1in
to the Four-Fermion Correlator in
\vglue.2in
the Intersection of Del Pezzo Surfaces~\footnote{Supported in part 
by the `Deutsche Forschungsgemeinschaft' and the `Volkswagen Stiftung'}}
\vglue.2in

Andrei Cheshel~\footnote{On leave from Department of Physics, Krasnojarsk
State University, 660028 Russia}

{\it Department of Theoretical Physics\\
     University of Kaiserslautern}\\
{\it 67653 Kaiserslautern, Germany}\\
{\sl cheshel@physik.uni-kl.de}

and

Sergei V. Ketov 

{\it Department of Physics\\
     Tokyo Metropolitan University\\
     1--1 Minami-osawa, Hachioji-shi\\
     Tokyo 192--0397, Japan}\\
{\sl ketov@comp.metro-u.ac.jp}
\end{center}
\vglue.2in
\begin{center}
{\Large\bf Abstract}
\end{center}

\noindent The Becker-Becker-Strominger formula, describing the string 
world-sheet instanton corrections to the four-fermion correlator in the 
Calabi-Yau compactified type-IIA superstrings, is calculated in the special 
case of the Calabi-Yau threefold realized in the intersection of two Del Pezzo 
surfaces. We also derive the selection rules in the supersymmetric GUT of the
Pati-Salam type associated with our construction.

\newpage

\section{Introduction}

One of the central problems in modern string and field theories is a 
calculation of strong-coupling effects. A calculation of the instanton 
corrections to various physical quantities is the important part of this 
problem. A study of the non-perturbative corrections in string theory due to 
the M-Theory branes was pioneered by Becker, Becker and Strominger  
\cite{bbs}. The simplest instantons are the so-called string world-sheet 
instantons whose contributions are independent upon the string coupling. The 
string world-sheet instantons were extensively studied in the past \cite{ww} 
even before ref.~\cite{bbs}. In the context of the IIA superstring 
compactification, the existence of the world-sheet string instantons can be 
related to the holomorphic curves in the internal {\it Calabi-Yau} (CY) space
 \cite{ww}.  

In the context of eleven-dimensional M-Theory \cite{m}, the ten-dimensional 
IIA superstring theory arises from the M-Theory compactification on a circle 
$S^1$, whereas the IIA superstrings themselves can be understood as the double
compactified (in spacetime as well as in worldvolume) M2-branes \cite{duff}. 

The M2-branes can be wrapped about the $S^1$ and a CY (supersymmetric) 
2-cycle ${\cal C}_2$. They give rise to instantons in four (uncompactified) 
spacetime dimensions, whose effects can be computed by the standard methods of
quantum field theory \cite{itep}. The low-energy effective 
four-dimensional field theory of the CY compactified type-IIA superstrings 
is given by the N=2 supergravity interacting with $h_{2,1}$ hypermultiplets and
$h_{1,1}$ vector N=2 multiplets, where $h_{2,1}$ and $h_{1,1}$ are the Hodge
numbers of CY \cite{pol}. The moduli space $\cm$ of the compactified theory is 
given by a direct product of the hypermultiplet moduli space $\cm_{H}$ and the
N=2 vector multiplet moduli space $\cm_V$, while the 
$S^1\times {\cal C}_2$-wrapped M2-branes correct the geometry of $\cm_V$ only.
The BPS (or the supersymmetric map) condition on these wrapped M2-brane
configurations just amounts to the holomorphy condition on the world-sheet
instantons \cite{ww}. The same conclusion was rederived in ref.~\cite{bbs} by
requiring the equivalence between a global supersymmetry transformation and a
kappa-transformation of the Green-Schwarz supersring action, 
$$ \pa X^{\bar m}=0~,\quad {\rm or}\quad \bar{\pa}X^m=0~,\eqno(1.1)$$
where $\pa$ is the holomorphic string world-sheet exterior derivative, 
and $X^m$ are the complex coordinates in CY, $m=1,2,3$.

The topological equation formally describing the string world-sheet instanton 
corrections to the four-point fermion (gaugino) correlator $\cf_{IJKL}$, where 
$I,J,K,L=1,2,\ldots, h_{1,1}$, was obtained by Becker, Becker and Strominger 
 \cite{bbs},
$$ \D_{{\cal C}_2}\cf_{IJKL}= Ne^{-\int_{{\cal C}_2}J-i\int_{{\cal C}_2}B}
\int_{{\cal C}_2}b_I\int_{{\cal C}_2}b_J\int_{{\cal C}_2}b_K
\int_{{\cal C}_2}b_L~~~~,\eqno(1.2)$$
where ${\cal C}_2$ is the homology class of the instanton, $\{b_I\}$ is the 
orthonormal basis of harmonic $(1,1)$ forms in CY, $J$ is the K\"ahler $(1,1)$
form of CY, $B$ is the (closed) Neveu-Schwarz $(1,1)$ form, and $N$ is the 
normalization factor independent upon $b_I$. 

Like any other $(1,1)$ form, the form $J+iB$ can be decomposed with respect to
the cohomology basis $\{b_I\}$,
$$ J+iB= \sum_{I=1}^{h_{1,1}} z^Ib_I~~,\eqno(1.3)$$
where the complex coefficients $\{z^I\}$ are called CY moduli. Integrating
eq.~(1.3) once with respect to the modulus $z^I$ yields the famous 
topological formula for the world-sheet instanton corrections to the Yukawa 
couplings $\cf_{IJK}$ \cite{pol}. In the case of Yukawa couplings, mirror 
symmetry is known to confirm the topological equation on them \cite{check}. 
This fact indirectly supports the more general equation (1.2) also \cite{bbs}.
   
Like the similar equation on the Yukawa couplings, the topological eq.~(1.2) 
is merely a formal equation since one still has to specify how the integrals 
on the right-hand-side of the equation are to be calculated. Their calculation
 for a generic CY space represents the important technical problem whose 
solution is unknown, to the best of our knowledge. There are, nevertheless, 
some explicit calculations of the Yukawa couplings in the literature for the 
special CY spaces realized as the complete intersections in a product of the 
projective spaces \cite{check}. 

Our main purpose in this paper is to calculate eq.~(1.2) in the case of the 
special CY to be defined in the intersection of Del Pezzo surfaces. The Del 
Pezzo surface is a manifold of complex dimension $2$ with a positive 
first Chern class \cite{manin}. We use the `old' geometrical methods first 
developed in 
ref.~\cite{dgkm} for computing the Yukawa couplings in a superstring model with
three generations of quarks and leptons, see also ref.~\cite{bh}. The 
geometrical approach is based on Poincar\'e duality and a knowledge of the 
homology group basis of CY.

Our paper is organized as follows: in sect.~2 we introduce into the special
CY spaces realized as the complete intersections of five quadrics in a
product of two projective spaces $\cp^4\times \cp^4$. The main body of our
paper is given by sect.~3 where we formulate the mathematical instruments 
allowing us to calculate the integrals in eq.~(1.2). Some explicit examples 
are given in sect.~4. We conclude with sect.~5 where some connections between 
our work and the recent literature are outlined.

\section{Quadrics, Del Pezzo and CY}

Let's consider the compact CY spaces realized as the complete intersections 
in a product of two projective spaces, $\cp^4\times \cp^4$, with the 
configuration matrix~\footnote{We use the standard notation \cite{cy}.}
$$ \left( \begin{array}{ccccccc} 4 & || & 2 & 2 & 0 & 0 & 1 \cr
 4 & || & 0 & 0 & 2 & 2 & 1 \cr \end{array} \right)~~~. \eqno(2.1)$$
To decipher this matrix, we introduce two sets of homogeneous 
coordinates, $x$ and $y$, in each $\cp^4$, define two Del Pezzo surfaces 
$K_x$ and $K_y$, and a  hypersurface $S$ in $\cp^4\times \cp^4$ by the 
following constraints \cite{dr}:
$$\eqalign{
K_x = \left\{ x\in \cp^4:\qquad P_1(x)=\sum^4_{i=0}x_i^2=0~,\quad 
P_2(x)=\sum_{i=0}^4 a_i x_i^2 =0 \right\}~~, \cr
K_y = \left\{ y\in \cp^4:\qquad P_3(y)=\sum^4_{i=0}y_i^2=0~,\quad 
P_4(x)=\sum_{i=0}^4 b_i y_i^2 =0 \right\}~~, \cr
S = \left\{ (x,y)\in \cp^4\times \cp^4:\qquad P_5(x,y)=\sum^4_{i,j=0}
c_{ij}x_iy_j=0 \right\}~~. \cr}\eqno(2.2)$$

The sum of entries in each line of the matrix (2.1) to the right of $||$ 
exceeds exactly by one the dimension of the embedding space $\cp^4$ to the left
of $||$, so that
$$ K_0= (K_x\times K_y)\cap S \eqno(2.3)$$
appears to be K\"ahler and of the vanishing first Chern class, i.e. $K_0$ is a
 CY space, in agreement with the theorem of Greene, Vafa and Warner \cite{gvw}.
 We assume that the real coefficients of the quadrics $P_2$, $P_4$ and $P_5$ 
in eq.~(2.2) are chosen to obey the transversality condition for all 
hypersurfaces in the definition (2.3) of $K_0$, i.e. 
$$ dP_1\wedge dP_2\wedge dP_3\wedge dP_4\wedge dP_5\neq 0~.\eqno(2.4)$$
This equation guarantees the smoothness of the simply connected manifold
$K_0$ \cite{jug}.

The first column in eq.~(2.1) thus indicates that we consider a CY in the 
product $\cp^4\times \cp^4$, whereas the other columns denote bi-powers of the
polynomials of $x$ and $y$ in eq.~(2.2). 

The non-trivial Hodge numbers of $K_0$ are given by 
$$ h_{2,1}(K_0)=28\quad {\rm and}\quad h_{1,1}(K_0)=12~,\eqno(2.5)$$
so that its Euler characteristic $\c(K_0)$ is 
$$\fracmm{1}{2}\c(K_0)= h_{1,1}- h_{2,1}=-16~.\eqno(2.6)$$ 

It is not easy to construct the complete intersection CY spaces with the 
physically interesting values of the Euler characteristic, $\c=\pm 6,\pm 8$. 
However, it is easily becomes possible through the so-called `orbifoldization' 
process \cite{cy}. In our case, we can introduce the quotient $K$ of the 
manifold $K_0$ with respect to a discrete symmetry subgroup 
$G_F={\bf Z}_2^2$ of $G$, which acts freely in $K_0$. This yields the CY 
space $K$ of the Euler characteristic $\c(K)=-8$. The $G$-group 
elements generating $G_F$ can be chosen as follows:
$$ g_1 = {\rm diag}(1,1,1,-1,-1)~,\quad
g_2 = {\rm diag}(1,1,-1, 1,-1)~,\eqno(2.7)$$
so that the action of $g_1$ reads
$$ g_1:~(x_0,x_1,x_2,x_3,x_4;y_0,y_1,y_2,y_3,y_4)\to
 (x_0,x_1,x_2,-x_3,-x_4;y_0,y_1,y_2,-y_3,-y_4)~,\eqno(2.8)$$
and similarly for $g_2$. The manifold $K_0$ has the hidden 
discrete symmetry group 
$G$ isomorphic to ${\bf Z}_2^5$, whose action is given by 
$$ \eqalign{
{\bf Z}_2(A):~ & A={\rm diag}(-1,1,1,1,1)~~,\cr
{\bf Z}_2(B):~ & B={\rm diag}(1,-1,1,1,1)~~,\cr
{\bf Z}_2(C):~ & C={\rm diag}(1,1,-1,1,1)~~,\cr
{\bf Z}_2(D):~ & D={\rm diag}(1,1,1,-1,1)~~,\cr
{\bf Z}_2(S):~ & S(x_i)=y_i~,\quad S(y_j)=x_j~~.\cr}
\eqno(2.9)$$

In the embedding space $\cp^4\times \cp^4$ we have
$$ g_1=ABC\quad {\rm and} \quad g_2=ABD~.\eqno(2.10)$$
The CY manifold $K_0$ is the simply connected covering space of the
CY space $K$. The latter still possesses some hidden symmetries that
survive after its factorization by $G_F$. These discrete symmetries are 
$$ G_H=\fracmm{{\bf Z}_2(A)\times{\bf Z}_2(B)\times{\bf Z}_2(C)
\times{\bf Z}_2(D)\times{\bf Z}_2(S)}{{\bf Z}_2(g_1)\times{\bf Z}_2(g_2)}
={\bf Z}_2(A)\times{\bf Z}_2(B)\times{\bf Z}_2(S)~~.\eqno(2.11)$$

In the context of the IIA superstring compactification, the CY space $K$ 
gives rise to the four-dimensional unified model with {\it four} generations
of quarks and leptons, and an $E_6$ gauge group. Further breaking of $E_6$
by the standard mechanism of the vacuum Wilson loops yields the 
Pati-Salam-like unified model with a gauge group $SU(4)_c\times SU(2)_L
\times SU(2)_R\times U(1)$. The Yukawa couplings in this four-generation 
superstring model were calculated in ref.~\cite{dr}. 
 
\section{Instantons in Del Pezzo}

Equation (1.1) implies that the CY-compactified IIA superstring world-sheet 
instantons are described by the isolated holomorphic curves in CY. A single 
instanton corresponds to a curve of genus zero. In the case of $K_0$, there 
are 256 holomorphic or $CP(1)$ curves. A derivation of this number was given,
for example, in ref.~\cite{bd} where it appeared as the leading term in the 
series expansion of the fundamental period as a solution to the Picard-Fuchs 
equation for the given CY. A geometrical derivation of the same result is  
given below in this section. However, first we need more information about the
 geometrical structure of the space $K_0$ defined by eq.~(2.3) and the 
topology of the Del Pezzo surfaces $K_x$ and $K_y$.

As is well known in algebraic geometry \cite{gh}, a smooth intersection of two
quadrics in $\cp^4$ is biholomorphic equivalent to the projective plane with
five different blown-up points. Since the Hodge number $h_{1,1}$ of $\cp^2$ 
is equal to one, after blowing up at five points $h_{1,1}$ equals to $1+5=6$,
while the other Hodge numbers remain unchanged. Next, the Del Pezzo surface
$K_x$ possesses exactly 16 complex lines  $\{C_x\}$ that can be described 
by the relations
$$\eqalign{ 
a_{42}a_{32}a_{10}x_2-\ve_1a_{40}a_{30}a_{42}x_0-i\ve_2a_{41}a_{31}a_{20}x_1
& =~0~,\cr
a_{43}a_{32}a_{10}x_3-\ve_3a_{40}a_{31}a_{20}x_0-\ve_4a_{41}a_{30}a_{21}x_1
& =~0~,\cr
a_{43}a_{42}a_{10}x_4-\ve_5a_{41}a_{30}a_{20}x_0-i\ve_6a_{40}a_{21}a_{31}x_1
& =~0~,\cr}
\eqno(3.1)$$
where $a_{kl}=\sqrt{a_k-a_l}$, $0\leq l<k\leq 4$, and $\ve_j=\pm 1$, 
$j=1,2,\ldots,6$. The sign coefficients $\ve_j$ and our notation for the 
complex lines on the Del Pezzo surfaces are collected in {\sl Table I.} 

The homology class of the K\"ahler form on Del Pezzo $K_x$ can be represented 
by the intersection of the hyperplane $S$ with $K_x$,
$$  H =\{x_0=0\}\subset K_x~~.\eqno(3.2)$$

Under the symmetry group $G$ the 16 lines on the Del Pezzo 
surface $K_x$ are naturally decomposed into {\it three} classes: (i) the five 
lines $E_i$, $i=1,2,3,4,5$, that (pairwise) do not intersect with each other 
and thus represent five linearly independent homology classes of 
$H_2(K_x,{\bf R})$; together with the hyperflat section $H$ 
(dual to a K\"ahler form of the Del Pezzo surface $K_x$) they form a basis in
 $H_2(K_x,{\bf R})$, (ii) ten lines $F_{ij}$
that have intersections only with $E_i$ and $E_j$, and (iii) one line $G$
intersecting with all $E_i$ (see {\sl Table 1} too). The 256 holomorphic 
curves are then decomposed with respect to the $G_F={\bf Z_2}\times{\bf Z}_2$ 
discrete symmetry group of order 4 into four classes that are cyclically 
symmetric with respect to their interchanging.
\vglue.2in

{\sl Table 1.} Complex lines in the Del Pezzo surface $K_x$ $(K_y)$ 
and the sign factors $\ve_j$.

\begin{tabular}{c|cccccc|c|cccccc}
\hline
line & $\ve_1$ &  $\ve_2$ &  $\ve_3$ &  $\ve_4$ &  $\ve_5$ &  $\ve_6$ &
line & $\ve_1$ &  $\ve_2$ &  $\ve_3$ &  $\ve_4$ &  $\ve_5$ &  $\ve_6$ \\
\hline
$E_1$ & $+$ &  $+$ &  $-$ &  $+$ &  $+$ &  $+$ &
$F_{14}$ & $-$ &  $-$ &  $-$ &  $+$ &  $+$ &  $+$ \\
$E_2$ & $-$ &  $+$ &  $+$ &  $+$ &  $+$ &  $-$ &
$F_{15}$ & $+$ &  $-$ &  $-$ &  $-$ &  $+$ &  $-$ \\
$E_3$ & $+$ &  $+$ &  $+$ &  $-$ &  $-$ &  $-$ &
$F_{23}$ & $-$ &  $+$ &  $-$ &  $-$ &  $+$ &  $-$ \\
$E_4$ & $-$ &  $-$ &  $-$ &  $+$ &  $-$ &  $-$ &
$F_{34}$ & $-$ &  $-$ &  $+$ &  $-$ &  $-$ &  $-$ \\
$E_5$ & $+$ &  $-$ &  $-$ &  $-$ &  $-$ &  $+$ &
$F_{25}$ & $-$ &  $-$ &  $+$ &  $-$ &  $+$ &  $+$ \\
$G$ & $+$ &  $+$ &  $-$ &  $+$ &  $-$ &  $-$ &
$F_{34}$ & $-$ &  $-$ &  $+$ &  $-$ &  $-$ &  $-$ \\
$F_{12}$ & $-$ &  $+$ &  $+$ &  $+$ &  $-$ &  $+$ &
$F_{35}$ & $+$ &  $-$ &  $+$ &  $+$ &  $-$ &  $+$ \\
$F_{13}$ & $+$ &  $+$ &  $+$ &  $-$ &  $+$ &  $+$ &
$F_{45}$ & $-$ &  $+$ &  $-$ &  $-$ &  $-$ &  $+$ \\ 
\hline
\end{tabular}
\vglue.2in

Accordingly, we get the following matrix of the intersection indices:
$$ \eqalign{
(E_i,E_j)~ =~ & -\d_{ij}~,\qquad (E_i,F_{jk})=\d_{ij}+\d_{ik}~,\cr
(E_i,G)~~ =~ & 1~, \quad (E_i,H)=1~,\quad  (H,H)=4~.\cr}\eqno(3.3)$$

Having obtained the holomorphic curves and the homology basis explicitly,  
it is not difficult to determine the action of the discrete symmetry group 
on the latter. The generating elements $(g_1,g_2,A,B)$ of the group $G$ act as
follows:
$$\eqalign{
g_1(E_1,E_2,E_3,E_4,E_5,H)~=~& (E_3,F_{45},E_1,F_{25},F_{24},H)~,\cr
g_2(E_1,E_2,E_3,E_4,E_5,H)~=~& (E_4,F_{35},F_{25},E_1,F_{23},H)~,\cr
A(E_1,E_2,E_3,E_4,E_5,H)~=~& (F_{12},G,F_{23},F_{24},F_{25},H)~,\cr
B(E_1,E_2,E_3,E_4,E_5,H)~=~& (F_{15},F_{25},F_{35},F_{45},G,H)~.\cr}
\eqno(3.4)$$
For example, to get $g_1(E_2)=F_{45}$, we choose for definiteness $a_0=0$,
$a_1=1$, $a_3=3$ and $a_4=4$ in eq.~(3.1) where the coefficients $\ve_j$ are 
given by {\sl Table 1}. We find
$$\eqalign{ 
g_1(E_2) & = g_1 \left\{ x_2 + \sqrt{6}\, x_0 -i\sqrt{6} 
= x_3-i4x_0 -3x_1 = x_4-3x_0+2x_1 =0\right\} \cr
& = \left\{ x_2+\sqrt{6}\,x_0-i\sqrt{6}\,x_1=x_3+i4x_0+3x_1=
 x_4 +3x_0 -i2x_1=0\right\}=F_{45}~.\cr}\eqno(3.5)$$

The curve $H$ is invariant under all these symmetries, whereas each line $E_i$
goes into one of the 16 lines lying in the intersection of quadrics. The 
symmetry transformations act independently on each factor $\cp^4$ in a  
product $\cp^4\times \cp^4$, so that it is enough to consider only one 
projective space $\cp^4$. The action of the $S$ symmetry of $G$ just replaces
each $(1,1)$ form on Del Pezzo $K_x$ by the corresponding $(1,1)$ form on 
$K_y$. We find the following decompositions:
$$ F_{ij} = \fracmm{1}{3}\left( \sum^5_{i=1} E_i +H\right) -E_i - E_j~~,
\eqno(3.6)$$ 
and
$$ G= \fracmm{1}{3}\left(2H-\sum^5_{i=1} E_i\right)~~.\eqno(3.7)$$
For example, to prove eq.~(3.6), we begin with a decomposition
$$ F_{ij} = \sum^5_{i=1}c_iE_i + c_6 H \eqno(3.8)$$
whose coefficients $(c_i,c_6)$ are to be determined. Let's now consider the
intersections of $F_{ij}$ with $H$, $E_i$, $E_j$ and 
$M=\sum^5_{i=1}E_i$ by using the index intersection matrix (3.3). We find
$$ \eqalign{
(F_{ij},H)= & c_1 + c_2 +c_3+c_4+c_5+4c_6 =1~,\cr
(F_{ij},E_i)= & -c_i + c_6=1~,\cr
(F_{ij},E_j)= & -c_j+c_6=1~,\cr
(F_{ij},\sum^5_{i=1}E_i) & = -(c_1+c_2+c_3 +c_4+c_5)+5c_6=2~.\cr}\eqno(3.9)$$
Hence, the coefficients in eq.~(3.8) are given by
$$ c_6=1/3\quad {\rm and}\quad c+i=c_j=-2/3~.\eqno(3.10)$$
Equation (3.7) is obtained similarly. 

In the Grand Unification Theories of the Pati-Salam type, based on the gauge 
group $E_6$ that is supposed to be broken by Wilson lines as
$$ E_6\to SU(4)_c\times SU(2)_L\times SU(2)_R\times U(1)~,\eqno(3.11)$$ 
the representation $\Bar{\bf 27}$ of $E_6$ is decomposed as follows: 
$$\eqalign{
\Bar{\bf{27}}=~& [ (q,l)=(4c,2L,1R)] +[q^c,l^c=(\bar{4}c,1L,\bar{2}R)]\cr
 ~& + [H=(1c,2L,2R)] + [g,g^c=(6c,1L,1R)]+[n=(1c,1L,1R)]~,\cr}\eqno(3.12)$$  
where $(q_{R,L},l_{R,L})$ stand for the quark-lepton families, $H$ are the new
leptons, $g$ are the new quarks and $n$ is the singlet.

As was demonstrated in ref.~\cite{jug}, the particle spectrum corresponding to
the $(2,1)$ forms in $K$ is given by
$$ h_{2,1}:\qquad 10(n,g,g^c) + 6(f,H)~,\eqno(3.13)$$
where $f$ stands for $(q,l,q^c,l^c)$. The anti-particles corresponding to the
$(1,1)$ forms in $K$ are given by
$$ h_{1,1}:\qquad 6(\bar{n},\bar{g},\bar{g}^c) + 2(\bar{f},H)~.\eqno(3.14)$$
The transformation properties of the fields, in accordance with the
decomposition (3.12), are collected in {\sl Table 2.}

Equations (3.4) also allow us to identify the special combinations of the 
basic $(1,1)$ homology elements that are invariant under the CY symmetry 
group $G_H$ of eq.~(2.11). They are
$$\eqalign{
F_i^{x,y}= & H^{x,y} + 6E_i^{x,y}- 2\sum^5_{i=1}E_i^{x,y}~~,\cr
H^{\pm}= & H^x\pm H^y~~,F_i^{\pm}= F_i^x\pm F^y_i~~,}\eqno(3.15)$$
where $i=1,2,3,4,5$. 

In the context of the CY superstring compactification, the invariant elements 
of the $(1,1)$ cohomology basis correspond to the physical matter 
fields transforming in $\Bar{\bf 27}$ of $E_6$ \cite{pol,cy}. 
A direct calculation yields the following set of twelve invariant combinations
 in the given homology basis of $H_2(K^x,{\bf R})$ dual to the cohomology group
$H^{1,1}$ \cite{jug}:
$$ \eqalign{
(\bar{n},\bar{g},\bar{g}^c)_1=~ & H^x + H^y~,\cr
(\bar{n},\bar{g},\bar{g}^c)_2=~ & F^+_2~,\quad 
(\bar{n},\bar{g},\bar{g}^c)_3= F^+_5~,\cr
(\bar{n},\bar{g},\bar{g}^c)_4=~ & H^x - H^y\equiv H^-~,\cr
(\bar{n},\bar{g},\bar{g})_5=~ & F^-_2~,\quad
(\bar{n},\bar{g},\bar{g}^c)_6=  F^-_5~,\cr
(\bar{g},\bar{l})_1=~ & F^+_3~,\quad
(\bar{g},\bar{l})_2=  F^-_3~,\cr
(\bar{g}^c,\bar{l}^c)_1=~ & F^+_4~,\quad
(\bar{g}^c,\bar{l}^c)_2=  F^-_4~,\cr
\bar{H}_1=~& F^+_1~,\quad \bar{H}_2=F_1^-~,\cr}\eqno(3.16)$$
where the quark-lepton families $(q,l)$ and extra leptons $(H)$ are merely 
considered here as the formal notation. There are no two different 
combinations of the cycles that would have the same transformation properties 
under the discrete symmetries. We verified this statement by a 
straightforward calculation  (see {\sl Table 2}). This means that our 
identification of cycles is unique.
\vglue.2in

{\sl Table 2.} The transformation properties of the $(1,1)$ forms 
in ${\Bar{\bf 27}}$ of $E_6$ under the discrete symmetries.
\begin{center}
\begin{tabular}{c|ccccc}
\hline
fields & $g_1$ &  $g_2$ &  $A$ &  $B$ &  $S$ \\
\hline
$(\bar{n},\bar{g},\bar{g}^c)_1$ & $1$ &  $1$ &  $1$ &  $1$ &  $1$ \\
$(\bar{n},\bar{g},\bar{g}^c)_2$ & $1$ &  $1$ &  $1$ &  $-1$ &  $1$ \\
$(\bar{n},\bar{g},\bar{g}^c)_3$ & $1$ &  $1$ &  $-1$ &  $1$ &  $1$ \\
$(\bar{n},\bar{g},\bar{g}^c)_4$ & $1$ &  $1$ &  $-1$ &  $1$ &  $-1$ \\
$(\bar{n},\bar{g},\bar{g})_5$ & $1$ &  $1$ &  $1$ &  $-1$ &  $-1$ \\
$(\bar{n},\bar{g},\bar{g}^c)_6$ & $1$ &  $1$ &  $-1$ &  $1$ &  $-1$ \\
$(\bar{g},\bar{l})_1$ & $-1$ &  $1$ &  $-1$ &  $-1$ &  $1$ \\
$(\bar{g},\bar{l})_2$ & $-1$ &  $1$ &  $-1$ &  $-1$ &  $-1$ \\
$(\bar{g}^c,\bar{l}^c)_1$ & $1$ &  $-1$ &  $-1$ &  $-1$ &  $1$ \\
$(\bar{g},\bar{l})_2$ & $1$ &  $-1$ &  $-1$ &  $-1$ &  $-1$  \\
$\bar{H}_1$ & $-1$ &  $-1$ &  $-1$ &  $-1$ &  $1$ \\
$\bar{H}_2$ & $-1$ &  $-1$ &  $-1$ &  $-1$ &  $-1$ \\ 
\hline
\end{tabular}
\end{center}
\vglue.2in

The instantons in the Del Pezzo intersection have the form 
$C_x\times C_y$ that yields $16\times 16=256$, as it should. The intersection
 of these 256 surfaces with the hyperplane $S$ in eq.~(2.3) yields 128 complex
 curves of genus zero on one of the Del Pezzo surfaces $\times$ point on the 
other Del Pezzo surface. Accordingly, there are two ways of choosing on which 
Del Pezzo surface we take the line to lie on, while there are four ways of  
choosing a point on the other Del Pezzo surface. This yields in total 
$2\times 4 \times 16=128$ different instantons of the type $line\times point$,
 and, in addition, 128 different instantons of the type $line\times line$. 
Unlike ref.~\cite{dgkm}, where a similar problem was solved in the case of the
 {\it cubic} Del Pezzo intersection in $\cp^3\times \cp^3$, we have 
a more degenerate (and more symmetric) situation. 

Each holomorphic curve corresponding to an instanton is thus given by an
intersection of $C_x\times C_y$ with the hyperplane $S$ in accordance with 
eq.~(2.3), where $C_x$ are 16 lines in the Del Pezzo surface $K_x$ and 
similarly for $K_y$,  
$$\cl =(C_x\times C_y)\cap S~~.\eqno(3.17)$$
There are four classes amongst the 256 instantons that are cyclically 
connected in our case. To calculate eq.~(1.2) we have to choose 
a representative $\cl$ from each class. The holomorphic curve $\cl$ is the 
image of the string world-sheet in the CY space under the instanton map. 

Now, on the one hand side, the integral of any closed form $\o$ of the maximal
degree over $K_0=(K_x\times K_y)\cap S$ can be represented by the value of 
the cohomology class of $\o$ on the cycle $K_0$. On the other hand side, 
the cycle
intersection in the homologies is dual to the exterior multiplication in the
cohomologies. Hence, we have
$$ \int_{K_0} \o = (w\wedge H)[K_x\times K_y]~,\eqno(3.18)$$
where we have introduced the class of cohomologies $w$ of $\o$, and the 
image $H$ of the cohomology class of the K\"ahler form in $\cp^{24}$ dual to 
the hyperplane $S$. The hyperplane $S$ represents the hypersurface $S$ after
Segre embedding \cite{gh} restricted on $K_x\times K_y$. The Segre embedding
in this case means the embedding of a product $\cp^4\times \cp^4$ into the 
projective space $\cp^{24}$ by the coordinate identification 
$w_{ij}=x_iy_j$ where $w_{ij}$ are the homogeneous coordinates of $\cp^{24}$.

It is worth mentioning that only one component $H^x(H^y)$ remains in what 
follows from $H=H^x+H^y$. Hence, our problem reduces to a calculation of the
intersection indices from the homology group $H_2(K_x,{\bf Z})$ only. 
In general, the Poincar\'e duality establishes the isomorphism between the
closed (DeRham) homologies and compact-dual cohomology classes, as well as the 
isomorphism between compact (DeRham) homologies and cohomologies \cite{bott}.
In the {\it compact} CY case we consider, there is no difference between 
the closed and compact classes.
 
Taken together, this allows us to replace the integral of the $(1,1)$ 
form $b_I$ along the curve $\cl$ in eq.~(1.2) by the intersection index of 
this curve $\cl$ with the cycle $F_I$ that is Poincar\'e dual to $b_I$, 
$$ \int_{\cl}b_I=(F_I\cdot \cl)~.\eqno(3.19)$$ 

As a result, the instanton correction to the four-fermion correlator 
in eq.~(1.2) is proportional to a product
$$\int_\cl b_I \int_\cl  b_J \int_\cl b_K \int_\cl b_L  = 
(F_I\cdot \cl)(F_J\cdot \cl)(F_K\cdot \cl)
(F_L\cdot \cl)~,\eqno(3.20)$$
where the brackets with dots stand for the intersection indices of the
corresponding cycles, which are to be determined from the matrix (3.3) 
of our basic curves $(E^x_i,H^x,E^y_i,H^y)$ --- see the next sect.~4 for
some explicit examples.

We label our 256 curves by the corresponding lines in one of the following 
form: $line\times line$, $point\times point$ and $line\times line$ in the 
cover $K_0$. The validity of the cycle intersection matrix for both Del Pezzo 
surfaces $K_x$ and $K_y$ is justified by the fundamental commutative diagramm:
\vglue.1in

$$
\let\labelstyle\textstyle
\Dg
K_x  & \lTo^{p_x}          & K_x\times K_y & \rTo^{p_y}         & K_y \\
     & \luTo_{p_x\circ i}  &   \uTo >i     & \ruTo_{p_y\circ i} &  \\
     &                     &    \cl     & & \\ 
\endDg
$$
\vglue.1in

In a more explicit notation we just get
$$ \int_{\cl} b_I^x\equiv \left.\int_{\cl}p^*_xb_I^x\right|_{\cl}
=\int_{\cl}p^*i^*b_I^x
=\int_{\cl}(p_x\circ i)^*b^x_I=\int_{(p_x\circ i)_*\cl}b_I^xx\equiv
\int_{\cl_x}b_I^x=(F_x,\cl_x)~,\eqno(3.21)$$
where the projection $p_x$ lifts the form $b_I$ on $K_x$ to 
$K_x\times K_y$ with its simultaneous restriction on the curve $\cl_x$. The
notation $(i)$ in eq.~(3.21) stands for the embedding of $\cl$ into 
$K_x\times K_y$. 
\vglue.2in

\section{Examples}

Equations (3.19) and (3.20) reduce a calculation of eq.~(1.2) to a calculation
of the homology intersection indices. In turn, this can be easily done when
 instantons are labelled by lines in the intersection of quadrics. As was
demonstrated in sect.~3, there are the 128 holomorphic curves of the type
$point\times line$ and $line\times point$, and the 128 curves of the type
$line\times line$. The intersection of these curves with our homology basis
is given by eq.~(3.3), in agreement with eq.~(3.18). The 16 lines on Del Pezzo
are divided into three classes: one $G$, five $E_i^x$ that intersect with $G$,
and ten $F_{ij}^x$ (see {\sl Table 1}). Hence, each type of lines receives 
further classification according to its class. It is worth mentioning that the 
instanton contributions of the type $point\times line$ and $line\times point$
do not always coincide.

As our first (simple) example, let's consider the correlator given by
$$ \int H^+  \int H^+  \int H^+  \int H^+ =(H^+\cdot \cl)^4 \eqno(4.1)$$
over $\cl=G^x\times G^y$. Let's recall (sect.~3) that $H^+=H^x+H^y$, while
$$ \int_{G^x\times\{point\}} H^+=\int_{\{point\}\times G^y} H^+=1\eqno(4.2)$$
and
$$ \int_{G^x\times\{point\}} H^-=1~,\quad
\int_{\{point\}\times G^y} H^-=-1~,\quad
 \int_{G^x\times G^y} H^-=0~.\eqno(4.3)$$
Therefore, we obtain 
$$ (H^+\cdot \cl)^4= \[ (H^x+H^y)\cdot (G^x+G^y)\]^4=
 (H^x\cdot G^x + H^y\cdot G^y)^4=(1+1)^4=2^4~.\eqno(4.4)$$

Similarly, we find
$$ \int_{G^x\times G^y} H^-
  \int_{G^x\times G^y} H^-  \int_{G^x\times G^y} H^-  
\int_{G^x\times G^y} H^- =(H^-\cdot [G^x\times G^y])^4= $$
$$ =\left[ (H^x-H^y)\cdot (G^x\times G^y)\right]^4=
( H^x\cdot G^x - H^y\cdot G^y)^4=(1-1)^4=0~.\eqno(4.5)$$ 

In fact, any correlator containing the factor $\int H^-$ vanishes. 
Thus we see that some fermionic correlators are equal to zero despite of the 
fact that the discrete symmetries allow non-vanishing values for them.

As another (less trivial) example, let's consider the factor
$$ \int _{F_{ij}\times\{point\}}H^+=\left( [H^x+H^y]\cdot 
[\fracm{1}{3}(\sum E_i +H)-E_i-E_j]^x\right)=
\fracmm{5}{3}+\fracmm{4}{3}-2=1~.\eqno(4.6)$$
Similarly, we find
$$ \int _{\{point\}\times F^y_{ij}}H^+= \int_{F_{ij}^x\times\{point\}}H^+=
1~,\quad
\int _{F^x_{ij}\times F^y_{ij}}H^+=2~,\quad\eqno(4.7)$$
and
$$ \int _{F^x_{ij}\times\{point\}}H^-=
\int _{\{point\}\times F^y_{ij}}H^-=-1~,\quad
\int _{F^x_{ij}\times F^y_{ij}}H^-=0~.\eqno(4.8)$$

A more complicated example is given by
$$\eqalign{
\int_{G^x\times \{point\}} F^+_5 & \equiv  (F_5^+,G^x\times \{point\})=
\left( (F^x_5+F^y_5),(G^x\times \{point\})\right) \cr
& =(F^x_5\cdot G^x)=(H^x+6E^x_5 -2\sum_iE^x_i)\cdot G^x \cr
& = (H^x+6E^x_5-2\sum_iE^x_i)\cdot \,\fracmm{1}{3}(2H^x-\sum_iE^x_i)\cr
& = \fracmm{2}{3}H^x\cdot H^x +4H^x\cdot E^x_{5}-\fracmm{4}{3}\sum_i E^x_i
\cdot H^x \cr
& -\fracmm{1}{3}H^x\cdot\sum_iE^x_i -2E^x_{5}\cdot \sum E^x_i +
\fracmm{2}{3}\sum_i E^x_i\cdot \sum_jE^x_j \cr
& = \fracmm{8}{3} + 4 - \fracmm{20}{3} -\fracmm{5}{3}+2-\fracmm{10}{3}
=-3~.\cr}\eqno(4.9)$$
Similarly, we find
$$ \int_{\{point\}\times G^y} F^+_5=-3~,\eqno(4.10)$$
and, hence, 
$$ \left(\int_{G^x\times G^y} F^+_5\right)^4=(-3-3)^4=6^4~.\eqno(4.11)$$

We find, in addition,
$$ \int_{G^x\times \{point\}} F^-_i =\int_{G^x\times \{point\}} F^+_i=-3~,
\eqno(4.12)$$
and
$$ \int_{\{point\}\times G^y} F^-_i=-
\int_{\{point\}\times G^y} F^+_i=+3~.\eqno(4.13)$$

The similar contributions are given by
$$  \int_{G^x\times G^y} F^-_i= \int_{F^x_{ij}\times F^y_{ij}} F^-_i=0~,
\eqno(4.14)$$
$$\int_{\{point\}\times F^y_{ij}} F^-_i=-
\int_{F^x_{ij}\times\{point\}} F^-_i= -7+6\d_{ij}~,\eqno(4.15)$$
and
$$ \int_{F^x_{ij}\times F^y_{ij}} F^+_i=14-12\d_{ij}~.\eqno(4.16)$$

The rest of the integrals is given by
$$ \int_{F^x_{ij}\times G^y} F^+_i= \int_{G^x\times F^y_{ij}} F^+_i=
4-6\d_{ij}~,$$
$$\int_{\{point\}\times E^y_i} F^-_i=-\int_{\{point\}\times E^y_i} F^+_i
=-\int_{E^x_i\times \{point\}} F^-_i=-\int_{E^x_i\times \{point\}} F^+_i=3~,$$
$$\int_{E^x_i\times E^y_i}F^+_i=9~,\quad 
\int_{E^x_i\times E^y_i}F^-_i=0~,\quad
\int_{E^x_j\times E^y_i}F^-_i=-6+6\d_{ij}~,$$
$$\int_{E^x_j\times\{point\}}F^+_i=-\int_{\{point\}\times E^y_j}F^-_i=
3-6\d_{ij}~,\quad \int_{E_j^x\times E^y_i}F^-_i=6-6\d_{ij}~,$$
$$\int_{E^x_i\times\{point\}}H^+=\int_{\{point\}\times E^y_i} H^+=
\int_{E^x_i\times\{point\}}H^-=-\int_{\{point\}\times E^y_j} H^-=1~,$$
$$\int_{E^x_i\times E^y_j}H^+=\int_{E^x_i\times G^y}H^+=
\int_{G^x\times E^y_i}H^+=2~,$$
$$ \int_{G^x\times F^y_{ij}}H^+=\int_{F^x_{ij}\times E^y_i}H^+=2~,$$
$$\int_{G^x\times F^y_{ij}}H^-=\int_{E^x_i\times F^y_{ij}}H^-=
\int_{F^x_{ij}\times E^y_{i}}H^-=0~.\eqno(4.17)$$

The fermionic correlators in eq.~(1.2) are given by various products of four 
factors calculated above. 

\section{Conclusion} 

It is surprising, from the mathematical viewpoint, that the fermionic 
correlators (1.2) are entirely determined by topology so that they can be 
explicitly calculated. Their physical significance is yet to be understood. At
the very least, however, all the fermionic correlators (1.2) vanish in the 
(classical) tree approximation by index theorems \cite{itep}, so that the 
instanton corrections obtained are actually the leading contributions to these
 correlators. This leads to the highly non-trivial selection rules for the 
physical processes described by the fermionic correlators in the CY 
compactified type-II strings. For example, some correlators exactly vanish 
(sect.~4) even though the discrete symmetries of CY allow non-vanishing values
 for them. 

Though our geometrical approach is similar to the one used earlier for other  
 string models in refs.~\cite{ww,dgkm}, there are also some conceptual
differences. We find it simpler to consider instantons in the simply connected
 covering space (CY) manifold $K_0$ of $K$, instead of the CY space $K$. Each 
instanton in $K$ has four representatives in $K_0$, which are all equivalent 
as regards the $G$-invariant real quantities.

We merely discussed the {\it one}-instanton corrections to the four-fermion 
correlators. One may expect the existence of the {\it multi}-instanton 
corrections from the maps of higher degree (more than one). Unfortunately, the
 status of multi-instantons in the context of type-IIA superstrings is not 
quite clear \cite{bbs} (see, however, ref.~\cite{din}). 

One might also think that the intersection of Del Pezzo surfaces is the very 
special case of the type-IIA string/M-Theory compactification. In fact, as was
 recently noticed in ref.~\cite{inv}, there is a non-trivial duality between 
toroidal compactifications of M-Theory and del Pezzo surfaces. According to 
ref.~\cite{inv}, a group of the classical symmetries of Del Pezzo (i.e. the
global diffeomorphisms preserving the canonical class of Del Pezzo) 
corresponds to the U-duality symmetries of the toroidally compactified
 M-Theory. The M-Theory (BPS) branes are mapped under this `mysterious duality'
to rational curves on Del Pezzo, so that the electric-magnetic duality of
M-Theory receives a nice geometrical description in terms of the Del Pezzo 
surfaces \cite{inv}. In particular, the bound states of the $(1/2)$-BPS branes
in M-Theory can be related to the intersections of spheres in Del Pezzo 
\cite{inv}. Further developments of this new duality require counting 
intersections of the holomorphic curves of higher genus in CY \cite{penn}. 

In the context of Horava-Witten theory \cite{hw}, similar calculations of 
instanton corrections are needed when one considers a torus-fibered CY 
threefold ${\cal Z}$  over the Del Pezzo base, with the non-trivial first 
homotopy group $\p_1({\cal Z})={\bf Z}_2$. When a gauge vacuum on the hidden 
 brane is trivial, the threefold ${\cal Z}$ admits three families of the 
semi-stable holomorphic vector bundles associated with an N=1 supersymmetric 
gauge theory having three (chiral) quark-lepton familes and the GUT group 
$SU(5)$ in the observable brane \cite{penn}. Both five-branes in this 
Horawa-Witten type construction are wrapped about holomorphic curves in  
${\cal Z}$ whose homology classes are exactly calculable. 

Our investigation is also relevant for studying the supersymmetric 
Pati-Salam-type models from intersecting  D-branes (see, e.g.,
ref.~\cite{phe}), and the non-perturbative flipped $SU(5)$ vacua in heterotic
M-Theory \cite{ovrutl,far}. 

\section*{Acknowledgements}

The authors would like to thank R. Manvelyan, A.S. Tichomirov and A.K. Tsikh 
for useful discussions and suggestions. 
\vglue.2in

\end{document}

%%%%%%%%%%%%%%%%%%%%%%%%%%%%%%%%%%%%%%%%%%%%%%%%%%%%%%%%%%%%%%%%%%%%%%%%%%%%%%